\documentclass[]{aa} 
\usepackage{graphicx}
\usepackage{subfigure}
\graphicspath{ {images/} }

\usepackage{txfonts}
\usepackage{booktabs}
\usepackage{amsmath}
\usepackage{epstopdf}
\usepackage{enumitem}
\usepackage{multicol}
\usepackage{multirow}
\usepackage{chngcntr}
\usepackage{chngcntr}
\usepackage{soul}
\usepackage{color}
\usepackage{epstopdf}
\usepackage{ulem}
\newcommand{\kms}{km s$^{-1}$\xspace}
\newcommand{\HI}{{\rm H\,{\scriptsize I}}\xspace}

\usepackage{natbib}
\usepackage[colorlinks, citecolor={blue}]{hyperref}

\begin{document}

\title{GMRT observations of OHM candidates from the ALFALFA survey }

   \author{Shouzhi Wang
          \inst{1}
          \and
          Zhongzu Wu\inst{1}\fnmsep\thanks{zzwu08@gmail.com}
           \and
          Bo Zhang\inst{2}
           \and
          Yu.~Sotnikova\inst{3,4}
         \and
          T.~Mufakharov\inst{3,4}
          \and
         Zhiqiang Shen\inst{2}
          \and
         Yongjun Chen\inst{2}
         \and
         Jianfeng Wu\inst{1}
        }

   \institute{College of Physics, Guizhou University, 550025 Guiyang, PR China \email{zzwu08@gmail.com}
              \and Shanghai Astronomical Observatory,
Chinese Academy of Sciences, 80 Nandan Road, Shanghai 200030, PR China               \and
              Special Astrophysical Observatory of RAS, Nizhny Arkhyz 369167, Russia
        \and
       \textbf{ Kazan Federal University, 18 Kremlyovskaya St, Kazan 420008, Russia}
         }
   \date{  }


  \abstract
   {We present the results of our observations using the Giant Meterwave Radio Telescope (GMRT) to investigate the radio continuum and OH line emission of 10 OHM candidates from the Arecibo Legacy Fast ALFA (ALFALFA) survey. Among these candidates, we have identified two sources, AGC115713 and AGC249507, which display compact OH line emission that are spatially associated with radio continuum emission. These characteristics align with the typical properties of OHM galaxies. Furthermore, the infrared (IR) properties of these two galaxies are consistent with those of known OHM galaxies. 
   Of the two sources, AGC 249507 has been confirmed through optical redshift, whereas AGC 115713 meets a WISE color selection criterion in the literature, providing additional support for this source being an OHM galaxy rather than a nearby \HI galaxy.
   On the contrary, no significant spectral line emission were detected in the remaining eight OHM candidates using our full GMRT dataset. This suggests that the spectral line emission initially detected by the ALFALFA survey may have been significantly resolved in our high-resolution observations. Additionally, the absence of radio continuum emission in 6 candidates also distinguishes them from known OHM galaxies documented in the literature.  These findings support the notion that OHM emission may be distributed on a subarcsecond scale, underscoring the utility of arcsecond-scale observations in confirming OHM candidates, particularly those lacking optical redshift data.

  }


  \authorrunning{Wang et al. }            
   \titlerunning{GMRT observations of OHM candidates}  
   \maketitle
%
\section{Introduction}
OH megamasers (OHMs) are a distinctive class of extragalactic maser sources that emit non-thermal radiation from hydroxyl (OH) molecules, with two main lines at 1665/1667 MHz and two satellite lines at 1612/1720 MHz \citep{2013ApJ...774...35M}. These OH maser emissions are believed to be stimulated by infrared radiation from their surroundings, with the OH line luminosity being closely tied to the far-infrared luminosity \citep{1985Natur.315...26B,2008ApJ...677..985L}. Consequently, OHMs have typically been identified through targeted observations of luminous infrared galaxies (LIRGs, $\rm L_{FIR}$ $>$ $10^{11}$ $\rm L_{\odot}$) or ultraluminous infrared galaxies (ULIRGs, $\rm L_{FIR}$ $>$ $\rm 10^{12}$ $L_{\odot}$) \citep[see][]{1985ApJ...298L..51B,2000AJ....119.3003D,2002AJ....124..100D,2012IAUS..287..345W,2011ApJS..193...18W,2014A&A...570A.110Z}.

High-sensitivity \HI surveys, such as the Arecibo Legacy Fast Arecibo L-band Feed Array (ALFALFA) survey \citep{2018ApJ...861...49H}, surveys conducted with the Five-hundred-meter Aperture Spherical radio Telescope (FAST) \citep{2019RAA....19...22Z}, and those involving the Square Kilometer Array (SKA) and its precursors \citep[see][and reference therein]{2021ApJ...911...38R}, have emerged as valuable tools for blind surveys aimed at detecting new OHMs. In these untargeted emission line surveys searching for neutral hydrogen, the redshifted OH lines at $\rm z_{OH}$ can mimic an \HI line with a redshift $\rm z_{\HI}$ if $\rm \frac{\nu_{\HI}}{1+z_{HI}}$=$\rm \frac{\nu_{OH}}{1+z_{OH}}$, where $\rm \nu_{HI}$ =1420.4 MHz and $\rm \nu_{OH}$=1667.4 MHz \citep{2021ApJ...911...38R}. OHMs discovered through these blind surveys offer opportunities to validate the Infrared (IR) luminosity selection criteria employed in prior targeted surveys and to explore new environments conducive to OHM production \citep{2016MNRAS.459..220S}.

The Arecibo ALFALFA survey represents the first endeavor to detect a sample of OHMs through a blind survey. This survey has reported nine confirmed OH megamasers and 10 OHM candidates with no optical redshift but a potential redshift range of 0.16 to 0.22 \citep[see][]{2018ApJ...861...49H}. Optical spectroscopy is a primary method for confirming new OHMs. However, distinguishing between \HI and OH in galaxies lacking spectroscopic redshifts poses challenges. Researchers such as \citet{2021ApJ...911...38R} and \cite{2016MNRAS.459..220S} have demonstrated that \textbf{the wide-field infrared survey Explorer (WISE)} magnitudes and colors can serve as a means of distinguishing between \HI and OHM host populations in the absence of spectroscopic redshifts. Moreover, \cite{2023A&A...669A.148W} have conducted investigations into the arcsecond-scale radio continuum and OH line emission of a sample of known OHM galaxies with z $>$ 0.15 and have found that OH line emissions in known OHM galaxies exhibit compactness in arcsecond-scale \textbf{Very Large Array (VLA)} observations, often coinciding with radio continuum emission. These findings are consistent with the characteristics of low-redshift OHM galaxies documented in the literature \citep[see][and references therein]{2005ARA&A..43..625L}.  In contrast, neutral hydrogen gases in galaxies are distributed on a wide range of scales, from sub-kpc to hundreds of kpc \citep{1999MNRAS.304..475M}. Therefore, \textbf{detection of \HI emission in nearby galaxies depends on brightness sensitivity in current radio interferometries}, even with long integrations using the VLA in 5 arcsec resolution \citep{2005dmgp.book..161P}.
As a result, the distinctive properties of OH and \HI line emission at arcsecond-scale observations support one hypothesis that arcsecond-scale spectral line emission observations can be utilized to identify new OHMs discovered in blind \HI surveys.

This study presents the results of observations conducted with the Giant Meterwave Radio Telescope (GMRT) involving 10 OHM candidates from the ALFALFA survey \citep{2018ApJ...861...49H}. The primary objective of this research is to explore the characteristics of radio continuum and OH line emission on the arcsecond scale for these galaxies. Through this analysis, we aim to identify potential differences compared to known OHM galaxies and potentially confirm new OHM candidates. This endeavor also aims to validate the IR luminosity selection criteria used in previous targeted surveys and potentially uncover new environments conducive to OHM production \citep{2016MNRAS.459..220S}. We provide details on sample selection, observations, and data reduction in Section 2. The results, discussions, and conclusions are presented in Sections 3 to 5. We have adopted a Hubble constant of $H_0$=70 \kms $Mpc^{-1}$, with $\Omega _{m}=0.3$ and $\Omega _{\Lambda}=0.7$, for calculating the luminosity distance of the targets in this study.

\section{Observations and data reduction}
\label{sect:Obs}
\subsection{GMRT observations}
 
We conducted a total of 15 hours of GMRT observations involving 10 OHM candidates selected from the ALFALFA survey (see Table \ref{table1}). We observed standard flux density calibrators (3C48, 3C147, and 3C286) every hour to correct for variations in amplitude and bandpass. These observational scans (each lasting about 6 minutes) are used for flux density, delay, and bandpass calibrations. Each target source was observed for about 50-70 minutes, and \textbf{a compact radio source was utilized as a phase calibrator, being scanned typically for six minutes before and after observations of each target source} (see Table \ref{table1}).The images were created with a resolution of approximately $\sim$2-3 arcseconds using the full GMRT array and 30-40 arcseconds with the compact central array (see Table \ref{tablea1}). No self-calibration was performed due to the low signal-to-noise ratio of both radio continuum and line emission images. 

These observations were divided into three sessions carried out on different dates, specifically on January 17, 20, and February 6, 2023. For this observation, we utilized band 5 of the GMRT, providing a total bandwidth of 400 MHz, spanning from 1060 to 1460 MHz. This entire bandwidth was further divided into 16,384 channels,which resulted in a frequency resolution of 24.414 kHz (6\kms). 
The selected frequency band allowed us to investigate a wide range of aspects, including broadband radio continuum emission and the redshifted OH lines at 1667 and 1665 MHz, as well as potential HI lines.
However, approximately 234 MHz of the \textbf{spectra}, particularly within the range from 1060 to 1294 MHz, was heavily impacted by Radio Frequency Interference (RFI), we were unable to study the potiential \HI emission of these galaxies. Consequently, for this study, we focused solely on the remaining 166 MHz band, spanning from 1294 MHz to 1460 MHz, which was relatively free from RFI interference, enabling us to analyze the radio continuum and potential OH lines effectively.

\subsection{The arcsecond-scale \HI 21-cm line emission of nearby galaxies}
\label{samplehi}

Generally, \textbf{OH masing regions are believed to be associated with star formation}, and the occurrence of OH megamaser activity is also correlated with IR colors \citep{2021A&ARv..29....2P}. Thus, OHM candidates linked to IR emissions are likely to demonstrate a high probability of OH line emission. At the same time, the detected line emission associated with IR emission could also originate from \HI emission in nearby star-forming galaxies or \textbf{(U)LIRGs}. Therefore, it is crucial to compare the arcsecond-scale properties of \HI emission in nearby star-forming galaxies and (U)LIRGs to distinguish OH megamasers from \HI line emission. 

In a study conducted by \cite{2015ApJ...805...31L}, a set of star-forming galaxies featuring accessible interferometric observations of \HI emission was assembled. Utilizing this sample, we gathered the integrated line flux density and \HI mass from single-dish observations in the literature (see Table \ref{tablea2}). Additionally, we computed the \HI mass within the central 2 kpc of the nuclei, relying on the provided \HI surface densities of these galaxies by \cite{2015ApJ...805...31L}.
However, there is a shortage of available \HI emission data for nearby (U)LIRGs due to the small atomic gas fraction, complex morphologies, and frequent observation of strong HI absorption features \citep[see][]{2015ApJ...805...31L}. To address this gap, we compiled VLA-A/B and GMRT data for the (U)LIRGs in the Great Observatories All-sky LIRG Survey (GOALS) sample\footnote{https://goals.ipac.caltech.edu/}. The GOALS sample, comprising 202 (U)LIRGs at z<0.088, is a well-defined collection for enhanced infrared emission in nearby galaxies \citep{2009PASP..121..559A}. Approximately 40 sources in the GOALS sample with archived GMRT or VLA A/B data, exhibiting similar resolution to our current observations, were identified. Subsequently, we searched for single-dish \HI observations of these 40 (U)LIRGs in the literature. Twelve sources were found to have \HI absorption spectra from single-dish observations. These 12 sources were then excluded, and the remaining 28 sources, listed in Table \ref{table4}, were selected for the study of arcsecond-scale \HI emission. The \HI mass and integrated line flux of galaxies in this work have been derived from each other using the available parameters, applying the equation from \cite{2015A&ARv..24....1G}:$M_{\HI}/M_{\odot}=2.356\times10^{5}S_{\HI}D^2$/(1+z), where $S_{\HI}$ is the integrated flux in \textbf{Jy \kms} and D is the luminosity distance in Mpc, z is the redshift. 


\setlength{\tabcolsep}{0.05in}
  \begin{table*}
       \caption{\textbf{Observation parameters}.} 
     \label{table1}
  \centering
  \resizebox{\linewidth}{!}{
  \begin{tabular}{c c c c c c c c c c c}     
  \hline\hline
  AGC Name & on-source&Phase calibrator&  peak. Coords&major& minor& PA&$F_{INT}$& $F_{peak}$ &rms&$L_{1.4 GHz} $ \\
    &(min)&  &(J2000) & (")&(") &($\circ$) &(mJy) & (mJy/beam)&(mJy/beam)& (W $Hz^{-1}$ )\\
    \hline
     102708& 62 & 2340+266&-&-&- &- & -&<0.066&0.022&<21.7\\
     102850&62 & 0040+331&-&-&- &- & -&<0.096&0.032&<21.9\\
      114732& 50& 0119+084&-&- &- &- &-&<0.084&0.028&<21.8\\
115713&50 & 0141+138&014135.19+165730.51&1.6$\pm$0.3 & 1.2$\pm$0.3&32&1.7$\pm$0.1&1.25&0.055&23.1\\
      115018&62 & 0149+059&- &-&- &-&-&<0.078&0.026&<22.0\\
     124351& 57& 0224+069&-&- &- &-&-&<0.087&0.029&<21.9\\
      749309SE& 69& 1013+248&101103.21+273924.36 &5.2$\pm$0.2& 0.5$\pm$0.4&173&2.9$\pm$0.1&1.09&0.058&23.5\\
      749309NW& ...&...&101100.41+274036.52&3.4$\pm$0.4 & 3.2$\pm$0.4&132&1.2$\pm$0.1&0.44&0.058&23.1\\
     219835 & 70& 1156+314& 113033.97+322149.19&- &-&-&364$\pm$22&80.8&0.293&25.3\\
      249507& 65 &1407+284&140340.33+295456.48 & 2.1$\pm$0.1&1.4$\pm$0.1&50&3.0$\pm$0.1&1.86&0.025&23.4\\
     322231& 50& 2232+117&-&- &-&- &-&<0.090&0.030&<21.8\\

   \hline
   \end{tabular}}
   
   \vskip 0.1 true cm \noindent Notes: \textbf{Col. 1}: the AGC name given by \cite{2018ApJ...861...49H}. Cols. 2-3: on-source time and phase calibrator for this project. Cols. 4-9: the fitted parameters of the radio continuum emission including the peak coordinates, Gaussian component size, peak, and integrated flux densities. Col. 10: the 1 $\sigma$ noise level. Col. 11: the radio luminosity of the radio continuum emission. 
   The redshift for AGC 249507 is sourced from the LAMOST survey, while the redshift for other galaxies was derived based on the peak frequency of the potential OH 1667 MHz line as reported in \cite{2018ApJ...861...49H}.
   
   \end{table*}
   

   \setlength{\tabcolsep}{0.05in}
  \begin{table*}
       \caption{OHM detected in our GMRT observations.  }
     \label{table2}
  \centering
  \resizebox{\linewidth}{!}{
  \begin{tabular}{c c c c c c c c c c c c c}     
  \hline\hline
   AGC Name & OH Position & major & minor & PA&$V_{peak}$ &$F_{peak}$ & FWHM   & $rms_{image}$  &$rms_{line}$ &  $\int$ S dV& OH luminosity\\
    &   (J2000)  & (")&(") &($\circ$) & (\kms) &   (mJy)              & (\kms)    & (mJy)    &(mJy)  & (Jy \kms)&(log $\textit{L}_\odot)$\\
   \hline
     115713 &014135.19+165730.73 &<1.3& <0.7&-&51065.2&14.39&186.7&1.03&1.70&1.66&3.15\\
     249507&140340.34+295456.39&1.24$\pm$0.23& 0.55$\pm$0.39&6.5 &53421.4&8.26&203.4&0.79&0.92&1.35&3.06\\
     
   \hline
   \end{tabular}}
   
   \vskip 0.1 true cm \noindent Notes: Cols. 2-5: the Gaussian fitted peak position and component size of the OH  emission from the image plane. Cols. 6-8: The fitted parameters of the line profile. 
   Cols. 9-10: the noise level from the image plane with a frequency width of about 24 kHz ($\sim$ 6 \kms) and the noise level from our extracted line as shown in Fig. \ref{OH}. Cols. 11-12: the integrated line flux densities and luminosities. 
   We have used the identical redshift for calculating the luminosity, as detailed in Table \ref{table1}.
   
   \end{table*}


\setlength{\tabcolsep}{0.05in}
  \begin{table*}
       \caption{Infrared properties of OHM candidates.}
     \label{table3}
  \centering
  \begin{tabular}{c c c c c c c c c}     
  \hline\hline
AGC Name &WISE Coords&WISE[3.4]&WISE[4.6]& WISE[12]&WISE[22]&$F_{60}$&$F_{100}$&$L_{FIR}$\\
    &&(mag)&(mag)&(mag)&(mag)&(Jy)&(Jy)&(log $\textit{L}_\odot$)\\
    
    \hline
    102708&J000336.00+253202.7
&17.78&16.40&12.65&9.06&0.015&0.048&10.35\\
     102850&J002958.27+305833.9
&14.24&14.01&10.17&8.08&0.018&0.034&10.34\\
      114732&J010107.16+094624.3
&16.61&16.12&11.83&8.69&0.019&0.054&10.43\\
     115713&J014135.21+165730.6
&14.73&14.04&9.64&6.90&0.109&0.308&11.19\\
      115018&J015847.20+073202.6
&16.60&16.71&12.64&8.79&0.012&0.035&10.41\\
     124351&J021750.86+072428.7
&16.84&16.48&12.73&9.26&0.007&0.019&10.0\\
      749309SE&J101103.18+273925.7
&17.25&16.58&11.84&9.06&0.019&0.060&10.58\\
 749309NW&J101100.37+274036.6
&14.39&14.10&12.02&8.73&0.003&0.005&9.63\\
     219835 &J113033.78+322139.3&14.89&14.50&11.97&8.70&0.005&0.009&9.91\\
    249507&J140340.36+295456.6
&14.84&14.24&10.33&7.85&0.29$^{*}$&0.70$^{*}$&11.63$^{*}$\\
     322231&J223605.38+095726.2
&14.66&14.35&12.38&8.71&0.004&0.006&9.62\\

   \hline
   \end{tabular}
   \vskip 0.1 true cm \noindent Notes: 
\textbf{Cols. 2-6: the WISE coordinate and four-band data}. \textbf{Cols. 7-8}: the far infrared flux at 60 and 100 $\mu$m band. The "*" stands for the infrared flux from the IRAS survey, others are derived from WISE data from cols. 3-6.  \textbf{Col. 9}: the integrated far-infrared luminosity using the method from \cite{2006A&A...449..559B} (see text for details). We have used the identical redshift for calculating the luminosity, as detailed in Table \ref{table1}.
   \end{table*}


\begin{figure*}
   \centering
   \includegraphics[width=0.49\textwidth]{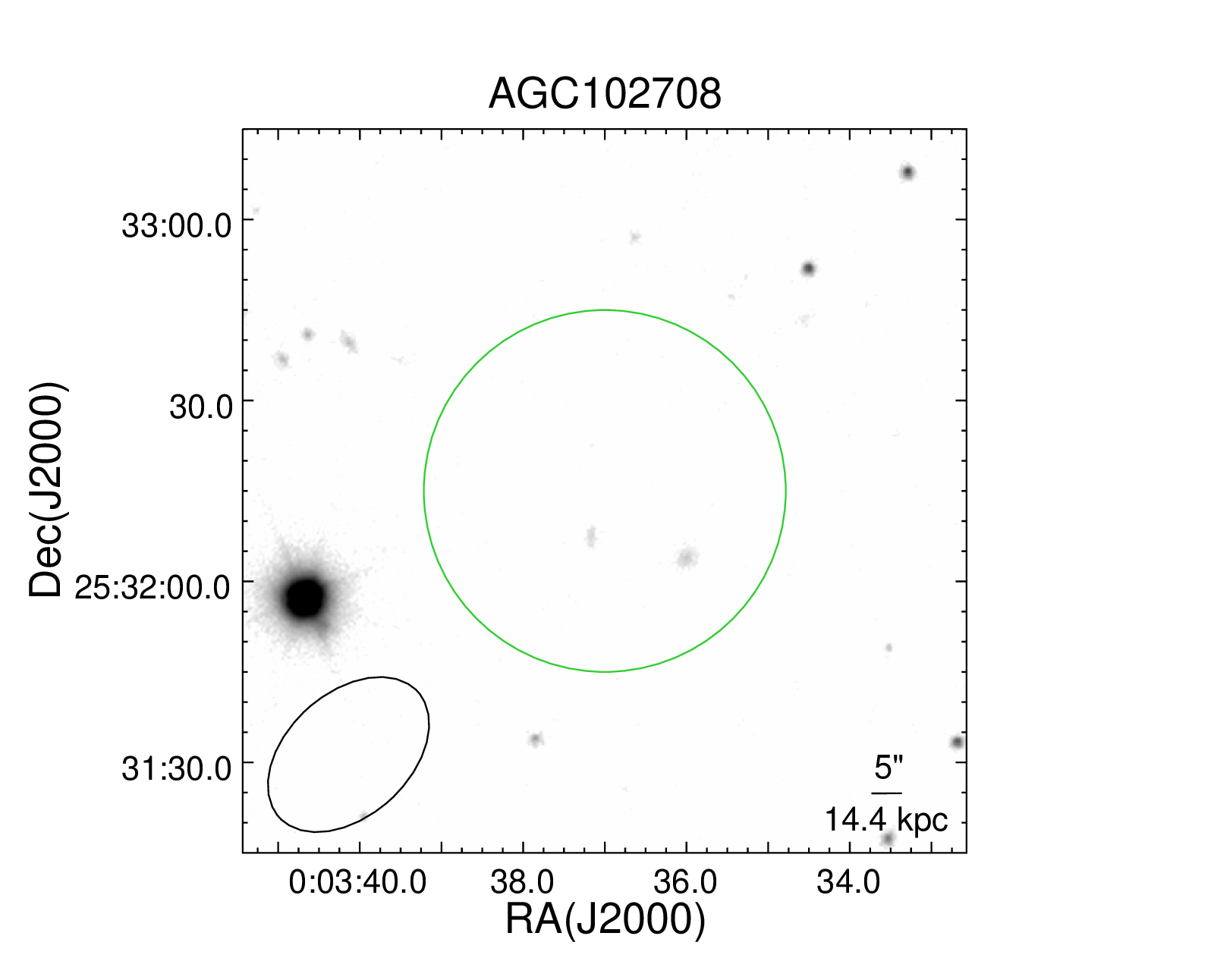}
   \includegraphics[width=0.49\textwidth]{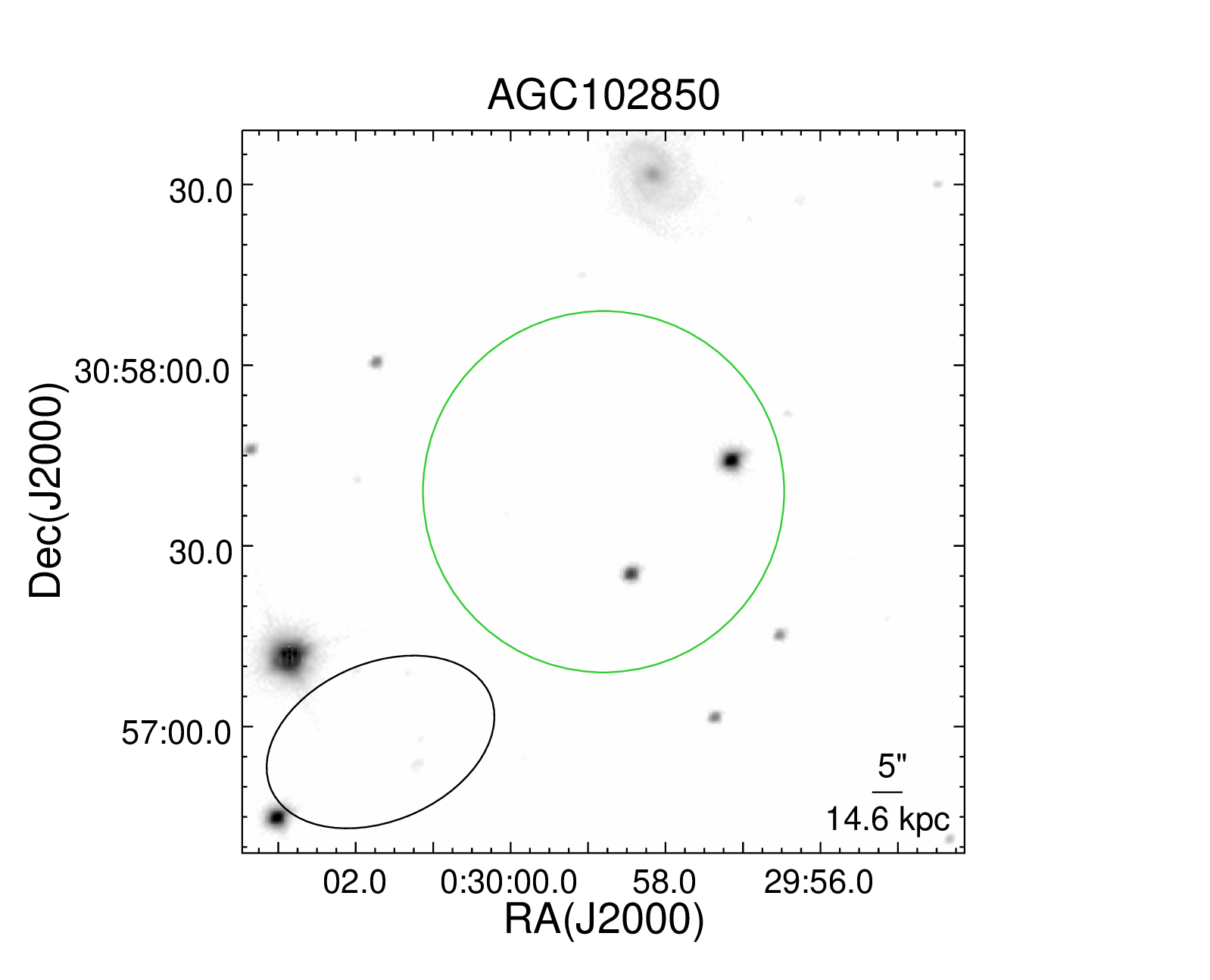}
    \includegraphics[width=0.49\textwidth]{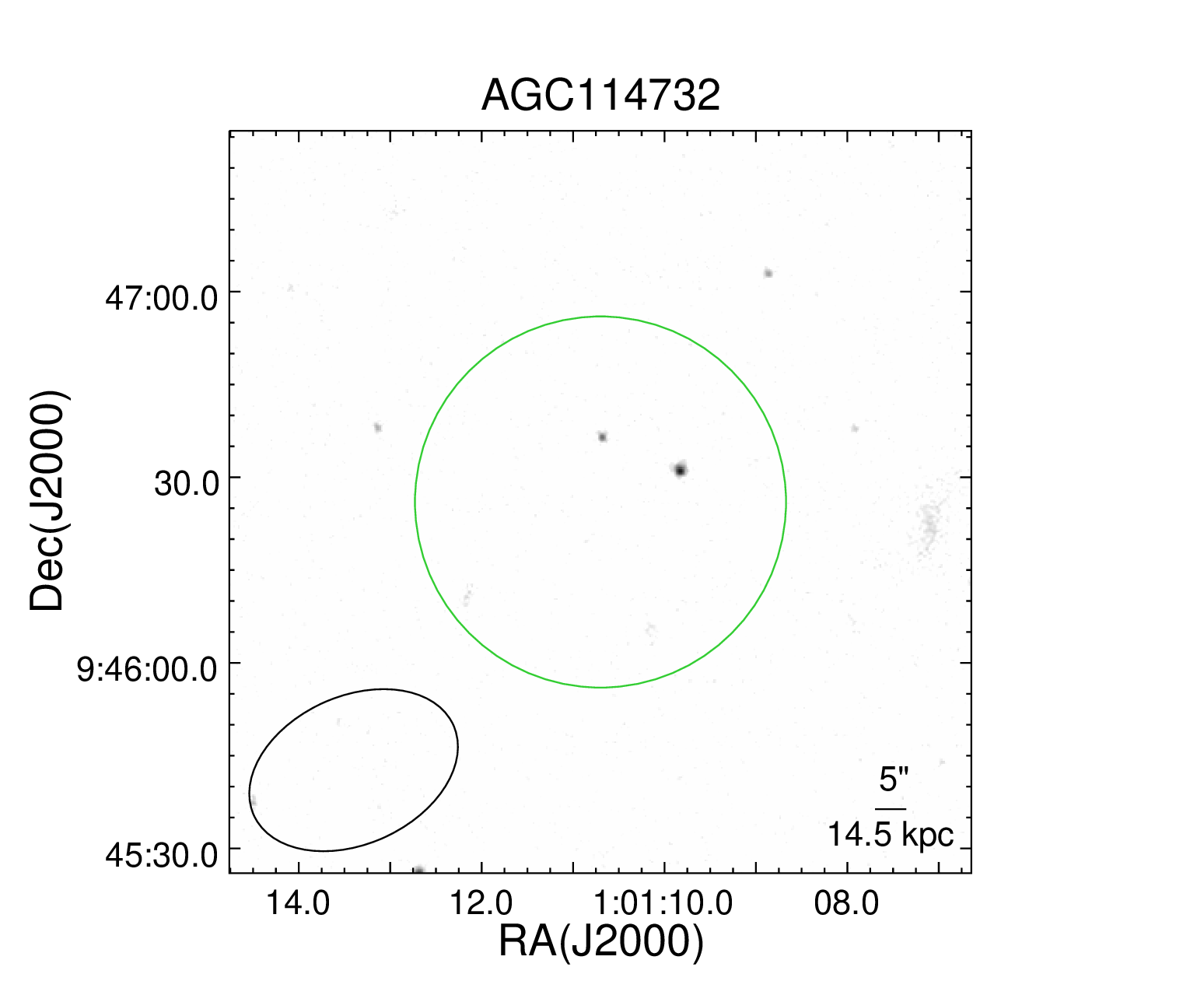}
    \includegraphics[width=0.49\textwidth]{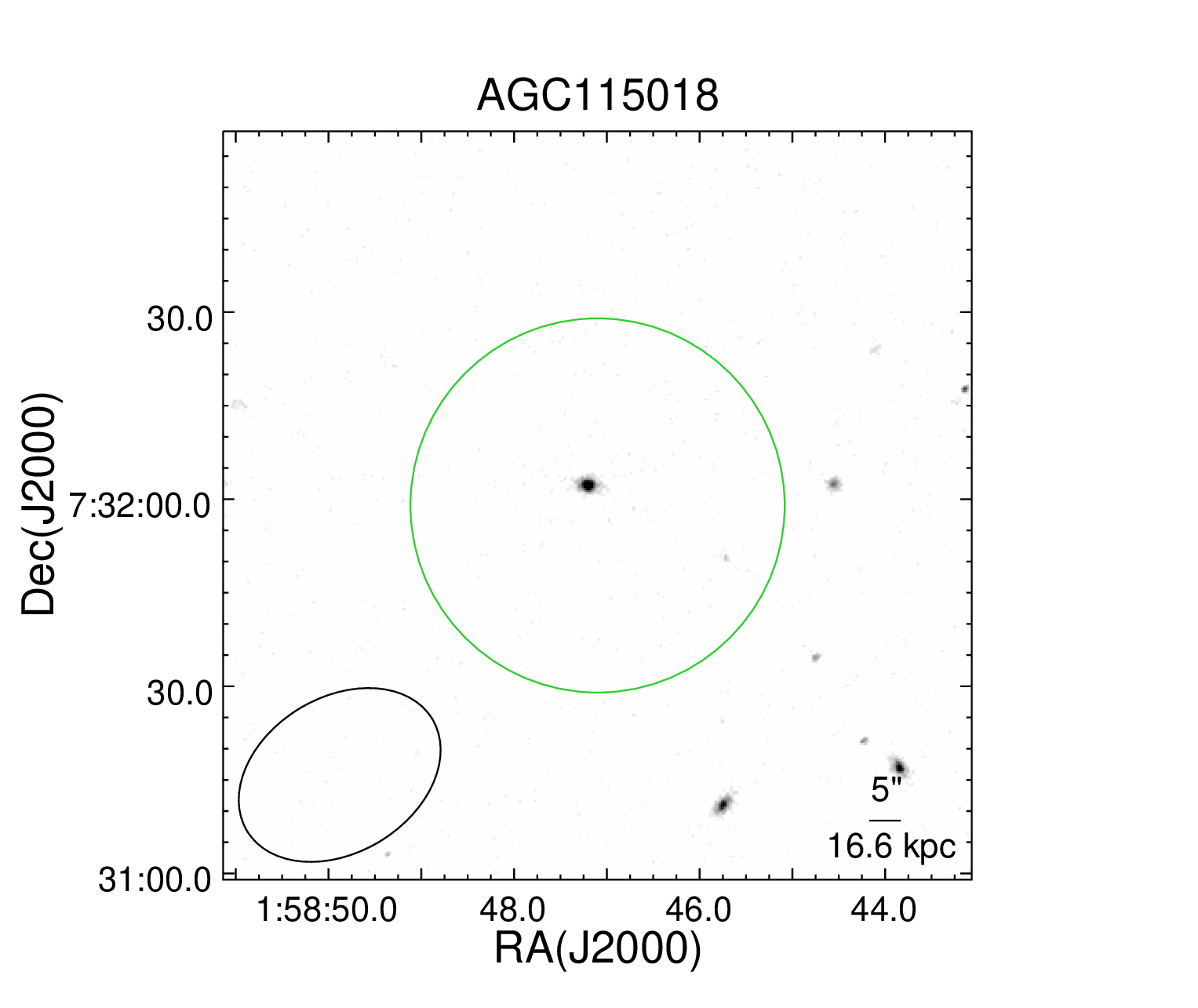}
    \includegraphics[width=0.49\textwidth]{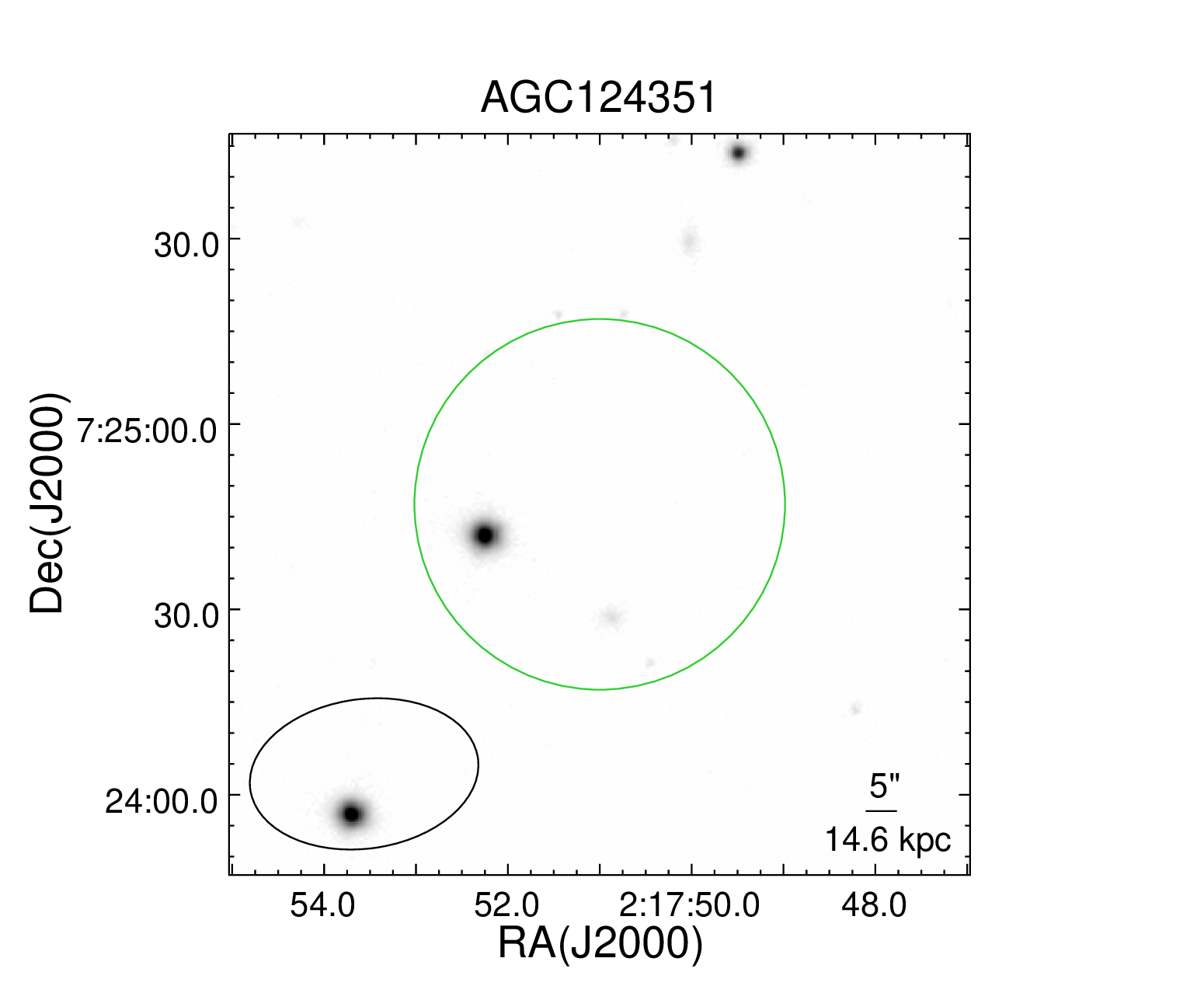}
    \includegraphics[width=0.49\textwidth]{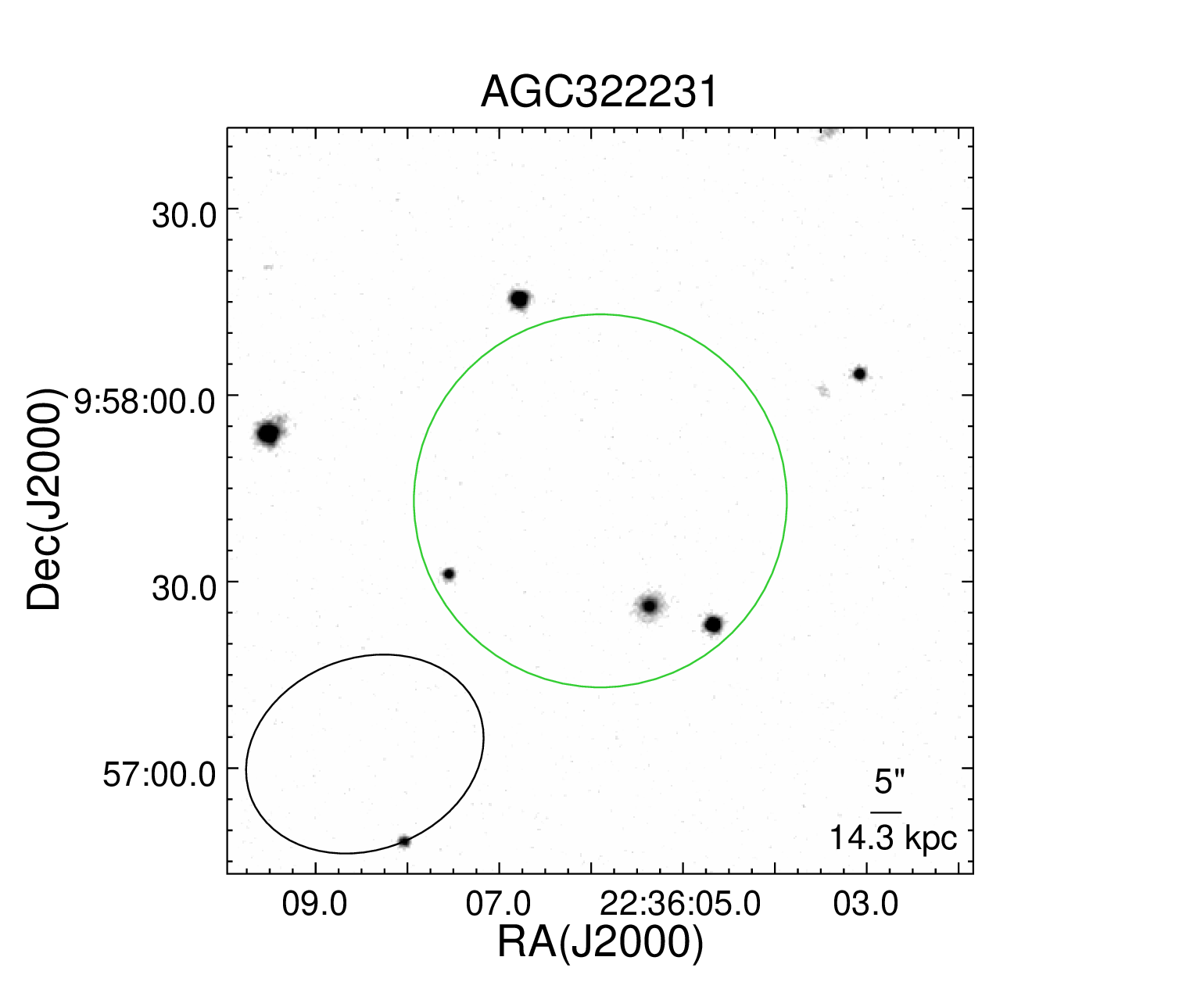}
    
\caption{OH line and radio continuum emission overlaid on SDSS R-band grey image. Green: The circular region for generating the emission of the OH line from GMRT-central data,the beam size of these sources and the parameters are listed in Table \ref{tablea1}. Magenta: The contour of OH line emission. The velocities range of the OH line emission: AGC115713: 50884.8-51155.4 \kms (see Fig. \ref{OH}), the contour levels are 3$\sigma$ (0.001 Jy/beam) $\times$ (1 2 4 ...).
AGC 249507: 53336.1-53665.1 \kms, the contour levels are 3$\sigma$ (0.000592 Jy/beam) $\times$ (1 2 4 ...). 
Blue: The L-band radio continuum emission and the parameters are listed in Table \ref{table1}. 
      }
      \label{contour}
\end{figure*}


\addtocounter{figure}{-1}

\begin{figure*}
   \centering
   \includegraphics[width=0.49\textwidth]{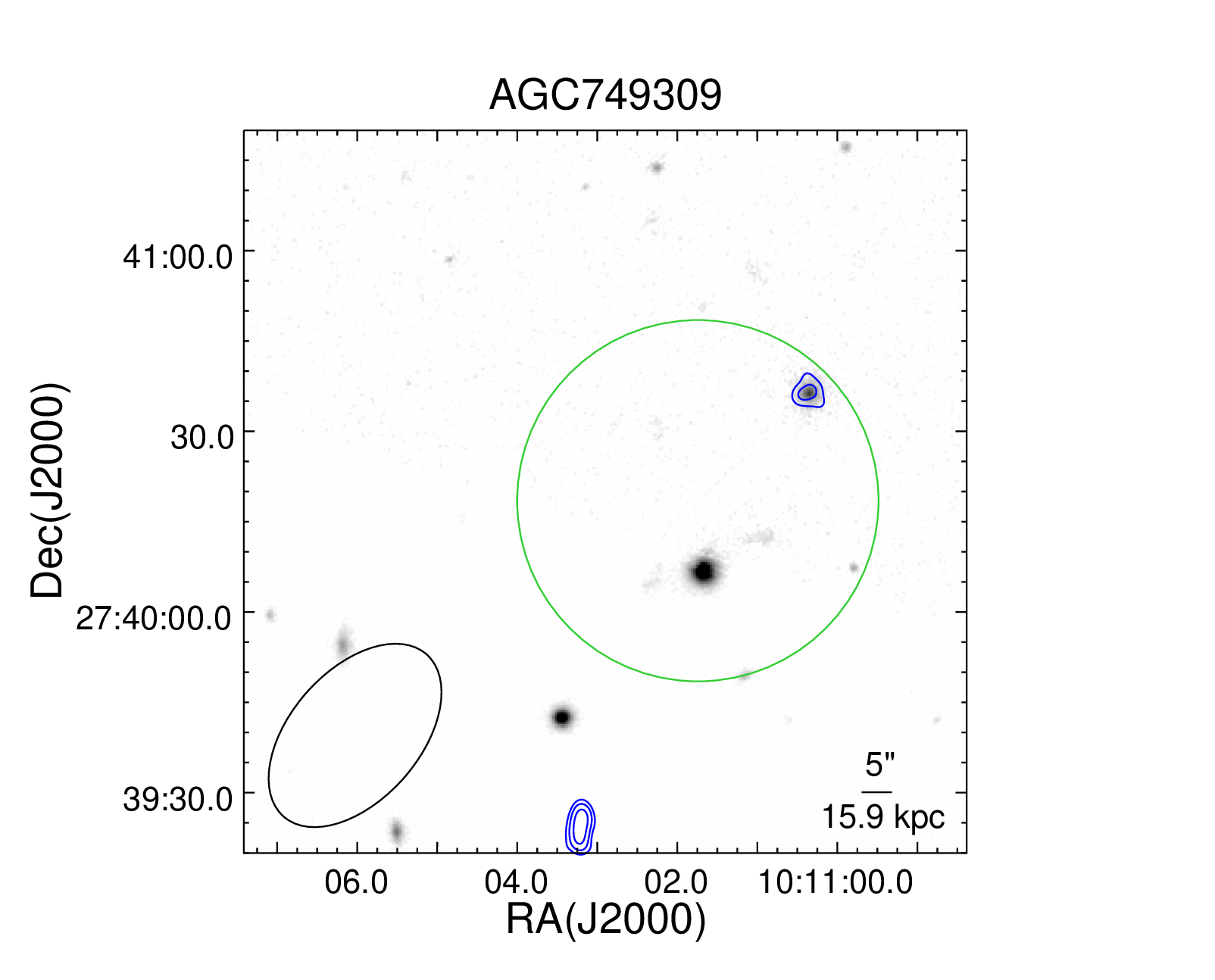}
    \includegraphics[width=0.49\textwidth]{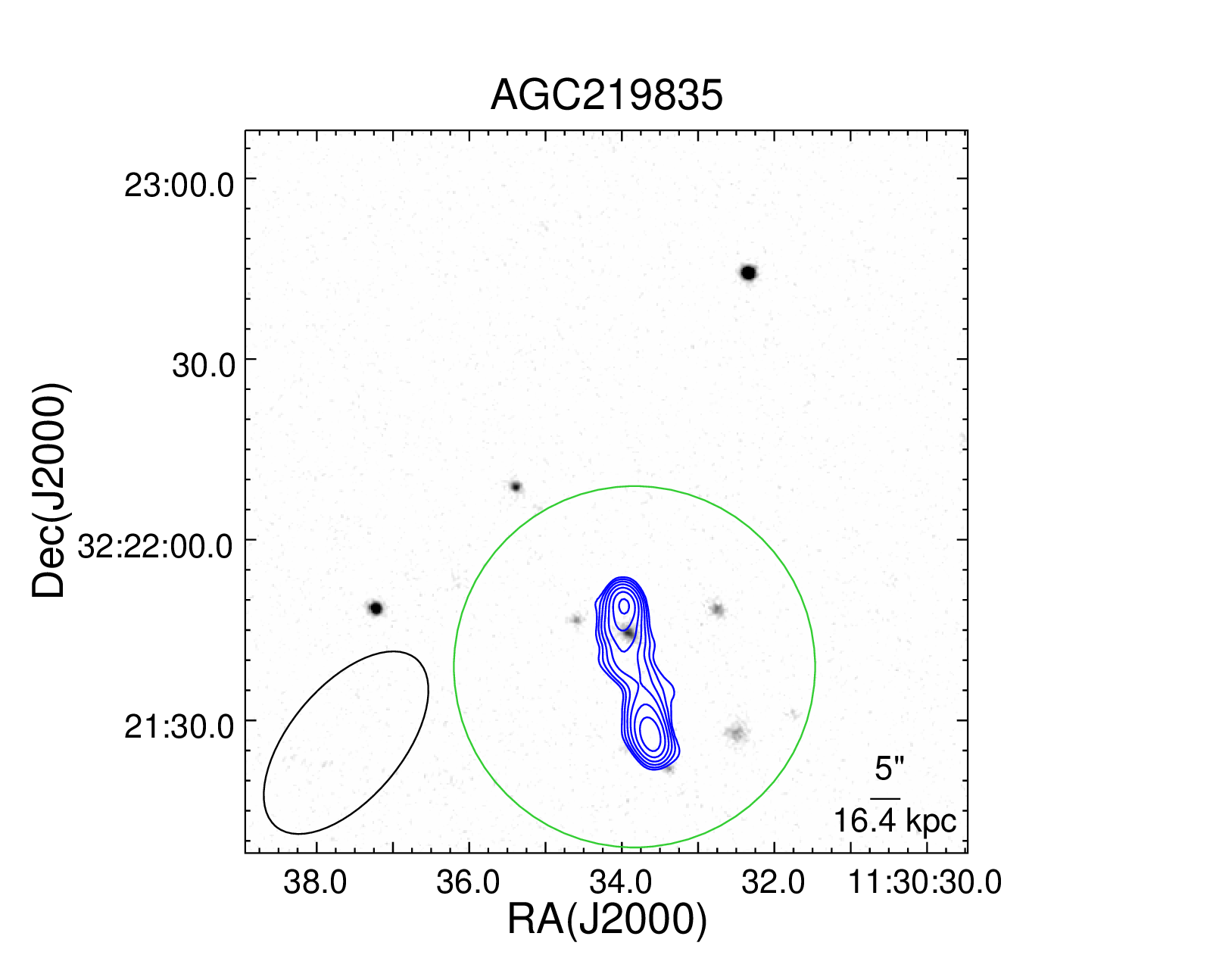}
   \includegraphics[width=0.49\textwidth]{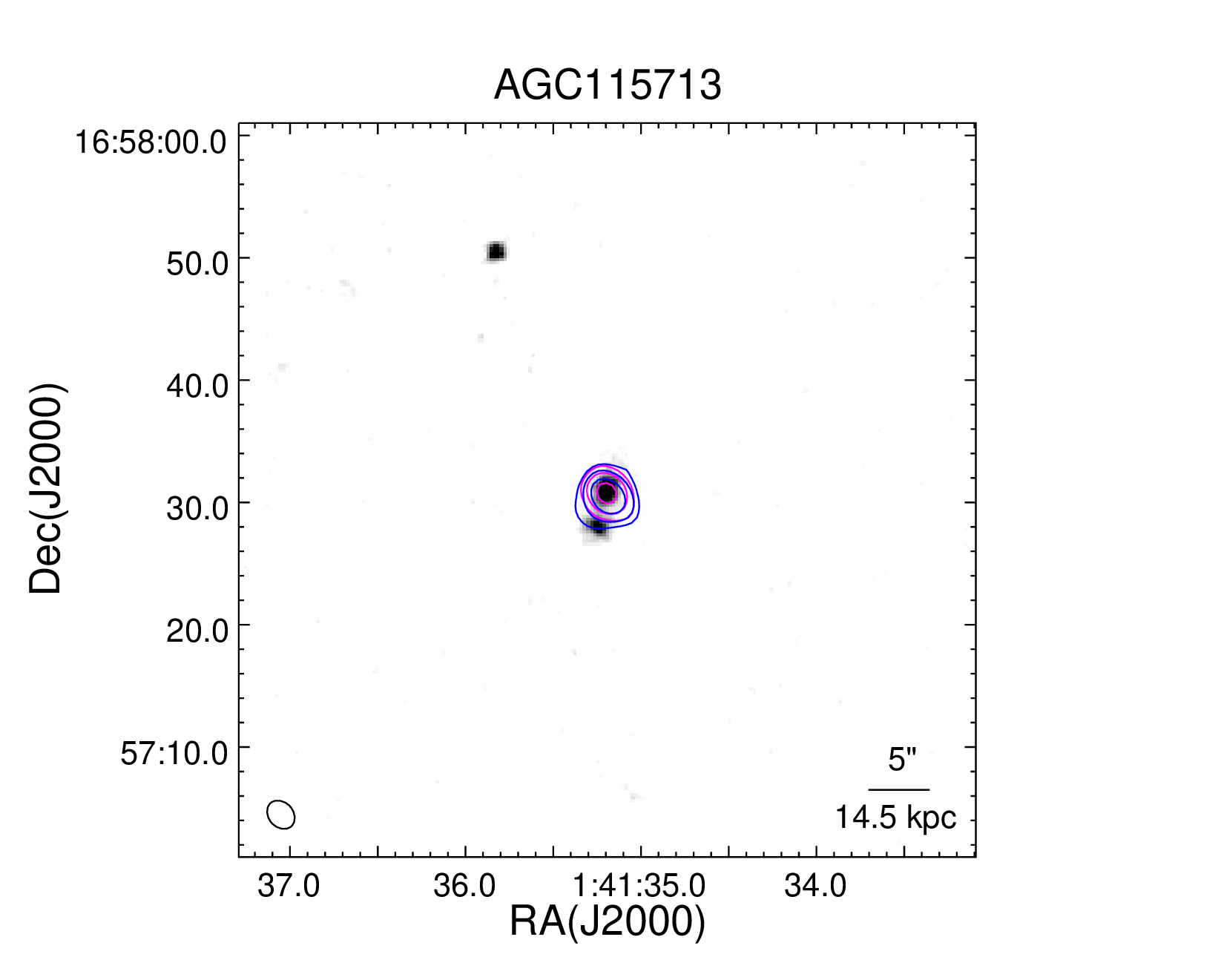}
   \includegraphics[width=0.49\textwidth]{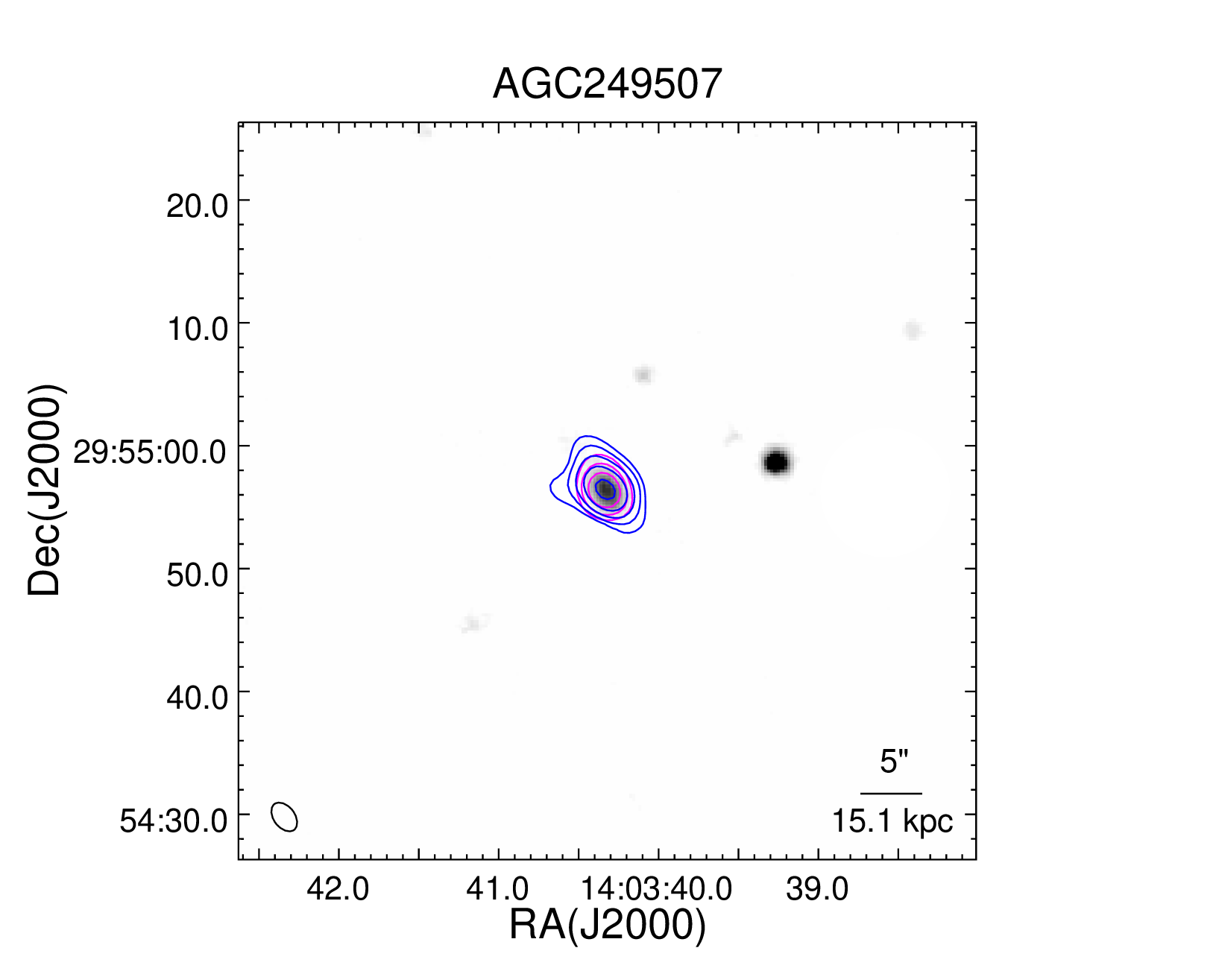}
 
\caption{continued
      }
      \label{contour}
\end{figure*}

\begin{figure*}
   \centering
 \includegraphics[width=0.49\textwidth]{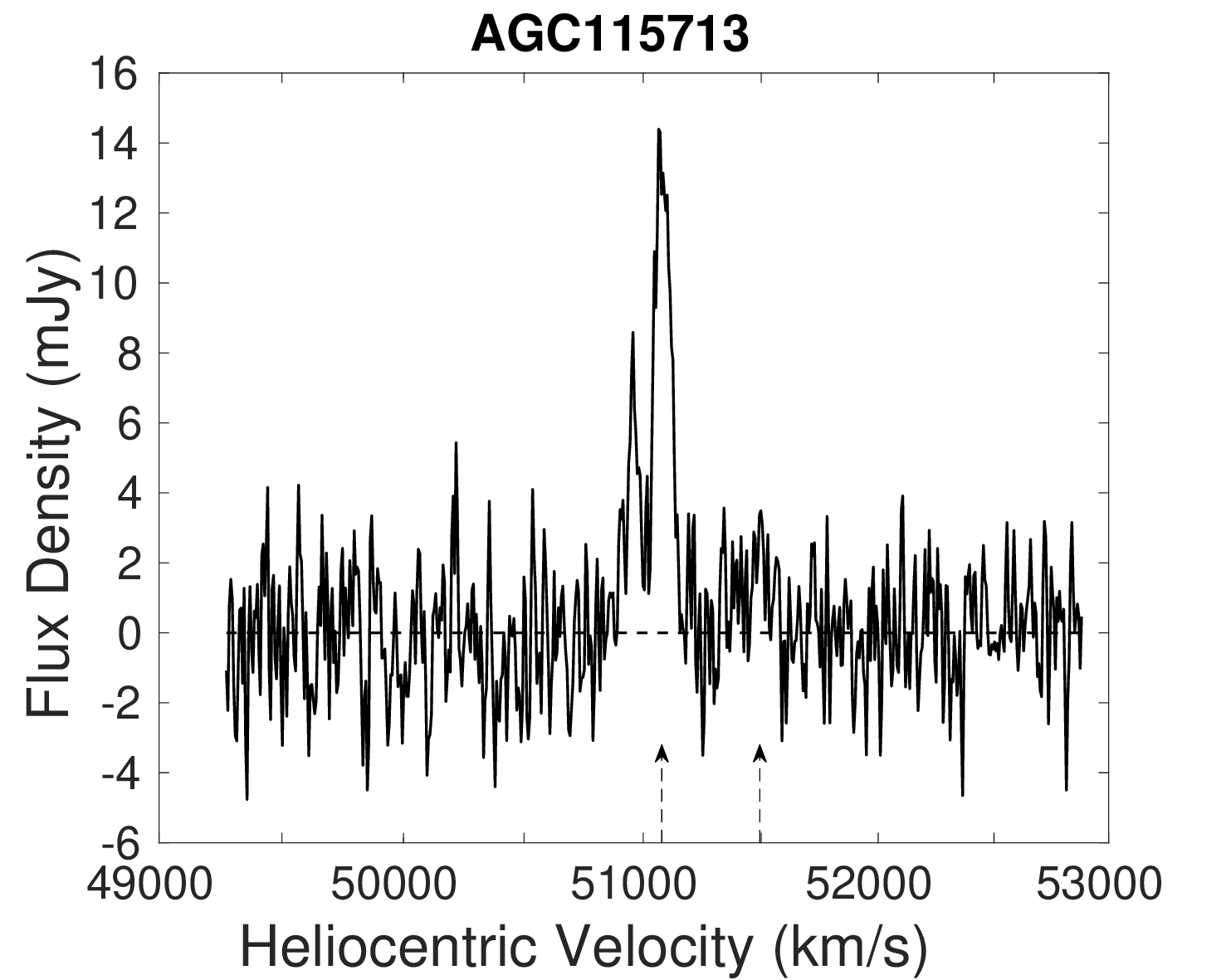}
 \includegraphics[width=0.49\textwidth]{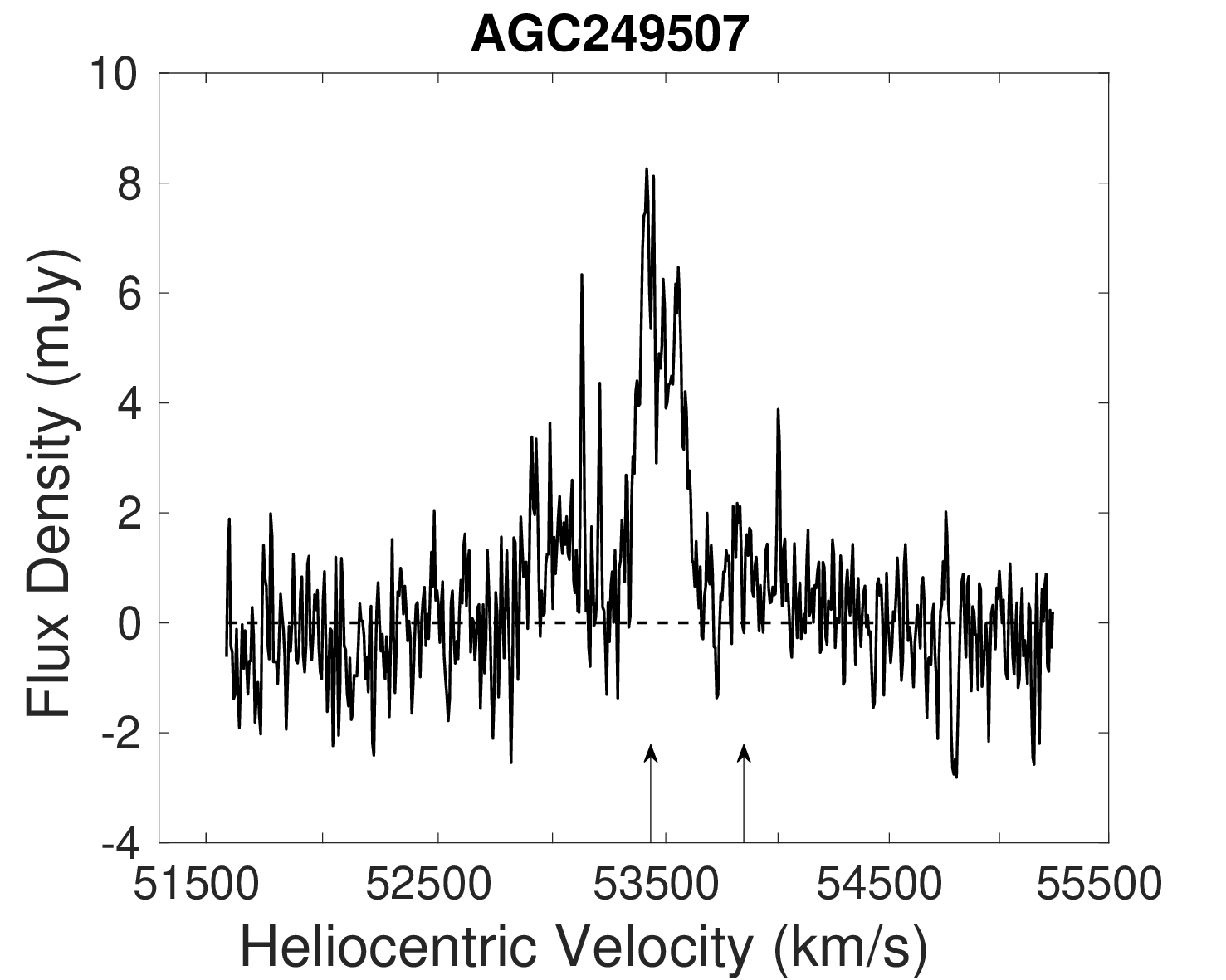}
      \caption{OH line profiles of AGC 115713 (left) and AGC 249507 (right). The spectra use the 1667.359 MHz line as the rest frequency for the velocity scale. Arrows indicate the expected velocity of the 1667.359 (left) and 1665.4018 (right) MHz lines, respectively. The arrow of the dotted line represents the expected velocity based on the redshift from OH line emission by \cite{2018ApJ...861...49H}. The arrow of the solid line is calculated by optical redshift from the LAMOST optical telescope survey. The OH line profiles of two sources were generated from a region about 7"$\times$7" and 5.65"$\times$5.26", respectively. The center coordinates and other parameters are listed in Table \ref{table2}.}

    \label{OH}%
\end{figure*}


\begin{figure*}
   \centering
   \includegraphics[width=0.47\textwidth]{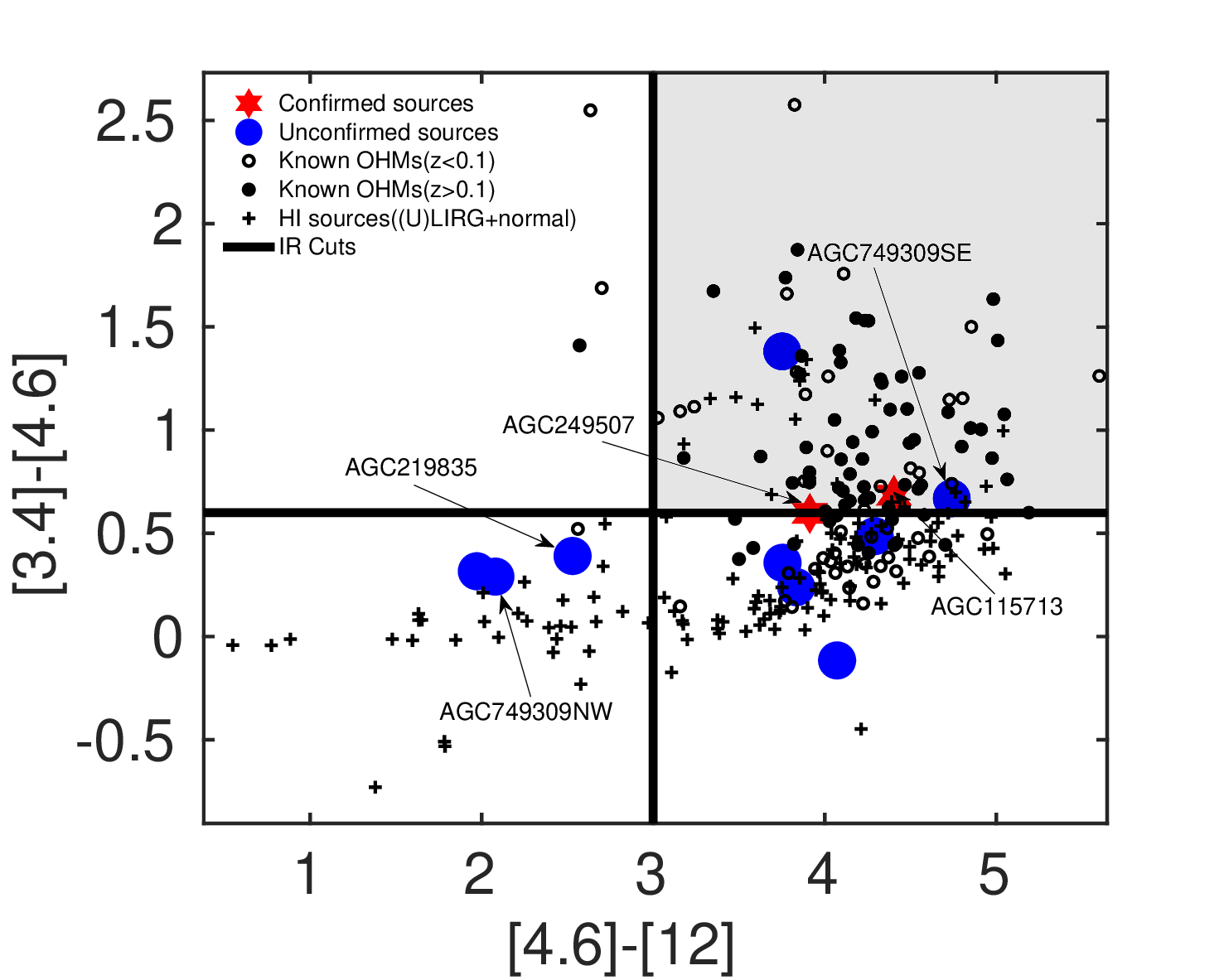}
   \includegraphics[width=0.47\textwidth]{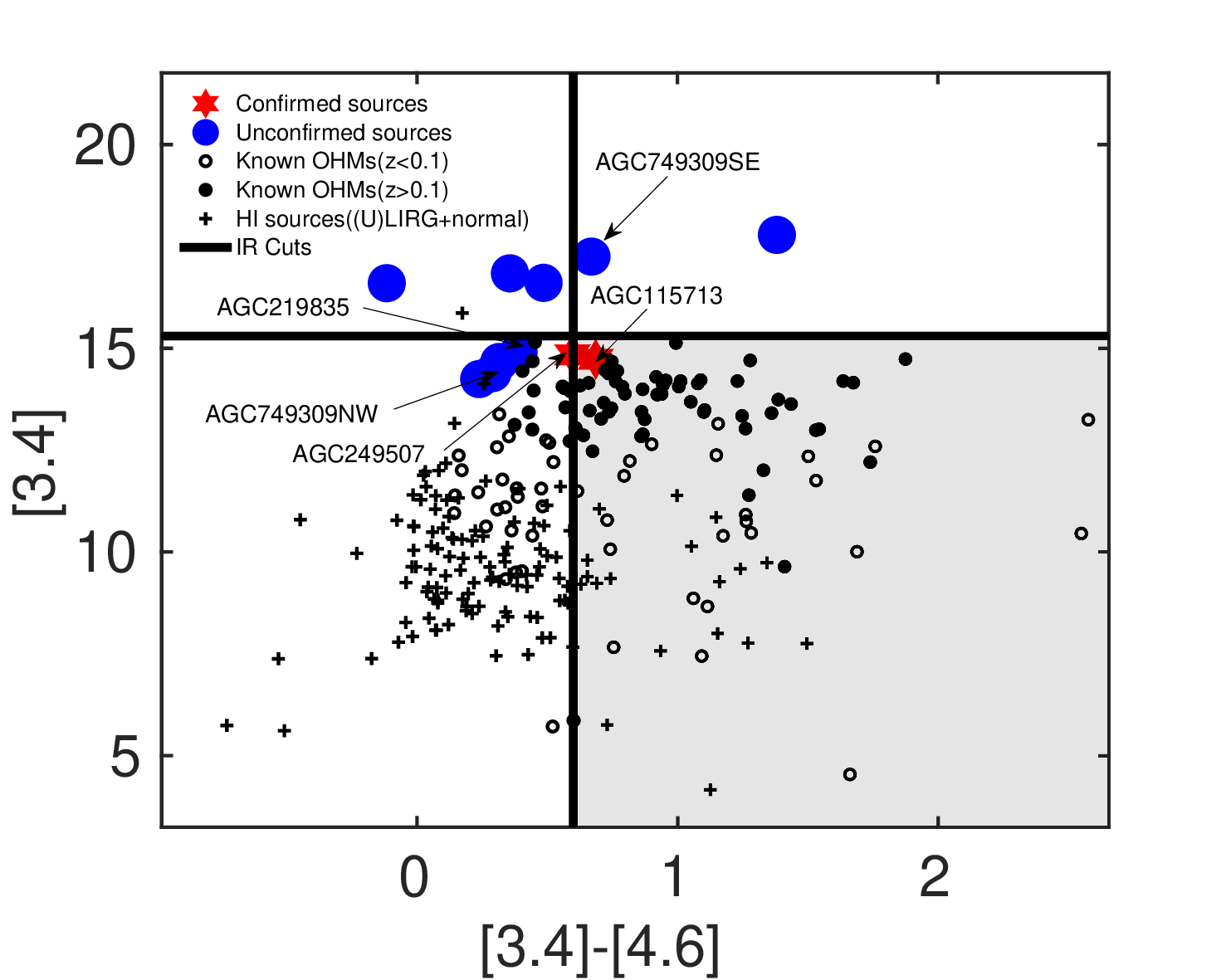}
   \includegraphics[width=0.47\textwidth]{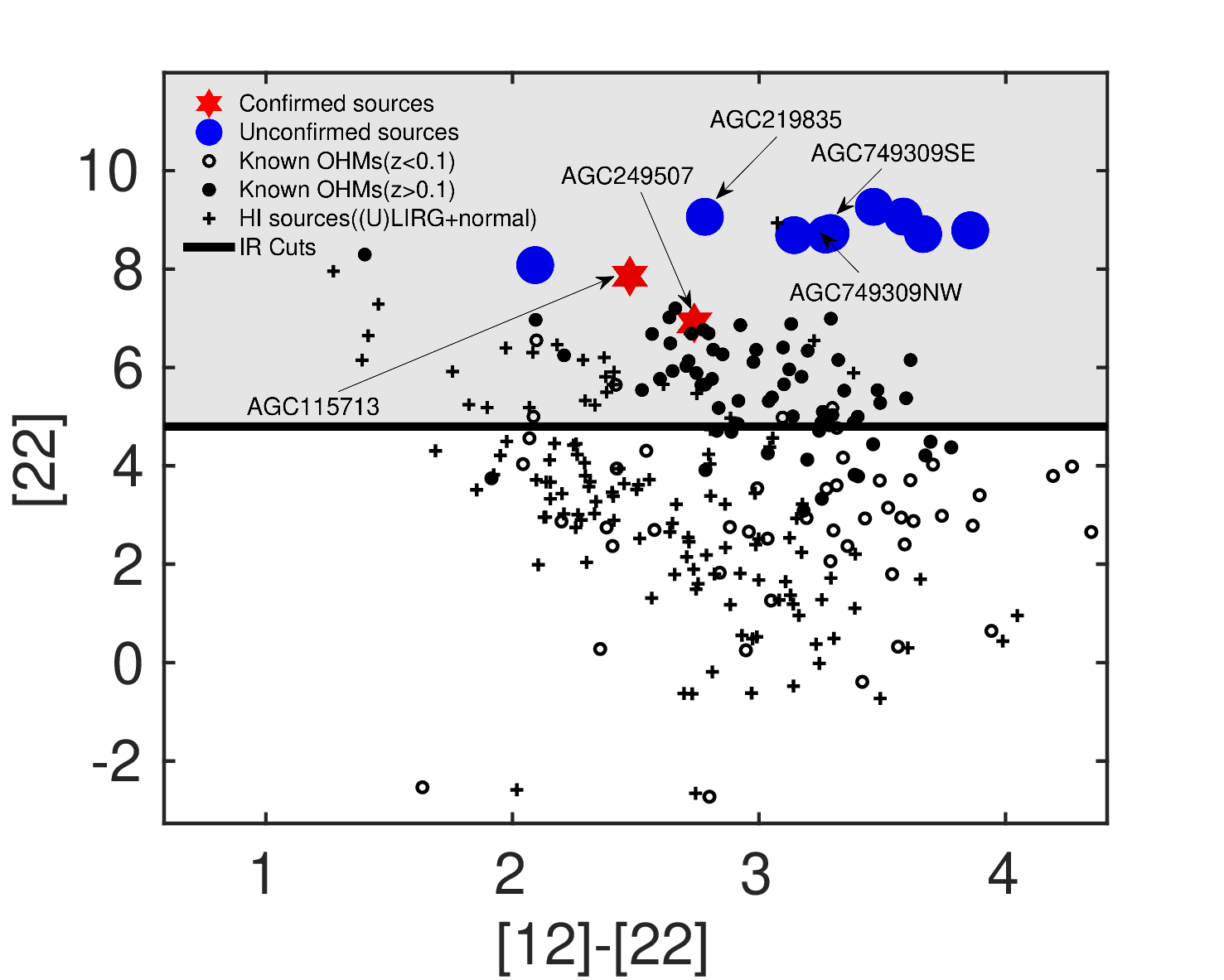}

\caption{Ten OHM candidates in the WISE color–magnitude and color–color space from \cite{2016MNRAS.459..220S}.  The red pentagrams are OHMs confirmed in this work (AGC115713 and AGC249507). The blue points represent the OHM candidates not confirmed in this work. The three zones (1-3) represent the infrared cuts to separate \HI and OH emitters \citep{2016MNRAS.459..220S}. 
      }
      \label{fig.K-S}
\end{figure*}


\begin{figure*}
   \centering
   \includegraphics[width=9cm]{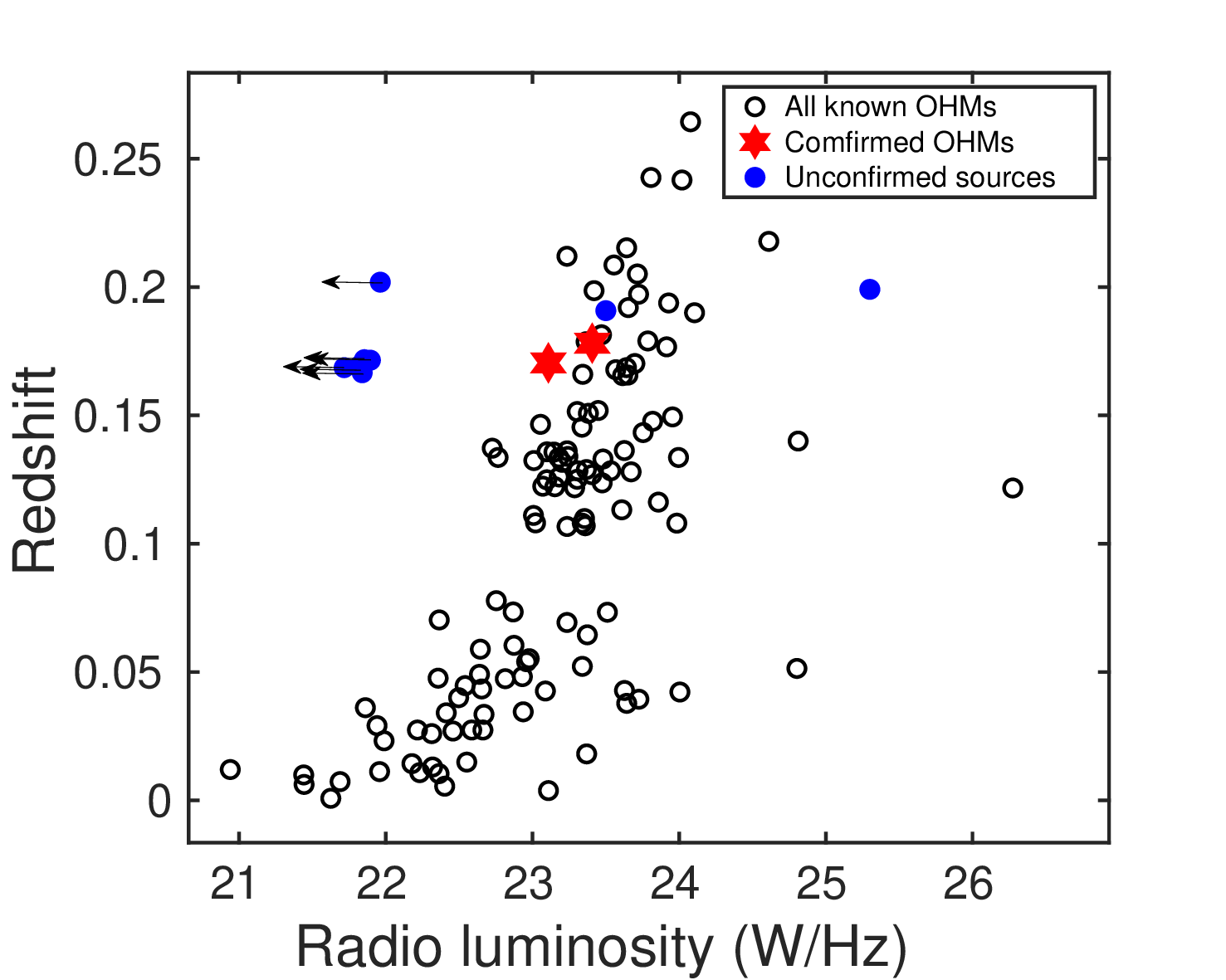}
\caption{L-band radio continuum luminosity distributions of OHM galaxies along the redshift.  The red stars and blue points stand for the confirmed and \textbf{unconfirmed} OHM candidates in this work.}
      \label{radioluminosity}
\end{figure*}


\begin{figure*}
   \centering
   \includegraphics[width=9cm]{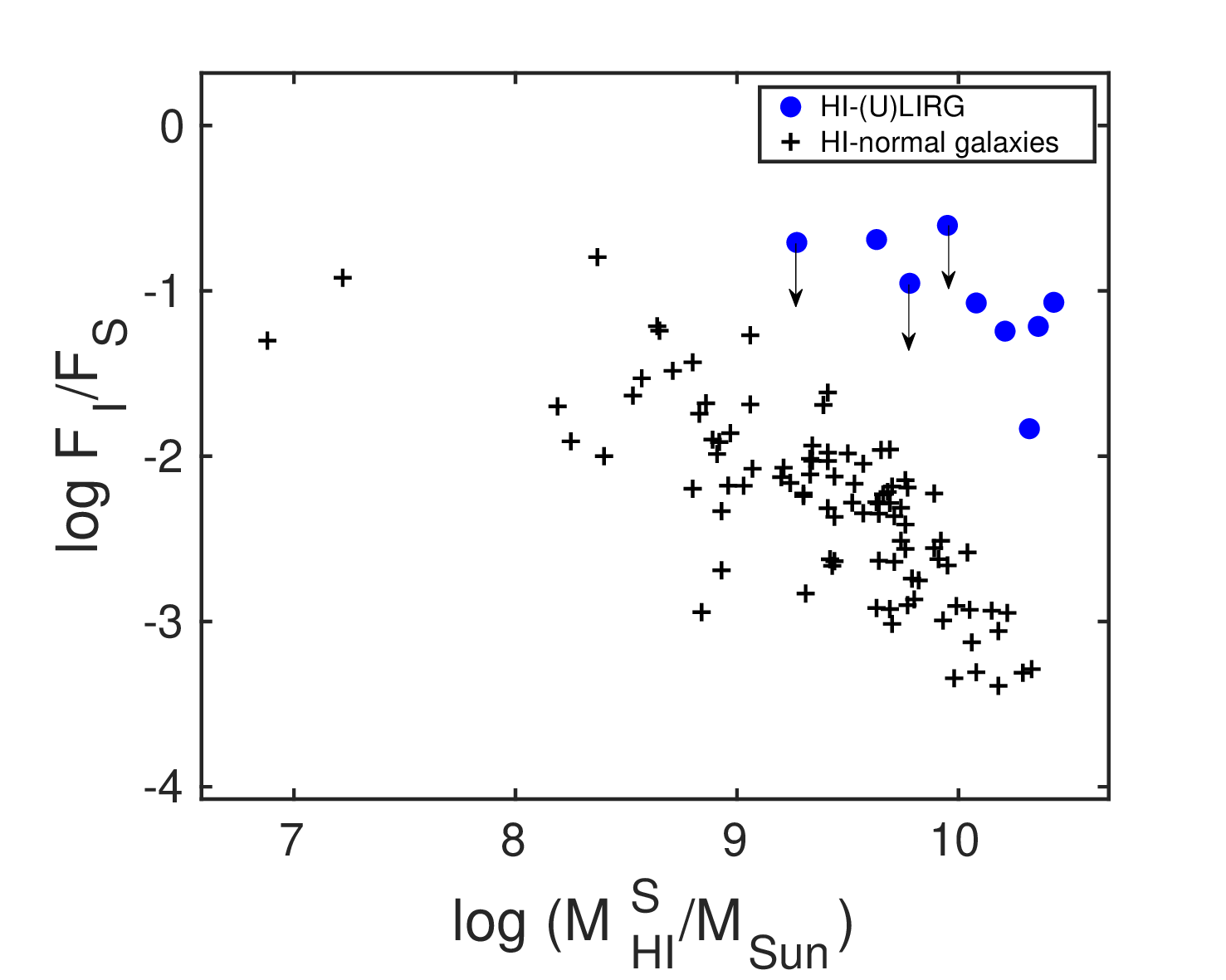}
\caption{\HI integrated line flux ratio ($F_{I}$/$F_{S}$) of nearby starforming galaxies and (U)LIRGs from Table \ref{table4} and \ref{tablea2}. $F_{S}$ and $F_{I}$ represent the \HI integrated line flux from the single-dish and interferometric array, respectively.}
      \label{HIcompare}
\end{figure*}


\subsection{Data reduction}
The GMRT and VLA archive data in this article were calibrated using the Common Astronomy Software Applications package (CASA\footnote{http://casa.nrao.edu}). The initial calibration of the GMRT data reduction process followed an online tutorial\footnote{http://www.ncra.tifr.res.in/\textasciitilde ruta/files/CASA\_spectral\-line\_analysis.pdf}.  The data reduction processes for GMRT observations and VLA archive data follow similar procedures. Initially, we examined the data and executed flagging using different modes, such as 'quack,' 'TFCROP,' and 'manual' modes. Subsequently, we conducted delay and bandpass calibration utilizing the flux calibrators.
Next, we calibrated the phase and amplitude of the standard flux calibrators (3C48, 3C147, and 3C286) and the phase calibrator, transferring the flux density scale from the flux calibrator to the phase calibrator. Following self-calibration of the flux and phase calibrators, we performed additional manual flagging of the bad data and then recalibrated the data. Finally, we applied the final calibration tables to both the calibrators and target sources.

Subsequently, all calibrated data were imported into the DIFMAP package \citep{1997ASPC..125...77S} to generate radio continuum and spectral line channel images. The radio continuum images were made using the RFI and line free channels. For the spectral line images, we subtracted the continuum emission based on a CLEANed map of the radio continuum emission. And since the high-resolution GMRT observations had the potential to resolve spectral line emissions, we also conducted calibration using data from the central GMRT array\footnote{http://www.gmrt.ncra.tifr.res.in/doc/gtac\_43\_status\_doc.pdf} (GMRT-central), which consisted of fourteen antennas covering an area of approximately 1 km. The fundamental parameters for both data sets are outlined in Table \ref{tablea1}.


\section{Results}

\subsection{Radio continuum emission from GMRT observations}
\label{3.1}
We have successfully detected radio continuum emission in 4 out of the 10 OHM candidate sources (see Table \ref{table1}). In Fig. \ref{contour}, we present arcsecond-scale radio continuum emission images of these sources. It is evident that the radio continuum emission appears to be compact in three of these sources, with the fitted image components (see Table \ref{table1}) being in close agreement with the beam Full Width at Half Maximum (FWHM) (see Table \ref{tablea1}). Notably, one source, AGC219835, exhibits a double-elongated structure.

Furthermore, we examined radio continuum images of our designated targets using three available VLA surveys: VLA Faint Images of the Radio Sky at Twenty centimeters (FIRST) \citep[FIRST][]{2015ApJ...801...26H}, NRAO VLA Sky Survey (NVSS) \citep[NVSS][]{1998AJ....115.1693C}, and the Very Large Array Sky Survey (VLASS) \citep[VLASS][]{2020PASP..132c5001L}. We retrieved the radio images of our sources from these surveys through the \textbf{Canadian Initiative for Radio Astronomy Data Analysis (CIRADA)} Image Cutout Web Service\footnote{http://cutouts.cirada.ca/}. Subsequently, we examined and analyzed these images using CASA software, determining that the noise levels were approximately 0.31-0.71 mJy for NVSS images (1.4GHz, resolution $\sim$45"), 0.09-0.19 mJy for VLA FIRST (1.4GHz, resolution $\sim$5"), and 0.07-0.09 mJy for VLASS images (3 GHz, resolution $\sim$2.5"). In conclusion, we observe a consistent alignment between the radio continuum images from these surveys and our high-sensitivity observations (20-30 $\mu$Jy/beam) of these OHM candidates. This supports the detection of radio continuum emission in four of the OHM candidates.

\subsection{OH lines emission from GMRT observations}
\label{result2}

We have achieved a significant detection of OH line emission from two of the ten OHM candidates using GMRT-full data: AGC 115713 and AGC 249507 (see Figs. \ref{contour} and \ref{OH}). The Signal-to-Noise Ratios (S/N) for both detections surpass 10 in both the image plane and the extracted line profiles with a frequency width of approximately 24 kHz (approximately 6 \kms) (see Table \ref{table2}). 
Additionally, we observed that the OH line emission in both sources are confined to single compact structures, with the size of the fitted components (see Table \ref{table2}) being smaller than the beam size of our GMRT observations (see Table \ref{tablea1}). 
The OH line profiles exhibit multiple peaks, probably corresponding to the prominent 1667 MHz line. However, it is difficult to 
identify the 1665 MHz line feature because of the blending 
of the two lines, which is commonly shown in known OHM 
galaxies \citep{2000AJ....119.3003D}.
 Although the OH 1612 MHz line falls within our frequency range, we found that its signal remains below the 3 $\sigma$ threshold (approximately 3 mJy/beam) in our observations of these sources.

We have centered at the peak frequency given by \cite{2018ApJ...861...49H} and binned more channels to detect potential signals for the \textbf{unconfirmed} sources.  Figs. \ref{figa1}, \ref{GMRTcontinuum1}, and \ref{GMRTcontinuum2} present the line profiles of \textbf{unconfirmed} sources, generated by binning the data every two channels at selected circular regions. In Fig. \ref{GMRTcontinuum1} and \ref{GMRTcontinuum2}, the dirty map displays the binning of all line channels within a bandwidth of 5 MHz centered at the peak frequency. In the case of AGC749309, we observed narrow line features that appear to be distributed in several regions around its potential host galaxy (see Section \ref{749309} and Fig. \ref{GMRTcontinuum1}). However, for the other seven candidates, we did not see any visible line features (see Fig. \ref{GMRTcontinuum2}). Subsequently, we utilized the data from antennas in the low-resolution GMRT-central and further detected three potential spectral line emission: AGC102850, AGC115018 and AGC749309 (see Fig. \ref{figa1}). These line profiles peak at a frequency similar to that observed in the ALFALFA survey by \cite{2018ApJ...861...49H}. However, these detections register at a noise level lower than 3 $\sigma$ in the image plane, preventing us from accurately determining the positions of these detections (see Section \ref{notes} and Fig. \ref{contour}).  


\section{Discussion}{\label{sec:discussion}}

\subsection{The confirmation of two OHM galaxies from GMRT observations}
The spatial correlation between OH masers and radio continuum emission has received substantial support from both physical models and arcsecond-scale radio observations \citep{2005ASPC..340..224P,2023A&A...669A.148W}. Since both radio continuum and OH line emission in OHM galaxies are primarily linked to starbursts in the nuclei region, typically within 100 pc \citep[see][]{2005ASPC..340..224P,2021A&A...647A.193H}, they tend to be compact and coexist at arcsecond-scale radio observations. Consequently, it is widely accepted that OHM emission are unlikely to be resolved at arcsecond-scale observations, and OHM emission are generally associated with radio continuum emission at this scale \citep{2023A&A...669A.148W}.

Our observations indicate that the OH line emission and radio continuum emission of the two sources, AGC115713 and AGC249507, are distributed within regions smaller than the beam size of the GMRT (see Table \ref{table1}, \ref{table2}, and Section \ref{result2}). This observation aligns well with the characteristics typically found in OHM galaxies observed at arcsecond-scale radio observations.

Generally, optical spectroscopy is a crucial method for confirming new OHMs. We made efforts to obtain redshift information from the literature and optical surveys, including the Sloan Digital Sky Survey (SDSS) and the Large Sky Area Multi-Object Fiber Spectroscopic Telescope (LAMOST) survey. We found that only AGC 249507 has an available redshift from the LAMOST survey \footnote{http://www.lamost.org/dr8/v2.0/spectrum/view?obsid=319315170}, and this redshift aligns with the system velocity derived from the OH line emission (see Fig. \ref{OH}).

Furthermore, \cite{2016MNRAS.459..220S} have proposed a method to distinguish OH megamasers from \HI line emitters without relying on optical spectroscopy. This method involves using the WISE infrared colors and magnitudes, incorporating four IR criteria: [3.4]-[4.6]>0.6, [4.6]-[12]>3.0, [22]>4.8, and [3.4]<15.3, which can be depicted as three zones in Fig. \ref{fig.K-S}. We have collected the WISE color data for our sources from the NASA/IPAC INFRARED science archive webpage \footnote{https://irsa.ipac.caltech.edu/cgi-bin/Gator/nph-scan?utf8=\%E2\%9C\%93\&mission=irsa\&projshort=WISE}. Using these criteria, we were only able to confirm AGC 115713 as an OHM galaxy (see Table \ref{table3} and Fig. \ref{fig.K-S}). While AGC 249507 does not entirely satisfy these IR criteria (see Table \ref{table3} and Fig. \ref{fig.K-S}), its confirmation through optical redshift information from the LAMOST survey demonstrates the effectiveness of our GMRT spectral line observations in confirming new OH megamasers in this sample.

Similarly, tests were conducted on the \HI galaxies listed in Tables \ref{tablea2} and \ref{table4} using the WISE four IR criteria. Our findings indicate that these criteria have excluded all of these \HI galaxies (see Fig. \ref{fig.K-S}). This outcome aligns with the results presented by \cite{2016MNRAS.459..220S}, demonstrating that these IR cuts can successfully eliminate 99.3\% of nearby \HI galaxies. Thus, the only OHM candidate (AGC 115713) passing the IR cuts is more likely to be an OHM rather than a nearby \HI galaxy. Additionally, the same IR cuts were applied to the known OHM galaxies documented in \cite{2014A&A...570A.110Z}. The results show that these IR criteria have also excluded all 51 known OHM galaxies at z<0.1. Among the known OHM galaxies at z>0.1, approximately 54\% of these sources successfully pass the IR cuts. Consequently, the IR cuts may not be as effective in confirming new OHM galaxies as in excluding nearby galaxies. However, it is noteworthy that the known OHM galaxies at z>0.1 are likely concentrated in regions close to the shaded areas shown in Fig. \ref{fig.K-S}. Furthermore, the two confirmed OH candidates in this study are also distributed in a region consistent with known OHM galaxies at z>0.1 (see Fig. \ref{fig.K-S}).

\subsection{Infrared properties of OHM candidates}
OH megamasers (OHMs) are generally recognized as highly luminous masers, typically found in [U]LIRGs. Consequently, all previously documented OHM galaxies are invariably linked to high infrared luminosities ($L_{FIR}$ $>$ $10^{11}$ $ L_{\odot}$), as determined from the 60 $\mu$m and 100 $\mu$m infrared flux densities provided by the IRAS survey \citep{2000AJ....119.3003D,2023AAS...24112202R}. Notably, it is worth mentioning that only one of the ten OHM candidates, AGC 249507, is accompanied by available 60 $\mu$m and 100 $\mu$m flux data in the IRAS survey (see Table \ref{table3}). 
\cite{2014A&A...570A.110Z} demonstrated that the MIR properties of OHMs and non-OHMs adhere to the power-law model rather than a single blackbody model. Therefore, for the remaining nine OHM candidates, we applied a power-law fit (F $\varpropto$ $\nu^{-\alpha}$) to the data from the four WISE bands for each source and subsequently extrapolated the far-infrared flux values at 60 $\mu$m and 100 $\mu$m, respectively. Subsequently, we calculated the far-infrared flux density of these sources using the equation from \cite{2006A&A...449..559B}: $S_{FIR}$=1.26 $\times$ $10^{-14}$ $(2.58 F_{60}+ F_{100}) W m^{-2}$, where $F_{60}$ and $F_{100}$ is in units of Jy, then we calculated the far-infrared luminosity $L_{FIR}$ based on this flux density using \textbf{equation $L_{FIR}$=4 $\pi$$D^{2}_{L}$$S_{FIR}$, where} $D_{L}$ is the luminosity distance and the results were presented in Table \ref{table3} .

Our analysis reveals a significant difference in infrared luminosities between the two confirmed OHM candidates and those that remain unconfirmed (see Table \ref{table3}). All ten galaxies have IR flux densities that fall below the selection criterion of $f_{60 {\mu}m}$ > 0.6 Jy, as originally defined by \cite{2002ApJ...572..810D} for targeted observations. Notably, the IR luminosities of the two confirmed OHM candidates also fall within the category of LIRGs, which are typically the host galaxies of known OHM galaxies. In contrast, the unconfirmed OHM candidates exhibit IR luminosities approximately one magnitude lower than those observed in known OHM galaxies.  Consequently,these \textbf{unconfirmed} OHM candidates cannot be categorized as (U)LIRGs, which are the characteristic host galaxies of known OHM galaxies. If these sources are ultimately validated through optical redshift measurements or the detection of a second radio spectral line, they might represent a new category of OHM galaxies that exist within different environmental conditions, distinct from (U)LIRGs.

\subsection{Radio continuum emission association of the OH line}
 
We have successfully detected radio continuum emission in four out of the ten OHM candidates, including two confirmed OHM galaxies.
In addition, we have compiled the VLA NVSS or FIRST flux densities for 115 previously discovered OH maser galaxies, as reported by \cite{2014A&A...570A.110Z}, and subsequently calculated the distributions of radio luminosity at various redshifts (see Fig. \ref{radioluminosity}). Our analysis revealed that the upper limit of radio luminosity for six sources without detectable radio continuum emission deviates significantly from the known OHM galaxies. Concurrently, we found that two confirmed OHM sources, AGC115713 and AGC249507, align well with the distribution exhibited by known OHM galaxies (see Fig. \ref{radioluminosity}). \cite{2022ApJ...931L...7G} have reported the first untargeted detection of an OHM at z > 0.5, which is also the highest redshift detection of such a system to date. The radio continuum emission is about 0.42 mJy at L-band, and the corresponding luminosity is about 23.6 W/Hz, which is also consistent with the OHM galaxies at low-redshift as shown in Fig. \ref{radioluminosity}.

From Fig. \ref{radioluminosity}, it is evident that known OHM galaxies with high radio luminosities ($L_{radio}$) are relatively rare, with only three sources (IRAS 01569-2939, IRAS 02483+4302, and IRAS 13451+232) falling in the range of 24.8 to 26.3 W/Hz. This finding supports the prevailing idea that radio continuum emission in OHM galaxies primarily arises from star formation processes \citep{2021A&ARv..29....2P}. We further investigated whether these galaxies represent genuine detections and found that, out of the three OHM galaxies with high $L_{radio}$, only IRAS 01569-2939 has been definitively confirmed as a detection, using the \textbf{Green Bank Telescope (GBT)} telescope \citep{2012IAUS..287..345W}. The other two sources, IRAS 02483+4302 and IRAS 13451+232, remain tentative, as they were only reported as potential detections by \cite{1989IAUC.4856....2K} and \cite{1990AJ....100.1457D}, respectively, and may require further confirmation \citep{2002AJ....124..100D}. Therefore, while sources with high radio luminosities may still be outliers, this consideration also applies to AGC 219835 in our sample, which also exhibits high radio luminosities (approximately 25.3 W/Hz, see Fig. \ref{radioluminosity}).

\subsection{The comparison of arcsecond-scale properties of \HI and OH line emission}

Typically, OH megamaser emission is thought to be confined within a 100 pc region of the nucleus \citep[see][and references therein]{2021A&A...647A.193H}. Consequently, the physical scale of OH megamaser emission results in distribution within a region with angular sizes smaller than 1", even for low redshift \textbf{OHM host galaxies} \citep{2005ARA&A..43..625L}. \cite{2023A&A...669A.148W} found that the OH line profiles of OHM galaxies at z>0.15, measured from VLA-A observations, exhibit no significant differences compared to those from single-dish Arecibo observations, except for variations in some narrow OH line features caused by interstellar scintillation due to their small sizes.
Given that OH candidates from the ALFALFA blind emission line survey generally have a redshift of $z_{OH}$ $\sim$ 0.2 \citep{2016MNRAS.459..220S}, these candidates might share similar properties with known OHM galaxies studied in \cite{2023A&A...669A.148W}. Specifically, the OH line emission is confined within a region smaller than 1", and the line profiles are consistent with those from single-dish observations.

Unlike OHM emission, \HI gas in nearby galaxies is believed to be distributed on scales ranging from sub-kpc to hundreds of kpc, mostly extended to a scale larger than 2 kpc \citep{2016MNRAS.460.2143W}. Consequently, arcsecond-scale observations of \HI emission may primarily capture only the mass distributed in the nuclear region of these galaxies, resulting in significant differences from \HI emission obtained from single-dish observations.
\textbf{We have derived the ratio of integrated \HI flux from the 2 kpc scale and
from single-dish observations of nearby star-forming galaxies and (U)LIRGs} (see section \ref{samplehi}). It is evident that nearly all of these galaxies exhibit \HI emission from the 2 kpc region, accounting for less than 10\% of the total \HI emission (see Fig. \ref{HIcompare} and Table \ref{tablea2}). We have obtained the \textbf{spectra} of the (U)LIRGs selected in section \ref{samplehi} either from literature or based on the archive data reduced in this work (see Table \ref{table4}). The results show that most nearby (U)LIRGs (19/28) display \HI absorption spectra at our selected interfermotric observations, with only six sources exhibiting diffused \HI emission. Meanwhile, the \HI line profiles extracted from the central 10"$\times$ 10" region (1.9 to 5 kpc; see Fig. \ref{HIcontour}) of these (U)LIRGs are also significantly different from those from single-dish observations, and the integrated line flux constitutes less than 10\% of the total emission \HI (see Fig. \ref{HIcompare}, \ref{HIcontour}, and \ref{HIline}).

The ALFALFA survey is a blind emission line survey for \HI galaxies at z$\leq$0.06 \citep{2018ApJ...861...49H}. The OH candidates identified by \cite{2018ApJ...861...49H} would actually have a redshift z< 0.024 if they were \HI emission rather than OH emission based on their peak line frequency.  It should be noted that the two confirmed OHM galaxies (AGC 115713 and AGC 249507) exhibit several line components (see Fig. \ref{OH}), and a direct comparison with the line profiles from ALFALFA is not possible, as they are not available in the literature. The brightest feature has an integrated flux density of about 1.66 and 1.35 \textbf{Jy \kms} (see Table \ref{table2}), while the integrated flux density from the ALFALFA survey is about 1.36 and 1.65 \textbf{Jy \kms}, respectively. The differences are approximately 15\% for the two sources. However, our line profiles also reveal weak line features, and the total integrated flux density from our observations for the two sources is about 1.97 \textbf{Jy \kms} and 2.29 \textbf{Jy \kms} (see section \ref{notes} for notes on each source), respectively. Therefore, the integrated flux density from the ALFALFA survey is approximately 70\% of the flux detected in our work.

This discrepancy could be attributed to our observations, which have nearly twice the sensitivity of the ALFALFA survey. Then, the velocity range may be different, as the weak line features of the two sources are lower than the noise level ($\sim$ 2 mJy) of the ALFALFA survey \citep[see Fig. \ref{OH} and][]{2018ApJ...861...49H}, along with uncertainties in the integrated flux density from both our observation and the ALFALFA survey. This result suggests that the line flux density of the two sources is not resolved with our arcsecond-scale observations, even for the weak line features. This is consistent with the properties of known OHM galaxies at similar redshifts, as shown in \cite{2023A&A...669A.148W}.

For the eight \textbf{unconfirmed} OHM candidates, the line emission is not detected with the GMRT-full data, indicating that the spectral line emission initially detected by the ALFALFA survey is possibly resolved. However, we estimate that the upper limit of the integrated flux with equation 3 (RMS$\times$FWHM) will not be ten times lower than the integrated flux given by \cite{2018ApJ...861...49H}, which is the typical ratio for \HI emission of nearby galaxies (see Fig. \ref{HIcompare}). This is due to the low signal-to-noise ratio (S/N) of these sources \citep[S/N$\sim$5][]{2018ApJ...861...49H}. Consequently, higher-sensitivity observations might be useful to further confirm whether these sources are significantly resolved at the central region of these galaxies, similar to nearby \HI galaxies shown in Fig. \ref{HIcompare}.

\subsection{Implications for confirming new OHMs from the blind survey}

Using GMRT observations, we have successfully confirmed two of the ten OHM candidates initially identified in the ALFALFA survey by \cite{2018ApJ...861...49H}. Our confirmation is based on arcsecond-scale radio continuum and OH line emission, as well as the characteristics exhibited by known OHM galaxies. Thus, the spectral line emission observed in the \textbf{unconfirmed} OHM candidates by the ALFALFA survey may not necessarily be OH line emissions.  They could potentially be \HI emissions in clusters or even some other unidentified type of radio line emission, as suggested in \cite{2015A&ARv..24....1G}. Therefore, conducting arcsecond-scale radio continuum and spectral line observations proves to be an effective approach for discovering new OHM galaxies that share common features with known OHM galaxies.

From the currently available VLA survey images (NVSS, FIRST, and VLASS), we have found that all OHM galaxies listed in \cite{2014A&A...570A.110Z} have been linked to radio continuum detections when they are located within the sky coverage of these surveys. This finding supports the idea that radio continuum emission plays a role in generating OH line emission \citep{2022AstBu..77..246S, 2022MNRAS.510.2495S}. Similarly, we can further confine that only AGC 115713, 249507, 749309, and 219835 as potential OHM candidates. This determination is based on their radio continuum detections in VLA surveys and the alignment of their radio luminosity with that of known OHM galaxies. This initial selection process using radio continuum emission enables us to reduce our sample size by 60\% before conducting high-resolution spectral line observations.  Hence, available radio continuum emission can be used as an initial screening criterion for OHM candidates, in agreement with the view of \cite{2020Ap&SS.365..118K}.

Among the ten OHM candidates, only the two LIRGs with the highest IR luminosity in this sample have been confirmed to be OHM galaxies (see Table \ref{table3}). \cite{2016MNRAS.459..220S} suggested that there is no previously unknown OHM producing population at z $\sim$ 0.2 and confirmed the validity of the IR selection method used in previous targeted surveys for OHM galaxies. These findings imply that IR luminosity could potentially serve as a preliminary selection criterion, akin to the previous targeted surveys, based on known OHM galaxies for potential OHM candidates. Although the available far-IR flux density and high far-IR luminosities cannot directly confirm the OHM galaxies found in blind \HI surveys \citep{2007ApJ...669L...9D}, they are well correlated with the OH luminosity \citep{2002AJ....124..100D} and play a crucial role in OHM formation \citep{2008ApJ...677..985L}. In cases where other \HI blind surveys identify OHM candidates without optical redshift information, the absence of associated infrared flux density detection or significantly lower calculated IR luminosity compared to known OHM galaxies (e.g., similar to eight out of ten of our OHM candidates with much lower IR luminosities than known OHM galaxies, as shown in Table \ref{table3}) may indicate a higher likelihood that they are not the OHM detections. Then, the number of OHM candidates need for further confimation might be significantly reduced.  

\textbf{Optical spectroscopic redshifts remain the most reliable method for confirming new OHM galaxies.} For \textbf{unconfirmed} galaxies, further spectral observations, either optical or radio, may be necessary to reach a definitive conclusion. Additionally, it is important to note that our current assessment of \textbf{unconfirmed} OHM galaxies in this study is based on existing knowledge about known OHM galaxies. We acknowledge that arcsecond-scale spectral line observations of known OHM galaxies, available in the literature, are still limited, particularly when dealing with diffused OHM emission. 
\textbf{The redshift range of known OHM galaxies spans from z=0.01 to z=0.267 \citep{2002AJ....124..100D}, with the addition of two recently discovered OHM galaxies at z=0.523 and z=0.709 by \cite{2022ApJ...931L...7G} and \cite{2023MNRAS.tmp.3695J}, respectively.}
Considering the redshift range of known OHM galaxies, arcsecond-scale observations may potentially reveal extended emissions, which could be attributed to intrinsic diffusion or the presence of multiple masing nuclei, as suggested by \cite{2000AJ....119.3003D}. These emissions should correspond to a physical scale ranging from several hundred parsecs to tens of kiloparsecs. However, high-sensitivity observations of known OHM galaxies are limited to less than two dozen in the literature. It appears that no diffuse OH maser emission has been detected in these observations, supporting the idea of the compact nature of OHM emission \citep{2023A&A...669A.148W}. 

 As the physical scale of \HI gas in nearby galaxies is predominantly extended to sizes larger than 2 kpc \citep{2016MNRAS.460.2143W}, we have adopted a 2 kpc physical scale for distinguishing between the \HI host and potential OHM candidates. This scale corresponds to about 1" at $z_{\HI}$ = 0.1 and a potential OHM candidate at $z_{OH}$ = 0.29. Given that the \HI surveys from ALFALFA are focused on \HI galaxies with $z < 0.06$, arcsecond-scale observations are suitable for this study. However, \cite{2021ApJ...911...38R} demonstrated that the upcoming \HI surveys will mostly extend to redshifts $z_{\HI}$ much higher than 0.1. In these galaxies, the 2 kpc scale will be smaller than 1", potentially necessitating sub-arcsecond and even mas-scale observations to distinguish \HI emission from OH emission.
Furthermore, \cite{2021ApJ...911...38R} indicated that the next generation \HI surveys will detect an unprecedented number of OHM galaxies, most of which will lack spectroscopic redshifts. Therefore, expanding high-sensitivity and high-resolution observations to a larger number of known OHM galaxies can serve to either confirm the compact nature of OHM emission or detect OHM maser emission with diffused characteristics, similar to the \textbf{unconfirmed} candidates in this study. These results can contribute to a deeper understanding of OH line emission properties and further confirming of OHM candidate from other \HI surveys.


\subsection{Notes on individual sources}
\label{notes}

\subsubsection{AGC102850}

From the data obtained from GMRT-central, we have identified potential spectral line emission with a central frequency of approximately 1421.5 MHz, which is slightly lower than the results obtained from the ALFALFA survey (see Fig. \ref{figa1}). 
It is worth noting that this spectral line emission appears to be diffuse, hindering our ability to pinpoint its optical counterpart (see Fig. \ref{contour}). Furthermore, we have not detected any radio continuum emission in this galaxy, with a noise level at approximately 0.032 mJy/beam. 

\subsubsection{AGC115018}

Utilizing the data acquired from GMRT-central, we have successfully detected potential spectral line profiles, as illustrated in Fig. \ref{figa1}. Specifically, the OH line emission was extracted from the vicinity of the optical counterpart as provided by \cite{2018ApJ...861...49H} (see Fig. \ref{contour}). 
The central frequency aligns with the findings of the ALFALFA survey (see Fig. \ref{figa1}). We have not detected any radio continuum emission, with a noise level approximately at 0.026 mJy/beam. 

\subsubsection{AGC115713}

The SDSS image of this source reveals the presence of two nuclei located in the southeast (SE) and northwest (NW) regions. Our GMRT observations have confirmed that the OH line emission and radio continuum emission spatially coincide with the NW nucleus, as illustrated in Figure \ref{contour}. This confirms the optical counterpart identification of SDSS 014135.2+165731, a selection made by \cite{2018ApJ...861...49H}. The OH line profile displays two distinct peaks corresponding to the OH 1667 MHz line. 
The two line features are about 1.66 Jy \kms and 0.31 Jy \kms, respectively. 
We have obtained a radio continuum flux of 1.23 mJy at 3 GHz from the VLASS survey and a flux of 1.73 mJy at 1.4 GHz observed by GMRT, resulting in a spectral index of -0.45.

\subsubsection{AGC249507}
This source has an available optical redshift of z=0.178139, obtained from LAMOST survey data release 8. Importantly, this redshift matches the OH line frequency, as shown in Figure \ref{OH}. Furthermore, both the radio continuum emission and OH line emission from this source exhibit spatial correlation, with the fitted component size being smaller than the beam size used in our GMRT observations (see Tables \ref{table1} and \ref{table2}). The OH line profile reveals the presence of multiple components with a resolution of approximately 6 km/s. The integrated line fluxes of the three features (see Fig. \ref{OH}), ranging from low to high velocities, are approximately 0.65, 1.35, and 0.29 Jy \kms. 

Additionally, our analysis indicates that the radio continuum flux for this source is approximately 2.4 mJy at 3 GHz from VLASS and 3.0 mJy at 1.4 GHz from our GMRT observations, resulting in a spectral index of -0.28.

\subsubsection{AGC749309}
\label{749309}
We have identified two radio continuum components near the potential OH emission source initially identified in the ALFALFA survey by \cite{2018ApJ...861...49H}. The integrated flux of these two components is 2.91 mJy and 1.2 mJy, respectively, with one situated in the northwest (NW) and the other in the southeast (SE)  (as illustrated in Fig. \ref{contour} and detailed in Table \ref{table1}). 
However, we have not detected significant spectral line emissions associated with these two radio continuum components. Instead, the spectral line emission appears to be distributed in the region between these two radio components (designated as D1-3, as shown in Fig. \ref{GMRTcontinuum1}) as well as in the vicinity of SDSS 101101.1+274012 (referred to as D2, see Fig. \ref{GMRTcontinuum1}), which is the optical counterpart identified by \cite{2018ApJ...861...49H}. Due to the diffuse distribution of the radio spectral line emission and the absence of any prominent components exceeding 3 $\sigma$ in the image plane, it is evident that higher sensitivity observations will be necessary to accurately pinpoint the precise location of the spectral line emission.

\cite{2011AJ....142..170H} initially classified the radio spectral line as \HI emission, while \cite{2018ApJ...861...49H} reclassified it as a potential OHM candidate. Despite our successful re-detection of the spectral line emission associated with this source, the radio spectral line emission is neither compact nor linked to radio continuum emission. Moreover, it lacks a clear association with prominent infrared emission, as indicated by available data from IRAS and WISE surveys. Consequently, these characteristics suggest that this source does not exhibit the typical properties associated with OH megamaser emission.

\subsubsection{AGC219835}
The radio continuum emission exhibits an elongated structure, spanning approximately 30", which likely exceeds the scale of the galaxy, as evident from the SDSS image (see Fig. \ref{contour}). The optical counterpart associated with the detected radio continuum emission is likely the quasar 4C+32.38. The radio continuum flux measures approximately 144.7 mJy at 3 GHz according to VLASS data, while GMRT observations report a flux of 364 mJy at 1.4 GHz.
Our analysis of the WISE data and the IRAS survey did not yield any 3 $\sigma$ detections. These characteristics do not align with the typical properties of an OHM galaxy, further strengthening our conclusion that this OHM candidate is not confirmed.

\section{Summary}
We present the results of the GMRT observations of 10 OHM candidates identified in the ALFALFA survey. Among these candidates, two sources (AGC115713 and AGC249507) exhibit compact OHM line emission and are associated with radio continuum emission, consistent with the typical characteristics of known OHM galaxies. Additionally, these two confirmed OHM galaxies display substantial Far-Infrared luminosity, aligning with the general model of OHM galaxies. Furthermore, AGC 249507 has an optical redshift recorded in the LAMOST survey and AGC115713 meets the IR selection criteria established in the literature, which also \textbf{supports our confirmation of the two sources as OHM host galaxies rather than nearby \HI galaxies.}

No significant spectral line emission were detected in the remaining 8 OHM candidates using our GMRT-full dataset. Furthermore, six of these candidates did not exhibit radio continuum emission, even with our high-sensitivity observations (approximately 30 $\mu$Jy/beam).
Upon analysis of the GMRT-central data, we found that three out of the eight candidates potentially exhibit spectral line emission, with central frequencies roughly consistent with those reported in the ALFALFA survey. 
Meanwhile, the absence of radio continuum emission and the presence of diffuse spectral line emission distinguish the \textbf{unconfirmed} OHM candidates from known OHM galaxies documented in the literature. 

These findings indicate that arcsecond-scale observations are an effective means of confirming OHM galaxy candidates. The spectral line emission observed in unconfirmed sources by the ALFALFA survey is more likely to be associated with \HI emission or other unidentified spectral lines \citep{2015A&ARv..24....1G}.
To further validate the nature of the radio line emission observed in the ALFALFA survey, additional measures such as optical spectroscopic redshift or line emission at other wavelength bands are necessary. These measures will help determine whether these unconfirmed candidates represent a new category of OHM galaxies displaying diffuse OH line emission that are resolved in arcsecond-scale observations.


\begin{acknowledgements}
 We thank the anonymous referee for the constructive comments and suggestions which substantially improved this paper.
  This work is supported by NSFC grants (Grant No. U1931203,12363001). The Giant Meterwave Radio telescope (GMRT) is the most sensitive synthetic aperture Radio telescope in the meter band. It is operated by a department of the Tata Institute of Basic Research - the National Radio Astronomical Center (NCRA) in the United States. Guoshoujing Telescope (the Large Sky Area Multi-Object Fiber Spectroscopic Telescope LAMOST) is a National Major Scientific Project built by the Chinese Academy of Sciences. Funding for the project has been provided by the National Development and Reform Commission. LAMOST is operated and managed by the National Astronomical Observatories, Chinese Academy of Sciences.

\end{acknowledgements}

\bibliography{wangsz}

\begin{thebibliography}{74}
\expandafter\ifx\csname natexlab\endcsname\relax\def\natexlab#1{#1}\fi

\bibitem[{{Allison} {et~al.}(2014){Allison}, {Sadler}, \& {Meekin}}]{2014MNRAS.440..696A}
{Allison}, J.~R., {Sadler}, E.~M., \& {Meekin}, A.~M. 2014, \mnras, 440, 696

\bibitem[{{Armus} {et~al.}(2009){Armus}, {Mazzarella}, {Evans}, {Surace}, {Sanders}, {Iwasawa}, {Frayer}, {Howell}, {Chan}, {Petric}, {Vavilkin}, {Kim}, {Haan}, {Inami}, {Murphy}, {Appleton}, {Barnes}, {Bothun}, {Bridge}, {Charmandaris}, {Jensen}, {Kewley}, {Lord}, {Madore}, {Marshall}, {Melbourne}, {Rich}, {Satyapal}, {Schulz}, {Spoon}, {Sturm}, {U}, {Veilleux}, \& {Xu}}]{2009PASP..121..559A}
{Armus}, L., {Mazzarella}, J.~M., {Evans}, A.~S., {et~al.} 2009, \pasp, 121, 559

\bibitem[{{Baan}(1985)}]{1985Natur.315...26B}
{Baan}, W.~A. 1985, \nat, 315, 26

\bibitem[{{Baan} {et~al.}(2007){Baan}, {Hagiwara}, \& {Hofner}}]{2007ApJ...661..173B}
{Baan}, W.~A., {Hagiwara}, Y., \& {Hofner}, P. 2007, \apj, 661, 173

\bibitem[{{Baan} \& {Haschick}(1990)}]{1990ApJ...364...65B}
{Baan}, W.~A. \& {Haschick}, A. 1990, \apj, 364, 65

\bibitem[{{Baan} {et~al.}(1985){Baan}, {Haschick}, \& {Schmelz}}]{1985ApJ...298L..51B}
{Baan}, W.~A., {Haschick}, A.~D., \& {Schmelz}, J.~T. 1985, \apjl, 298, L51

\bibitem[{{Baan} \& {Kl{\"o}ckner}(2006)}]{2006A&A...449..559B}
{Baan}, W.~A. \& {Kl{\"o}ckner}, H.~R. 2006, \aap, 449, 559

\bibitem[{{Beswick} {et~al.}(2005){Beswick}, {Aalto}, {Pedlar}, \& {H{\"u}ttemeister}}]{2005A&A...444..791B}
{Beswick}, R.~J., {Aalto}, S., {Pedlar}, A., \& {H{\"u}ttemeister}, S. 2005, \aap, 444, 791

\bibitem[{{Bicay} \& {Giovanelli}(1986)}]{1986AJ.....91..732B}
{Bicay}, M.~D. \& {Giovanelli}, R. 1986, \aj, 91, 732

\bibitem[{{Condon} {et~al.}(1998){Condon}, {Cotton}, {Greisen}, {Yin}, {Perley}, {Taylor}, \& {Broderick}}]{1998AJ....115.1693C}
{Condon}, J.~J., {Cotton}, W.~D., {Greisen}, E.~W., {et~al.} 1998, \aj, 115, 1693

\bibitem[{{Courtois} \& {Tully}(2015)}]{2015MNRAS.447.1531C}
{Courtois}, H.~M. \& {Tully}, R.~B. 2015, \mnras, 447, 1531

\bibitem[{{Courtois} {et~al.}(2009){Courtois}, {Tully}, {Fisher}, {Bonhomme}, {Zavodny}, \& {Barnes}}]{2009AJ....138.1938C}
{Courtois}, H.~M., {Tully}, R.~B., {Fisher}, J.~R., {et~al.} 2009, \aj, 138, 1938

\bibitem[{{Darling}(2007)}]{2007ApJ...669L...9D}
{Darling}, J. 2007, \apjl, 669, L9

\bibitem[{{Darling} \& {Giovanelli}(2000)}]{2000AJ....119.3003D}
{Darling}, J. \& {Giovanelli}, R. 2000, \aj, 119, 3003

\bibitem[{{Darling} \& {Giovanelli}(2002{\natexlab{a}})}]{2002AJ....124..100D}
{Darling}, J. \& {Giovanelli}, R. 2002{\natexlab{a}}, \aj, 124, 100

\bibitem[{{Darling} \& {Giovanelli}(2002{\natexlab{b}})}]{2002ApJ...572..810D}
{Darling}, J. \& {Giovanelli}, R. 2002{\natexlab{b}}, \apj, 572, 810

\bibitem[{{Davis} \& {Seaquist}(1983)}]{1983ApJS...53..269D}
{Davis}, L.~E. \& {Seaquist}, E.~R. 1983, \apjs, 53, 269

\bibitem[{{Dickel} \& {Rood}(1978)}]{1978ApJ...223..391D}
{Dickel}, J.~R. \& {Rood}, H.~J. 1978, \apj, 223, 391

\bibitem[{{Dickey}(1986)}]{1986ApJ...300..190D}
{Dickey}, J.~M. 1986, \apj, 300, 190

\bibitem[{{Dickey} {et~al.}(1990){Dickey}, {Planesas}, {Mirabel}, \& {Kazes}}]{1990AJ....100.1457D}
{Dickey}, J.~M., {Planesas}, P., {Mirabel}, I.~F., \& {Kazes}, I. 1990, \aj, 100, 1457

\bibitem[{{Dutta} {et~al.}(2018){Dutta}, {Srianand}, \& {Gupta}}]{2018MNRAS.480..947D}
{Dutta}, R., {Srianand}, R., \& {Gupta}, N. 2018, \mnras, 480, 947

\bibitem[{{Dutta} {et~al.}(2019){Dutta}, {Srianand}, \& {Gupta}}]{2019MNRAS.489.1099D}
{Dutta}, R., {Srianand}, R., \& {Gupta}, N. 2019, \mnras, 489, 1099

\bibitem[{{Fern{\'a}ndez} {et~al.}(2014){Fern{\'a}ndez}, {Petric}, {Schweizer}, \& {van Gorkom}}]{2014AJ....147...74F}
{Fern{\'a}ndez}, X., {Petric}, A.~O., {Schweizer}, F., \& {van Gorkom}, J.~H. 2014, \aj, 147, 74

\bibitem[{{Gallimore} {et~al.}(1994){Gallimore}, {Baum}, {O'Dea}, {Brinks}, \& {Pedlar}}]{1994ApJ...422L..13G}
{Gallimore}, J.~F., {Baum}, S.~A., {O'Dea}, C.~P., {Brinks}, E., \& {Pedlar}, A. 1994, \apjl, 422, L13

\bibitem[{{Giovanelli} \& {Haynes}(2015)}]{2015A&ARv..24....1G}
{Giovanelli}, R. \& {Haynes}, M.~P. 2015, \aapr, 24, 1

\bibitem[{{Glowacki} {et~al.}(2022){Glowacki}, {Collier}, {Kazemi-Moridani}, {Frank}, {Roberts}, {Darling}, {Kl{\"o}ckner}, {Adams}, {Baker}, {Bershady}, {Blecher}, {Blyth}, {Bowler}, {Catinella}, {Chemin}, {Crawford}, {Cress}, {Dav{\'e}}, {Deane}, {de Blok}, {Delhaize}, {Duncan}, {Elson}, {February}, {Gawiser}, {Hatfield}, {Healy}, {Henning}, {Hess}, {Heywood}, {Holwerda}, {Hoosain}, {Hughes}, {Hutchens}, {Jarvis}, {Kannappan}, {Katz}, {Kere{\v{s}}}, {Korsaga}, {Kraan-Korteweg}, {Lah}, {Lochner}, {Maddox}, {Makhathini}, {Meurer}, {Meyer}, {Obreschkow}, {Oh}, {Oosterloo}, {Oppor}, {Pan}, {Pisano}, {Randriamiarinarivo}, {Ravindranath}, {Schr{\"o}der}, {Skelton}, {Smirnov}, {Smith}, {Somerville}, {Srianand}, {Staveley-Smith}, {Tanaka}, {Vaccari}, {van Driel}, {Verheijen}, {Walter}, {Wu}, \& {Zwaan}}]{2022ApJ...931L...7G}
{Glowacki}, M., {Collier}, J.~D., {Kazemi-Moridani}, A., {et~al.} 2022, \apjl, 931, L7

\bibitem[{{Haynes} {et~al.}(2018){Haynes}, {Giovanelli}, {Kent}, {Adams}, {Balonek}, {Craig}, {Fertig}, {Finn}, {Giovanardi}, {Hallenbeck}, {Hess}, {Hoffman}, {Huang}, {Jones}, {Koopmann}, {Kornreich}, {Leisman}, {Miller}, {Moorman}, {O'Connor}, {O'Donoghue}, {Papastergis}, {Troischt}, {Stark}, \& {Xiao}}]{2018ApJ...861...49H}
{Haynes}, M.~P., {Giovanelli}, R., {Kent}, B.~R., {et~al.} 2018, \apj, 861, 49

\bibitem[{{Haynes} {et~al.}(2011){Haynes}, {Giovanelli}, {Martin}, {Hess}, {Saintonge}, {Adams}, {Hallenbeck}, {Hoffman}, {Huang}, {Kent}, {Koopmann}, {Papastergis}, {Stierwalt}, {Balonek}, {Craig}, {Higdon}, {Kornreich}, {Miller}, {O'Donoghue}, {Olowin}, {Rosenberg}, {Spekkens}, {Troischt}, \& {Wilcots}}]{2011AJ....142..170H}
{Haynes}, M.~P., {Giovanelli}, R., {Martin}, A.~M., {et~al.} 2011, \aj, 142, 170

\bibitem[{{Helfand} {et~al.}(2015){Helfand}, {White}, \& {Becker}}]{2015ApJ...801...26H}
{Helfand}, D.~J., {White}, R.~L., \& {Becker}, R.~H. 2015, \apj, 801, 26

\bibitem[{{Hess} {et~al.}(2021){Hess}, {Roberts}, {D{\'e}nes}, {Adebahr}, {Darling}, {Adams}, {de Blok}, {Kutkin}, {Lucero}, {Morganti}, {Moss}, {Oosterloo}, {Schulz}, {van der Hulst}, {Coolen}, {Damstra}, {Ivashina}, {Loose}, {Maan}, {Mika}, {Mulder}, {Norden}, {Oostrum}, {Ruiter}, {van Leeuwen}, {Vermaas}, {Vohl}, {Wijnholds}, \& {Ziemke}}]{2021A&A...647A.193H}
{Hess}, K.~M., {Roberts}, H., {D{\'e}nes}, H., {et~al.} 2021, \aap, 647, A193

\bibitem[{{Huchtmeier} \& {Richter}(1989)}]{1989gcho.book.....H}
{Huchtmeier}, W.~K. \& {Richter}, O.~G. 1989, {A General Catalog of HI Observations of Galaxies. The Reference Catalog.}

\bibitem[{{Hutchings} {et~al.}(1988){Hutchings}, {Neff}, \& {van Gorkom}}]{1988AJ.....96.1227H}
{Hutchings}, J.~B., {Neff}, S.~G., \& {van Gorkom}, J.~H. 1988, \aj, 96, 1227

\bibitem[{{Jarvis} {et~al.}(2023){Jarvis}, {Heywood}, {Jewell}, {Deane}, {Kl{\"o}ckner}, {Ponomareva}, {Maddox}, {Baker}, {Bianchetti}, {Hess}, {Roberts}, {Rodighiero}, {Ruffa}, {Sinigaglia}, {Varadaraj}, {Whittam}, {Adams}, {Baes}, {Murphy}, {Pan}, \& {Vaccari}}]{2023MNRAS.tmp.3695J}
{Jarvis}, M.~J., {Heywood}, I., {Jewell}, S.~M., {et~al.} 2023, \mnras [\eprint[arXiv]{2312.04345}]

\bibitem[{{Kazes} {et~al.}(1989){Kazes}, {Mirabel}, \& {Combes}}]{1989IAUC.4856....2K}
{Kazes}, I., {Mirabel}, I.~F., \& {Combes}, F. 1989, \iaucirc, 4856, 2

\bibitem[{{Koribalski} {et~al.}(2004){Koribalski}, {Staveley-Smith}, {Kilborn}, {Ryder}, {Kraan-Korteweg}, {Ryan-Weber}, {Ekers}, {Jerjen}, {Henning}, {Putman}, {Zwaan}, {de Blok}, {Calabretta}, {Disney}, {Minchin}, {Bhathal}, {Boyce}, {Drinkwater}, {Freeman}, {Gibson}, {Green}, {Haynes}, {Juraszek}, {Kesteven}, {Knezek}, {Mader}, {Marquarding}, {Meyer}, {Mould}, {Oosterloo}, {O'Brien}, {Price}, {Sadler}, {Schr{\"o}der}, {Stewart}, {Stootman}, {Waugh}, {Warren}, {Webster}, \& {Wright}}]{2004AJ....128...16K}
{Koribalski}, B.~S., {Staveley-Smith}, L., {Kilborn}, V.~A., {et~al.} 2004, \aj, 128, 16

\bibitem[{{Koribalski} {et~al.}(2020){Koribalski}, {Staveley-Smith}, {Westmeier}, {Serra}, {Spekkens}, {Wong}, {Lee-Waddell}, {Lagos}, {Obreschkow}, {Ryan-Weber}, {Zwaan}, {Kilborn}, {Bekiaris}, {Bekki}, {Bigiel}, {Boselli}, {Bosma}, {Catinella}, {Chauhan}, {Cluver}, {Colless}, {Courtois}, {Crain}, {de Blok}, {D{\'e}nes}, {Duffy}, {Elagali}, {Fluke}, {For}, {Heald}, {Henning}, {Hess}, {Holwerda}, {Howlett}, {Jarrett}, {Jones}, {Jones}, {J{\'o}zsa}, {Jurek}, {J{\"u}tte}, {Kamphuis}, {Karachentsev}, {Kerp}, {Kleiner}, {Kraan-Korteweg}, {L{\'o}pez-S{\'a}nchez}, {Madrid}, {Meyer}, {Mould}, {Murugeshan}, {Norris}, {Oh}, {Oosterloo}, {Popping}, {Putman}, {Reynolds}, {Rhee}, {Robotham}, {Ryder}, {Schr{\"o}der}, {Shao}, {Stevens}, {Taylor}, {van{\^A} der Hulst}, {Verdes-Montenegro}, {Wakker}, {Wang}, {Whiting}, {Winkel}, \& {Wolf}}]{2020Ap&SS.365..118K}
{Koribalski}, B.~S., {Staveley-Smith}, L., {Westmeier}, T., {et~al.} 2020, \apss, 365, 118

\bibitem[{{Lacy} {et~al.}(2020){Lacy}, {Baum}, {Chandler}, {Chatterjee}, {Clarke}, {Deustua}, {English}, {Farnes}, {Gaensler}, {Gugliucci}, {Hallinan}, {Kent}, {Kimball}, {Law}, {Lazio}, {Marvil}, {Mao}, {Medlin}, {Mooley}, {Murphy}, {Myers}, {Osten}, {Richards}, {Rosolowsky}, {Rudnick}, {Schinzel}, {Sivakoff}, {Sjouwerman}, {Taylor}, {White}, {Wrobel}, {Andernach}, {Beasley}, {Berger}, {Bhatnager}, {Birkinshaw}, {Bower}, {Brandt}, {Brown}, {Burke-Spolaor}, {Butler}, {Comerford}, {Demorest}, {Fu}, {Giacintucci}, {Golap}, {G{\"u}th}, {Hales}, {Hiriart}, {Hodge}, {Horesh}, {Ivezi{\'c}}, {Jarvis}, {Kamble}, {Kassim}, {Liu}, {Loinard}, {Lyons}, {Masters}, {Mezcua}, {Moellenbrock}, {Mroczkowski}, {Nyland}, {O'Dea}, {O'Sullivan}, {Peters}, {Radford}, {Rao}, {Robnett}, {Salcido}, {Shen}, {Sobotka}, {Witz}, {Vaccari}, {van Weeren}, {Vargas}, {Williams}, \& {Yoon}}]{2020PASP..132c5001L}
{Lacy}, M., {Baum}, S.~A., {Chandler}, C.~J., {et~al.} 2020, \pasp, 132, 035001

\bibitem[{{Le Reste} {et~al.}(2022){Le Reste}, {Hayes}, {Cannon}, {Herenz}, {Melinder}, {Menacho}, {{\"O}stlin}, {Puschnig}, {Rivera-Thorsen}, {Kunth}, \& {Velikonja}}]{2022ApJ...934...69L}
{Le Reste}, A., {Hayes}, M., {Cannon}, J.~M., {et~al.} 2022, \apj, 934, 69

\bibitem[{{Liu} {et~al.}(2015){Liu}, {Gao}, \& {Greve}}]{2015ApJ...805...31L}
{Liu}, L., {Gao}, Y., \& {Greve}, T.~R. 2015, \apj, 805, 31

\bibitem[{{Lo}(2005)}]{2005ARA&A..43..625L}
{Lo}, K.~Y. 2005, \araa, 43, 625

\bibitem[{{Lockett} \& {Elitzur}(2008)}]{2008ApJ...677..985L}
{Lockett}, P. \& {Elitzur}, M. 2008, \apj, 677, 985

\bibitem[{{Martin} {et~al.}(1991){Martin}, {Bottinelli}, {Dennefeld}, \& {Gouguenheim}}]{1991A&A...245..393M}
{Martin}, J.~M., {Bottinelli}, L., {Dennefeld}, M., \& {Gouguenheim}, L. 1991, \aap, 245, 393

\bibitem[{{McBride} {et~al.}(2013){McBride}, {Heiles}, \& {Elitzur}}]{2013ApJ...774...35M}
{McBride}, J., {Heiles}, C., \& {Elitzur}, M. 2013, \apj, 774, 35

\bibitem[{{Mirabel}(1982)}]{1982ApJ...260...75M}
{Mirabel}, I.~F. 1982, \apj, 260, 75

\bibitem[{{Mirabel} \& {Wilson}(1984)}]{1984ApJ...277...92M}
{Mirabel}, I.~F. \& {Wilson}, A.~S. 1984, \apj, 277, 92

\bibitem[{{Mundell} \& {Shone}(1999)}]{1999MNRAS.304..475M}
{Mundell}, C.~G. \& {Shone}, D.~L. 1999, \mnras, 304, 475

\bibitem[{{Noordermeer} {et~al.}(2005){Noordermeer}, {van der Hulst}, {Sancisi}, {Swaters}, \& {van Albada}}]{2005A&A...442..137N}
{Noordermeer}, E., {van der Hulst}, J.~M., {Sancisi}, R., {Swaters}, R.~A., \& {van Albada}, T.~S. 2005, \aap, 442, 137

\bibitem[{{Obreschkow} \& {Rawlings}(2009)}]{2009MNRAS.394.1857O}
{Obreschkow}, D. \& {Rawlings}, S. 2009, \mnras, 394, 1857

\bibitem[{{Olsson} {et~al.}(2010){Olsson}, {Aalto}, {Thomasson}, \& {Beswick}}]{2010A&A...513A..11O}
{Olsson}, E., {Aalto}, S., {Thomasson}, M., \& {Beswick}, R. 2010, \aap, 513, A11

\bibitem[{{Pedlar}(2005)}]{2005dmgp.book..161P}
{Pedlar}, A. 2005, in Dense Molecular Gas Around Protostars and in Galactic Nuclei, ed. W.~A. {Baan}, Y.~{Hagiwara}, \& H.~J. {van Langevelde}, 161

\bibitem[{{P{\'e}rez-Torres} {et~al.}(2021){P{\'e}rez-Torres}, {Mattila}, {Alonso-Herrero}, {Aalto}, \& {Efstathiou}}]{2021A&ARv..29....2P}
{P{\'e}rez-Torres}, M., {Mattila}, S., {Alonso-Herrero}, A., {Aalto}, S., \& {Efstathiou}, A. 2021, \aapr, 29, 2

\bibitem[{{Peterson} \& {Shostak}(1974)}]{1974AJ.....79..767P}
{Peterson}, S.~D. \& {Shostak}, G.~S. 1974, \aj, 79, 767

\bibitem[{{Pihlstr{\"o}m}(2005)}]{2005ASPC..340..224P}
{Pihlstr{\"o}m}, Y.~M. 2005, in Astronomical Society of the Pacific Conference Series, Vol. 340, Future Directions in High Resolution Astronomy, ed. J.~{Romney} \& M.~{Reid}, 224

\bibitem[{{Reif} {et~al.}(1982){Reif}, {Mebold}, {Goss}, {van Woerden}, \& {Siegman}}]{1982A&AS...50..451R}
{Reif}, K., {Mebold}, U., {Goss}, W.~M., {van Woerden}, H., \& {Siegman}, B. 1982, \aaps, 50, 451

\bibitem[{{Richter} {et~al.}(1994){Richter}, {Sackett}, \& {Sparke}}]{1994AJ....107...99R}
{Richter}, O.~G., {Sackett}, P.~D., \& {Sparke}, L.~S. 1994, \aj, 107, 99

\bibitem[{{Roberts} \& {Darling}(2023)}]{2023AAS...24112202R}
{Roberts}, H. \& {Darling}, J. 2023, in American Astronomical Society Meeting Abstracts, Vol.~55, American Astronomical Society Meeting Abstracts, 122.02D

\bibitem[{{Roberts} {et~al.}(2021){Roberts}, {Darling}, \& {Baker}}]{2021ApJ...911...38R}
{Roberts}, H., {Darling}, J., \& {Baker}, A.~J. 2021, \apj, 911, 38

\bibitem[{{Rots}(1980)}]{1980A&AS...41..189R}
{Rots}, A.~H. 1980, \aaps, 41, 189

\bibitem[{{Schneider} {et~al.}(1992){Schneider}, {Thuan}, {Mangum}, \& {Miller}}]{1992ApJS...81....5S}
{Schneider}, S.~E., {Thuan}, T.~X., {Mangum}, J.~G., \& {Miller}, J. 1992, \apjs, 81, 5

\bibitem[{{Shepherd}(1997)}]{1997ASPC..125...77S}
{Shepherd}, M.~C. 1997, in Astronomical Society of the Pacific Conference Series, Vol. 125, Astronomical Data Analysis Software and Systems VI, ed. G.~{Hunt} \& H.~{Payne}, 77

\bibitem[{{Sotnikova} {et~al.}(2022{\natexlab{a}}){Sotnikova}, {Mufakharov}, {Mikhailov}, {Stolyarov}, {Wu}, {Mingaliev}, {Semenova}, {Erkenov}, {Bursov}, \& {Udovitskiy}}]{2022AstBu..77..246S}
{Sotnikova}, Y.~V., {Mufakharov}, T.~V., {Mikhailov}, A.~G., {et~al.} 2022{\natexlab{a}}, Astrophysical Bulletin, 77, 246

\bibitem[{{Sotnikova} {et~al.}(2022{\natexlab{b}}){Sotnikova}, {Wu}, {Mufakharov}, {Mikhailov}, {Mingaliev}, {Erkenov}, {Semenova}, {Bursov}, {Udovitskiy}, {Stolyarov}, {Tsybulev}, {Chen}, {Zhang}, {Shen}, \& {Jiang}}]{2022MNRAS.510.2495S}
{Sotnikova}, Y.~V., {Wu}, Z., {Mufakharov}, T.~V., {et~al.} 2022{\natexlab{b}}, \mnras, 510, 2495

\bibitem[{{Springob} {et~al.}(2005){Springob}, {Haynes}, {Giovanelli}, \& {Kent}}]{2005ApJS..160..149S}
{Springob}, C.~M., {Haynes}, M.~P., {Giovanelli}, R., \& {Kent}, B.~R. 2005, \apjs, 160, 149

\bibitem[{{Staveley-Smith} \& {Davies}(1987)}]{1987MNRAS.224..953S}
{Staveley-Smith}, L. \& {Davies}, R.~D. 1987, \mnras, 224, 953

\bibitem[{{Suess} {et~al.}(2016){Suess}, {Darling}, {Haynes}, \& {Giovanelli}}]{2016MNRAS.459..220S}
{Suess}, K.~A., {Darling}, J., {Haynes}, M.~P., \& {Giovanelli}, R. 2016, \mnras, 459, 220

\bibitem[{{Tarchi} {et~al.}(2004){Tarchi}, {Greve}, {Peck}, {Neininger}, {Wills}, {Pedlar}, \& {Klein}}]{2004ASPC..320..112T}
{Tarchi}, A., {Greve}, A., {Peck}, A., {et~al.} 2004, in Astronomical Society of the Pacific Conference Series, Vol. 320, The Neutral ISM in Starburst Galaxies, ed. S.~{Aalto}, S.~{Huttemeister}, \& A.~{Pedlar}, 112

\bibitem[{{Theureau} {et~al.}(1998){Theureau}, {Bottinelli}, {Coudreau-Durand}, {Gouguenheim}, {Hallet}, {Loulergue}, {Paturel}, \& {Teerikorpi}}]{1998A&AS..130..333T}
{Theureau}, G., {Bottinelli}, L., {Coudreau-Durand}, N., {et~al.} 1998, \aaps, 130, 333

\bibitem[{{Theureau} {et~al.}(2017){Theureau}, {Coudreau}, {Hallet}, {Hanski}, \& {Poulain}}]{2017A&A...599A.104T}
{Theureau}, G., {Coudreau}, N., {Hallet}, N., {Hanski}, M.~O., \& {Poulain}, M. 2017, \aap, 599, A104

\bibitem[{{Wang} {et~al.}(2016){Wang}, {Koribalski}, {Serra}, {van der Hulst}, {Roychowdhury}, {Kamphuis}, \& {Chengalur}}]{2016MNRAS.460.2143W}
{Wang}, J., {Koribalski}, B.~S., {Serra}, P., {et~al.} 2016, \mnras, 460, 2143

\bibitem[{{Willett}(2012)}]{2012IAUS..287..345W}
{Willett}, K.~W. 2012, in Cosmic Masers - from OH to H0, ed. R.~S. {Booth}, W.~H.~T. {Vlemmings}, \& E.~M.~L. {Humphreys}, Vol. 287, 345--349

\bibitem[{{Willett} {et~al.}(2011){Willett}, {Darling}, {Spoon}, {Charmandaris}, \& {Armus}}]{2011ApJS..193...18W}
{Willett}, K.~W., {Darling}, J., {Spoon}, H. W.~W., {Charmandaris}, V., \& {Armus}, L. 2011, \apjs, 193, 18

\bibitem[{{Wu} {et~al.}(2023){Wu}, {Sotnikova}, {Zhang}, {Mufakharov}, {Zhu}, {Jiang}, {Chen}, {Shen}, {Sun}, {Peng}, \& {Wu}}]{2023A&A...669A.148W}
{Wu}, Z., {Sotnikova}, Y.~V., {Zhang}, B., {et~al.} 2023, \aap, 669, A148

\bibitem[{{Zhang} {et~al.}(2019){Zhang}, {Li}, {Wang}, {Zhu}, \& {Li}}]{2019RAA....19...22Z}
{Zhang}, J.-S., {Li}, D., {Wang}, J.-Z., {Zhu}, Q.-F., \& {Li}, J. 2019, Research in Astronomy and Astrophysics, 19, 022

\bibitem[{{Zhang} {et~al.}(2014){Zhang}, {Wang}, {Di}, {Zhu}, {Guo}, \& {Wang}}]{2014A&A...570A.110Z}
{Zhang}, J.~S., {Wang}, J.~Z., {Di}, G.~X., {et~al.} 2014, \aap, 570, A110

\end{thebibliography}

\begin{appendix}
\section{Online materials}
\twocolumn

 \begin{table*}
       \caption{\textbf{Parameters of the OHM candidates from ALFALFA}.}
     \label{tablea1}
  \centering
  \begin{tabular}{c c c c c c c c c  }     
  \hline\hline
   AGC Name & OH Position & $beam_{F}$  & $PA_{F}$ & $beam_{C}$&$PA_{C}$& $\Delta F$ & $\sigma_{F}$ &$\sigma_{C}$\\
    & (J2000)&(") $\times$ (") &($\circ$) &(") $\times$ (") &($\circ$) & (kHz) &   (mJy)              & (mJy)\\
   \hline
    102708 &000337.0+253215 &2.64$\times$2.39&35.6 &35.0$\times$26.6&-45.7 & 48&0.82 & 1.07 \\
    102850 &002958.8+305739 &4.35$\times$2.85& 37&39.6$\times$27.1& -55.9& 48& 1.07&  1.53\\
    114732 &010110.7+094626 &3.66$\times$2.79& 42.3&40.2$\times$31.9&-73.1& 48& 0.93& 1.50 \\
    115713 &014135.19+165730.73* &2.55$\times$2.02& 41.8&36.4$\times$28.2&-33.9& 48& 1.08& 1.81 \\
    115018 &015847.1+073159 &3.12$\times$2.32& 52.1&34.6$\times$32.5&-67.8 & 48& 0.74& 1.45 \\
    124351 &021751.0+072447 &4.14$\times$2.60& 63&46.0$\times$30.3&80 &48 & 0.72& 1.42 \\
    749309 &101102.9+274020&2.66$\times$2.32& 31.9&40.9$\times$25.5& -30.4  &48 & 0.70& 1.13 \\
    219835 &113034.2+322208 &2.84$\times$2.39&31 &42.9$\times$24.4& -42&48 & 1.33& 3.15 \\
    249507 &140340.34+295456.39* &2.64$\times$1.72& 37.3&36.2$\times$23.6&-22.8& 48& 0.94& 1.46 \\
    322231 &223605.9+095743 &3.31$\times$2.53& 55.1&39.3$\times$30.8& -71.2& 48& 0.87& 1.34 \\
     
   \hline
   \end{tabular}
   \vskip 0.1 true cm \noindent Notes: \textbf{Col. 2}: the OH position from the ALFALFA survey by \cite{2018ApJ...861...49H}. The '*' sign stands for the OH position and is obtained by this work (see Fig. \ref{contour}). \textbf{Cols. 3-6}: the beam FWHM using GMRT full array and GMRT central array, respectively. \textbf{Col. 7}: the channel width for binned 2 channels for improving the sensitivity. \textbf{Cols. 8-9}: the noise level achieved for the channel images with the bandwidth listed in \textbf{Col. 7}.
   \end{table*}


      \setlength{\tabcolsep}{0.05in}
  \begin{table*}
       \caption{\textbf{\HI host galaxies from some nearby star-forming galaxies}.} 
     \label{tablea2}
  \centering
  \resizebox{\linewidth}{!}{
  \begin{tabular}{l c c c c c c l c c c c c c}     
  \hline\hline
  Name &Distance&  $\int$ S dV & log $M_\HI^S$&log $\sum_\HI$& log $M_\HI^I$ & Reference& Name &Distance&  $\int$ S dV & log $M_\HI^S$&log $\sum_{\HI}$& log $M_{\HI}^{I}$ & Reference\\
    &(Mpc) & (\textbf{Jy \kms})&($M_\odot$)&($M_\odot$ $pc^{-2}$ )&($M_\odot$) & & &(Mpc) & (\textbf{Jy \kms})&($M_\odot$)&($M_\odot$ $pc^{-2}$)&($M_\odot)$ &\\
    \hline
NGC 0134&21.38&139.5&10.18&0.29&6.78&1&NGC 4299*&15.9&19.08&9.06&1.29&6.90&2\\
NGC 0253*&4.11&692.9&9.44&0.82&7.14&1&NGC 4303&19.8&84.71&9.89&0.84&7.34&2\\
NGC 0520*&30.59&24.88&9.74&0.73&6.32&2&NGC 4321&19.47&48.86&9.64&0.51&7.01&2\\
NGC 0598*&0.81&999.9&8.19&0.89&6.49&2&NGC 4394&19.08&8.09&8.84&-0.6&5.90&2\\
NGC 0628&10.55&424.3&10.05&0.62&7.12&2&NGC 4402*&18.6&7.73&8.80&0.87&6.64&2\\
NGC 0660*&12.00&148.39&9.70&0.19&6.14&2&NGC 4414&8.48&49.27&8.92&0.51&7.01&2\\
NGC 0772&34.52&68.99&10.29&0.48&6.98&2&NGC 4501&19.41&29.1&9.41&0.6&7.10&2\\
NGC 0891&9.17&213.2&9.63&0.21&6.71&3&NGC 4526*&18.79&0.2&7.22&0.71&6.56&2\\
NGC 0925&9.10&263.04&9.71&0.85&7.35&2&NGC 4535&19.75&71.66&9.82&0.57&7.07&2\\
NGC 1022*&19.48&3.8&8.53&0.4&5.69&4&NGC 4548&18.79&10.18&8.93&-0.26&6.24&2\\
NGC 1055&13.63&324.1&10.15&0.72&7.22&2&NGC 4561&19.42&27.44&9.39&1.2&7.70&2\\
NGC 1058&8.80&64.84&9.07&0.5&7.00&5&NGC 4569*&18.8&11.15&8.97&0.61&7.11&2\\
NGC 1097&16.87&142.6&9.98&0.14&6.64&1&NGC 4579&19.32&9.67&8.93&0.1&6.60&2\\
NGC 1530&37.01&50.89&10.22&0.77&7.27&5&NGC 4631&7.35&461.13&9.77&0.37&6.87&2\\
NGC 1569*&4.11&62.99&8.40&1.26&6.40&6&NGC 4639&18.97&19.0&9.21&0.64&7.14&2\\
NGC 1808*&12.56&57.2&9.33&0.72&7.22&7&NGC 4647&19.1&7.86&8.83&0.59&7.09&2\\
NGC 2273*&28.54&13.74&9.42&0.3&4.86&8&NGC 4651&19.12&56.9&9.69&0.91&7.41&2\\
NGC 2276&36.96&9.84&9.50&1.02&7.52&5&NGC 4654&19.55&50.59&9.66&0.93&7.43&2\\
NGC 2403&4.11&1448.1&9.76&0.85&7.35&9&NGC 4689&19.79&8.36&8.89&0.49&6.99&2\\
NGC 2764*&41.36&6.32&9.41&0.93&7.31&10&NGC 4698&19.51&30.06&9.43&0.27&6.78&2\\
NGC 2841&13.42&150.06&9.80&0.44&6.94&5&NGC 4713&19.43&47.65&9.69&1.17&7.67&2\\
NGC 2903&8.53&192.51&9.52&0.74&7.24&2&NGC 4736&4.37&98.89&8.65&0.91&7.41&5\\
NGC 2976*&4.11&59.2&8.37&0.9&6.73&11&NGC 4826*&16.3&40.67&9.41&0.88&6.85&2\\
NGC 3031*&4.11&266.8&9.03&0.35&6.85&12&NGC 4945*&5.2&319.1&9.31&-0.02&6.40&1\\
NGC 3032*&25.22&1.18&8.25&-0.16&6.15&2&NGC 5005&14.1&13.6&8.80&0.11&6.61&11\\
NGC 3079&19.51&129.02&10.06&0.44&6.94&5&NGC 5033&12.63&216.53&9.91&0.79&7.29&5\\
NGC 3147&43.05&27.82&10.08&0.28&6.78&5&NGC 5055&7.44&439.8&9.76&0.7&7.20&3\\
NGC 3310&17.06&70.4&9.68&0.97&7.47&5&NGC 5194&7.53&119.7&9.20&0.58&7.08&12\\
NGC 3338&19.12&114.44&9.99&0.59&7.09&2&NGC 5236&4.7&1640&9.93&0.44&6.94&13\\
NGC 3351*&8.71&50.45&8.96&0.28&6.05&5&NGC 5347&32.18&11.4&9.44&0.58&7.08&2\\
NGC 3368&10.42&78.64&9.30&0.58&7.08&2&NGC 5457&4.94&301.7&9.24&0.58&7.08&12\\
NGC 3486&8.41&132.31&9.34&0.91&7.41&5&NGC 5678&32.23&20.52&9.70&1.02&7.52&5\\
NGC 3504*&26.36&5.01&8.91&0.43&6.17&2&NGC 5713&30.26&50.8&10.04&0.96&7.46&12\\
NGC 3521&8.28&231.56&9.57&1.03&7.53&2&NGC 5775&27.61&83.34&10.18&0.62&7.12&5\\
NGC 3556&12.09&168.06&9.76&1.12&7.62&5&NGC 6207&17.55&30.48&9.34&0.82&7.32&5\\
NGC 3627&7.34&40.49&8.71&0.73&7.23&2&NGC 6503&6.27&77.76&8.86&0.68&7.18&14\\
NGC 3628&8.67&238.13&9.63&0.85&7.35&2&NGC 6643&25.79&37.28&9.77&1.08&7.58&5\\
NGC 3675&11.72&61.38&9.30&0.56&7.06&5&NGC 6814&21.5&34.32&9.57&0.73&7.23&5\\
NGC 3726&14.02&95.34&9.64&0.8&7.30&5&NGC 6946*&4.16&839.1&9.53&0.87&6.72&9\\
NGC 3893&16.38&70.6&9.65&1.19&7.69&12&NGC 6951&23.01&39.48&9.69&0.27&6.77&5\\
NGC 3938&12.4&59.4&9.33&0.82&7.32&12&NGC 7331&14.38&170.52&9.92&0.91&7.41&2\\
NGC 4030&20.14&64.15&9.79&0.55&7.05&5&NGC 7465*&27.90&24.01&9.64&0.86&6.77&2\\
NGC 4150*&4.11&0.63&6.40&-0.31&4.34&11&NGC 7479&33.56&33.43&9.95&0.79&7.29&2\\
NGC 4178&19.36&62.2&9.74&0.93&7.43&2&NGC 7582&21.66&25.06&9.44&0.31&6.81&14\\
NGC 4189&20.06&12.23&9.06&0.88&7.38&2&He 2–10* &12&12.9&8.64&0.93&7.43&1\\
NGC 4254&16.3&77.05&9.68&0.96&7.46&5&IC 342&4.32&4907.5&10.33&0.32&6.82&9\\
NGC 4258&6.85&460.1&9.71&0.57&7.07&3&Maffei 2*&4.1&93&8.57&0.54&7.04&11\\
NGC 4294*&19.19&29.79&9.41&1.3&7.49&2&             \\

   \hline
   \end{tabular}}
   
   \vskip  0.1 true cm \noindent Notes: Col. 1: source name Col. 2: the luminosity distance in (Mpc). \textbf{Cols. 3-4: The integrated \HI line flux and corresponding \HI mass from single-dish observations in the literature.} Col. 5: the surface density of \HI gas from \cite{2015ApJ...805...31L}. Col. 6: the \HI mass estimated from the surface density listed in Col. 5 with a region of 2 Kpc in the diamater. '*' stands for these sources whose diameter is smaller than 2 Kpc, the \HI mass of them originates from their angular scale presented in \cite{2015ApJ...805...31L}. \textbf{Col. 7}: the references for the single-dish \HI mass listed in \textbf{Col. 4}: (1) \cite{2004AJ....128...16K};
  (2) \cite{2018ApJ...861...49H};
  (3) \cite{1987MNRAS.224..953S};
  (4) \cite{1998A&AS..130..333T};
  (5) \cite{2005ApJS..160..149S};
  (6) \cite{1992ApJS...81....5S};
  (7) \cite{1982A&AS...50..451R};
  (8) \cite{2005A&A...442..137N};
  (9) \cite{1980A&AS...41..189R};
  (10) \cite{1986AJ.....91..732B};
  (11) \cite{2009MNRAS.394.1857O};
  (12) \cite{1978ApJ...223..391D};
  (13) \cite{2014MNRAS.440..696A};
  (14) \cite{2009AJ....138.1938C};
Cols. 8-14 are duplications of \textbf{Cols. 1-7}. 
 \end{table*}


\setlength{\tabcolsep}{0.05in}
  \begin{table*}
       \caption{\textbf{\HI host galaxies from GOALs sample}.} 
     \label{table4}
  \centering
  \resizebox{\linewidth}{!}{
  \begin{tabular}{l c c c c c c c c c c c c}     
  \hline\hline
  IRAS Name & RA&Dec&  z&Project& Antenna array& \textbf{spectra type}& Distance&  $\int$ S dV&log $M_\HI^S$ & log $M_\HI^I$ &$Ref^{S}$& $Ref^{I}$ \\
    &(J2000)&(J2000)  & & &&& (Mpc)&(\textbf{Jy \kms}) &( $\textit{M}_\odot$)& ( $\textit{M}_\odot$)&&\\
    \hline
 F00085-1223&00 11 06.5&-12 06 26&0.0196&-&VLA-B&absorption&85.3&10.22*&10.24&-&a&1\\
  F00163-1039&00 18 50.5&-10 22 09&0.0272&AW585&VLA-B&diffuse emission&118.7&4.91&10.21&8.97&b&2\\
   F00506+7248&00 54 03.6&+73 05 12&0.0157&-&GMRT-full&absorption&68.0&6.66&9.86&-&c&3\\
  F01053-1746&01 07 47.2&-17 30 25&0.0201&18-011&GMRT-full&absorption&87.3&3.29&9.77&-&b&2\\
 F01173+1405&01 20 02.7&+14 21 43&0.0315&-&GMRT-full&absorption&138.3&1.13&9.71&-&d&4\\
 F02152+1418&02 17 59.6&+14 32 39&0.0131&04DAR01&GMRT-full&diffuse emission&56.5&35.78&10.43&9.36&d&2\\
F02401-0013&02 42 40.7&-00 00 48&0.0038&-&VLA-A&absorption&16.3&29.84&9.27&-&d&5\\
F04315-0840&04 33 59.7&-08 34 44&0.0159&-&VLA-A&absorption&69.1&0.97&9.04&-&e&6\\
  F06107+7822&06 18 37.7&+78 21 25&0.0030&-&VLA-A&absorption&12.8&61.1&9.37&-&f&7\\
 F06295-1735 &06 31 47.2&-17 37 17&0.0210&18-011&GMRT-full&non-detection&91.4&4.58*&9.95&<9.35&c&2\\
 F07256+3355 &07 28 53.4&+33 49 09&0.0138&04DAR01&GMRT-full&non-detection&59.7&7.11&9.78&<8.82&g&2\\
 F09126+4432&09 15 55.1&+44 19 55&0.0393&-&GMRT-full&absorption&173.2&3.26&10.36&-&h&4\\
 F11011+4107 &11 03 53.2&+40 50 57&0.0345&-&GMRT-full&absorption&151.8&-&-&-&&4\\
  F11257+5850&11 28 32.3&+58 33 43&0.0104&-&VLA-A&absorption&44.9&6.50&9.49&-&h&8\\
 F12116+5448&12 14 12.9&-47 13 42&0.0083& &MERLIN&absorption&36.0&4.70*&9.16&-&i&9\\
   F13001-2339 &13 02 52.3&-23 55 18&0.0217&18-011&GMRT-full&absorption&94.5&-&-&-&&2\\
    F13362+4831&13 38 17.5&+48 16 37&0.0279&-&VLA-A&absorption&121.9&-&-&-&&10\\
  F13373+0105&13 39 55.3&00 50 07&0.0225&08DVT02&GMRT-full&diffuse emission&98.0&10.17&10.36&9.15&b&2\\
 F15163+4255 &15 18 06.3&+42 44 37&0.0402&-&GMRT-full&absorption&177.4&-&-&-&&4\\
 F15276+1309&15 30 00.8&+12 59 22&0.0133&06AHA01&GMRT-full&diffuse emission&57.5&5.48&9.63&8.94&d&2\\
 F15437+0234  &15 46 16.3&+02 24 56&0.0128&27-067&GMRT-full&non-detection&55.2&2.60&9.27&<8.56&d&2\\
F16104+5235&16 11 40.7&+52 27 24&0.0293&-&VLA-A&absorption&128.3&3.19&10.09&-&b&11\\
F16504+0228&16 52 58.9&+02 24 03&0.0243&-&VLA-A&absorption&106.0&3.78&10.00&-&j&12\\
F16577+5900&16 58 31.4&+58 56 11&0.0183&15CSG01&GMRT-full&absorption&79.7&-&-&-&&2\\
 F20221-2458&20 25 06.6&-24 48 33&0.0106&10JLB01&GMRT-full&diffuse emission&45.6&42.29&10.32&8.48&c&2\\
 F20550+1655&20 57 23.9&+17 07 39&0.0361& &GMRT-full&absorption&158.9&2.20*&10.12&-&c&4\\
 F23254+0830&23 27 56.7&+08 46 45&0.0290&-&VLA-B&absorption&127.1&6.67&10.40&-&k&13\\
  F23488+1949&23 51 24.8&+20 06 42&0.0145&06AHA01&GMRT-full&diffuse emission&62.6&13.14&10.08&9.01&d&2\\
         
   \hline
   \end{tabular}}
   
   \vskip 0.1 true cm \noindent Notes: Col. 1: the IRAS name. Cols. 2-3: the J2000 coordinates. Col. 4: redshift. Cols. 5-6: the project code and the corresponding array. Col. 7: the type of the \HI line profiles from the project listed in Col. 5 or from the reference in Col. 13. Col. 8: the luminosity distance in (Mpc). Cols. 9-10: the \HI line flux in \textbf{Jy \kms} and mass from single-dish observations, the ’*’ sign stands for the \HI line flux that we fit from the line profiles in the literature. \textbf{Col. 11}: the \HI mass estimated using line profiles presented in Fig. \ref{HIline}. Col. 12: the References for the single-dish \HI line flux and mass:
   (a) \cite{1984ApJ...277...92M}; (b) \cite{1991A&A...245..393M}; (c) \cite{2015MNRAS.447.1531C}; (d) \cite{2018ApJ...861...49H};
   (e) \cite{2017A&A...599A.104T}; (f) \cite{1983ApJS...53..269D};
(g) \cite{1986AJ.....91..732B}; (h) \cite{1989gcho.book.....H}; (i) \cite{1974AJ.....79..767P}; (j) \cite{1994AJ....107...99R}; (k) \cite{1982ApJ...260...75M};
   Col. 13: the references for the spectral type in Col. 7 and \HI mass in Col. 11: (1) \cite{2014AJ....147...74F}; (2) this work; (3) \cite{2018MNRAS.480..947D}; (4) \cite{2019MNRAS.489.1099D}; (5) \cite{1994ApJ...422L..13G}; (6) \cite{2010A&A...513A..11O}; (7) \cite{2004ASPC..320..112T}; (8) \cite{1990ApJ...364...65B}; (9) \cite{2005A&A...444..791B}; (10) \cite{1988AJ.....96.1227H};
(11) \cite{2022ApJ...934...69L}; (12) \cite{2007ApJ...661..173B}; (13) \cite{1986ApJ...300..190D};

   \end{table*}


\counterwithin{figure}{section}

\begin{figure*}
   \centering
 \includegraphics[width=0.4\textwidth]{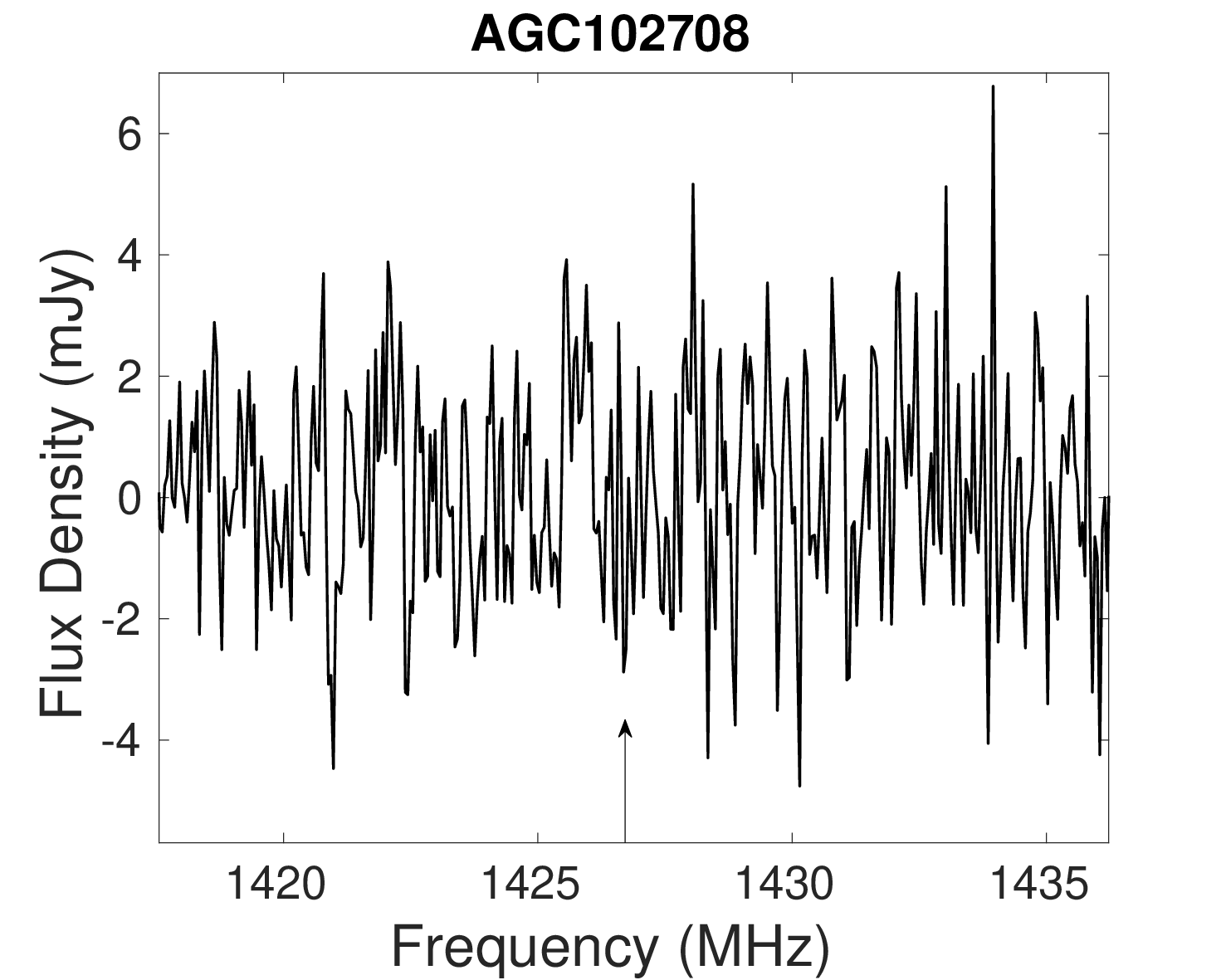}
 \includegraphics[width=0.4\textwidth]{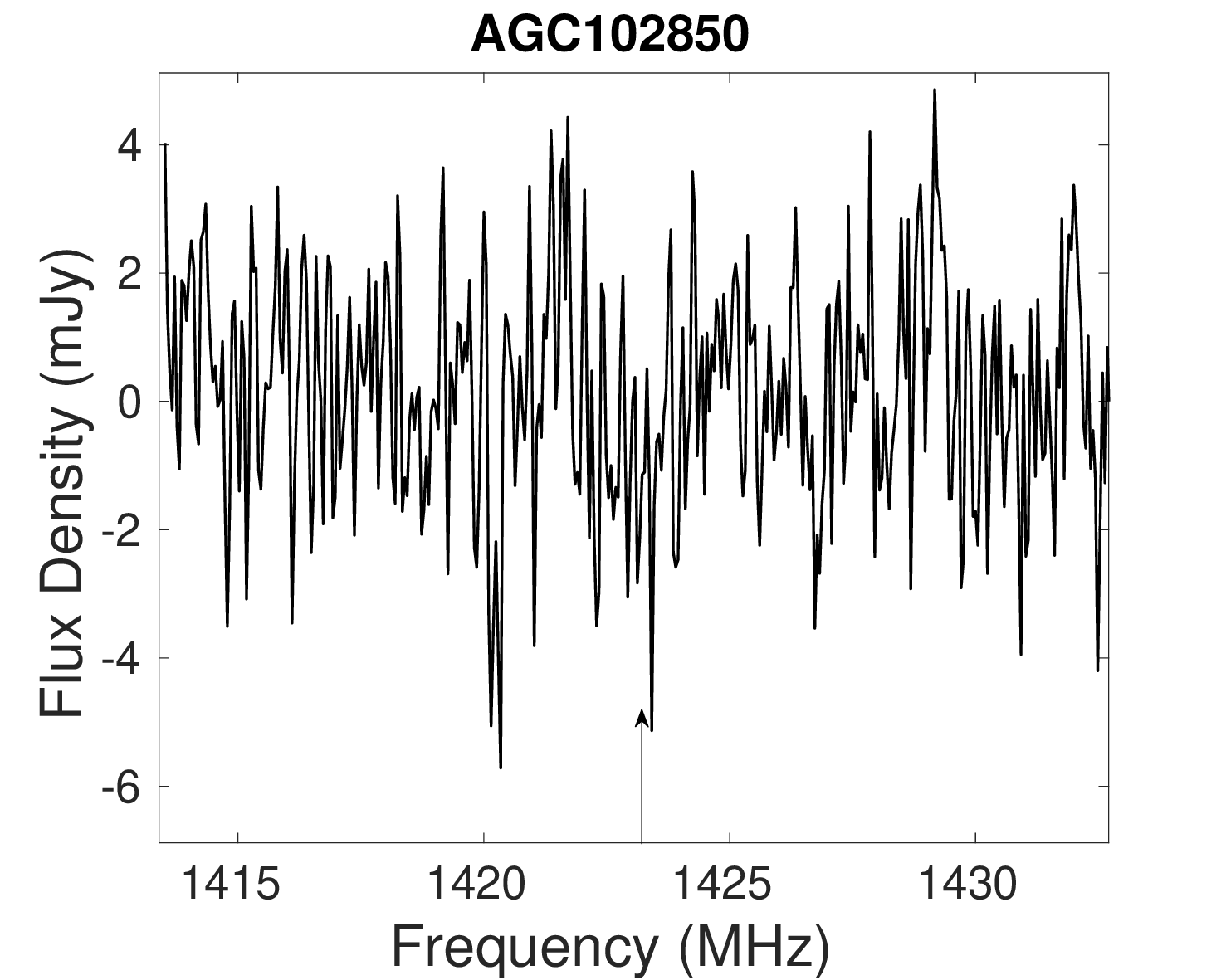}
 \includegraphics[width=0.4\textwidth]{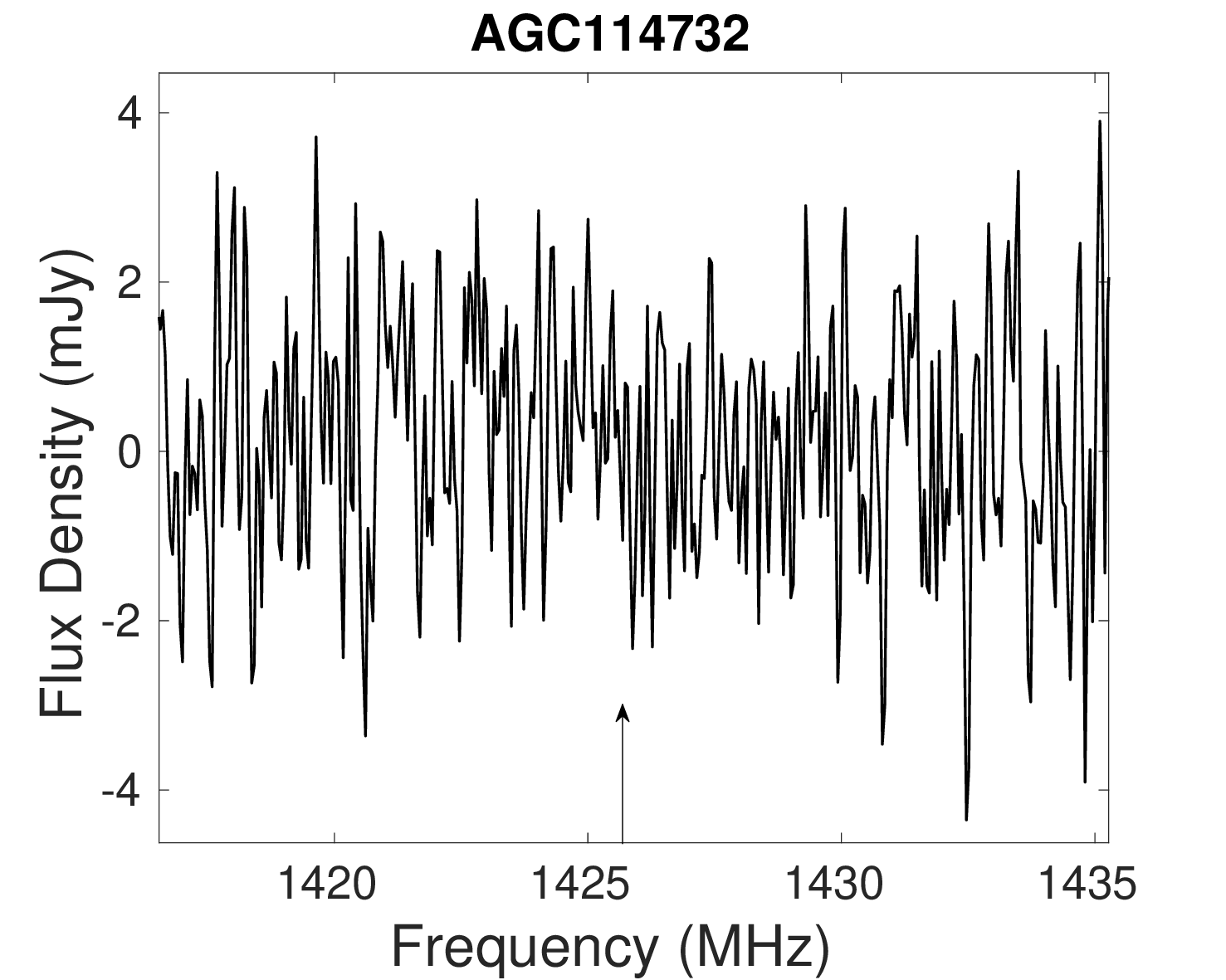}
 \includegraphics[width=0.4\textwidth]{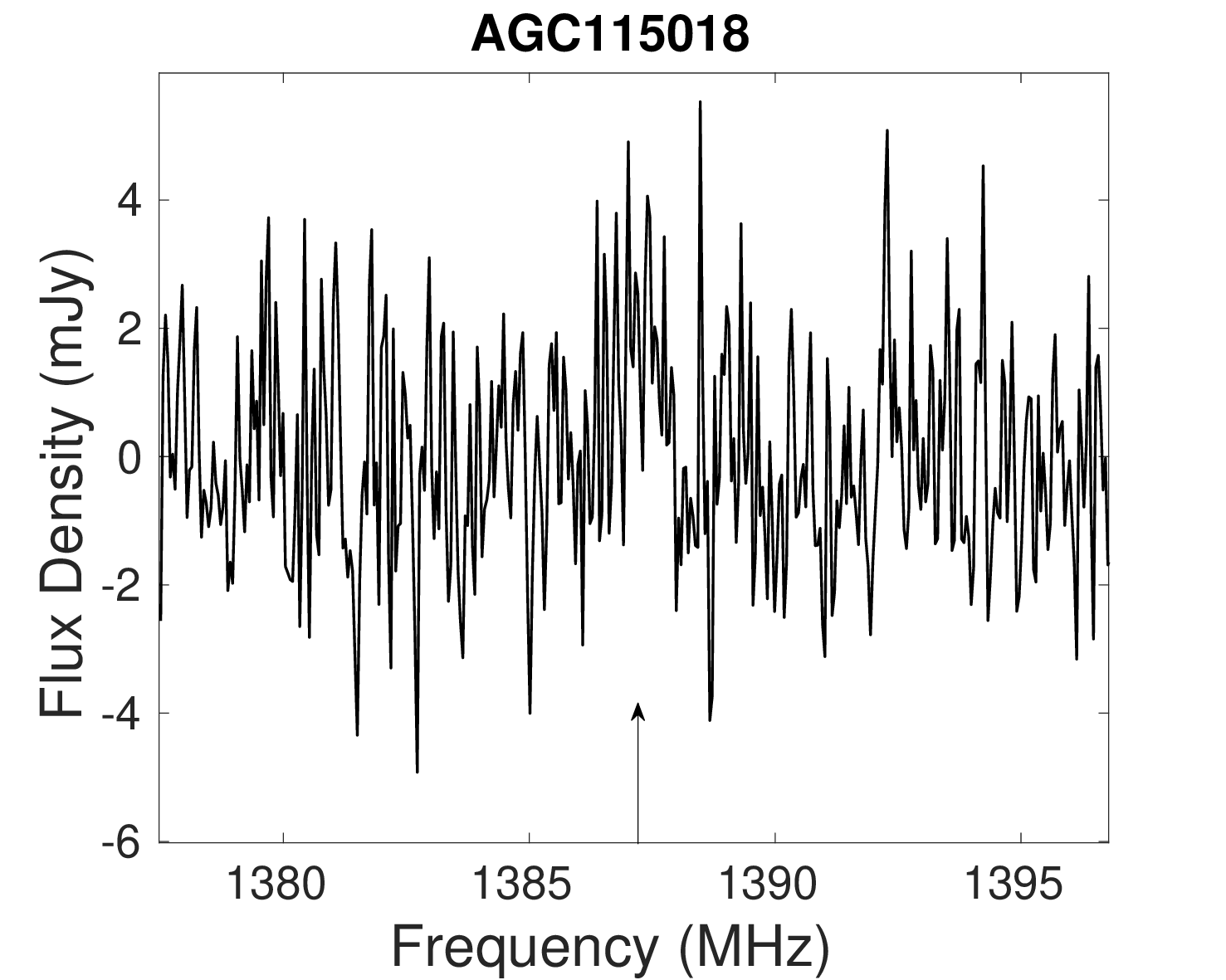}
  \includegraphics[width=0.4\textwidth]{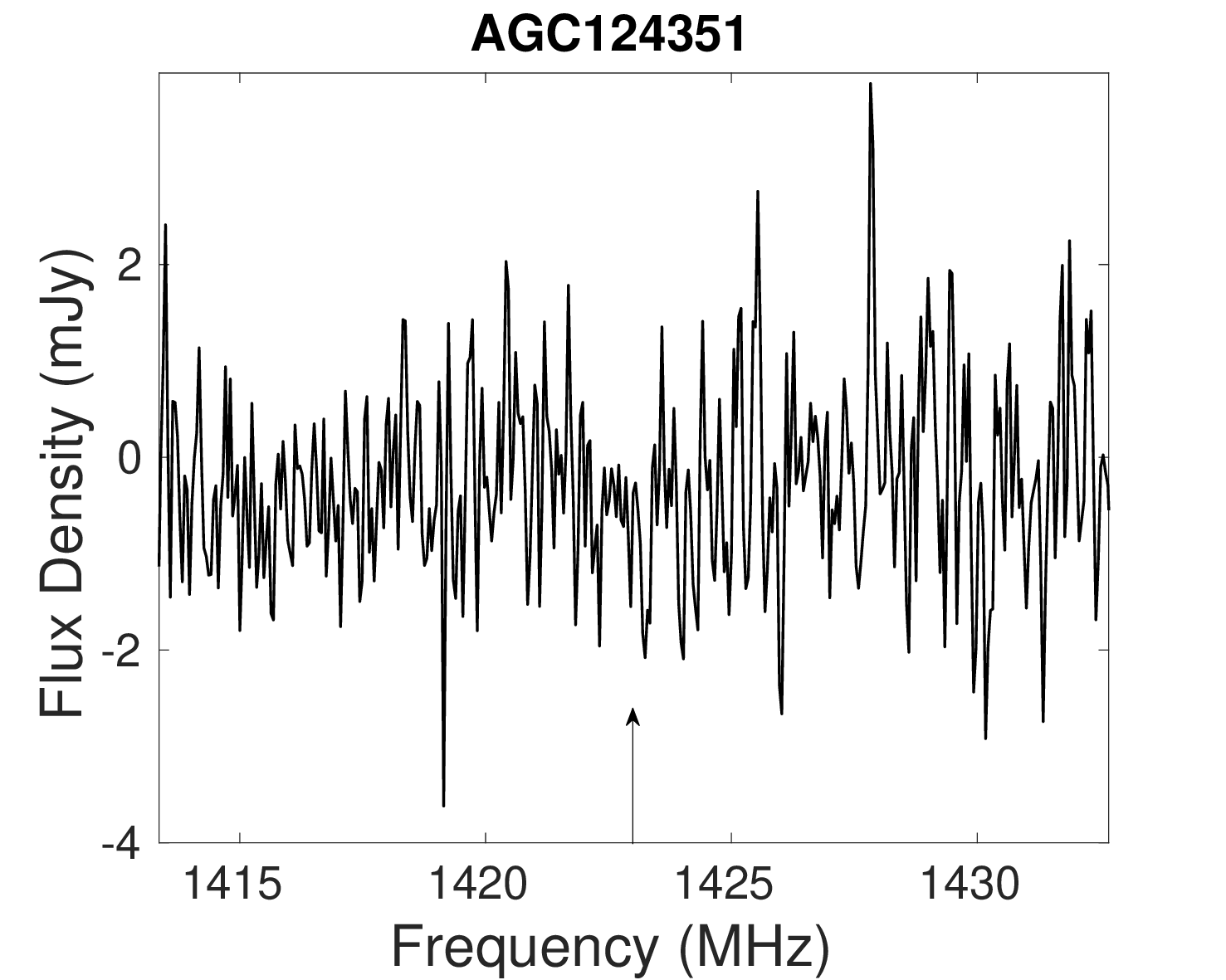}
 \includegraphics[width=0.4\textwidth]{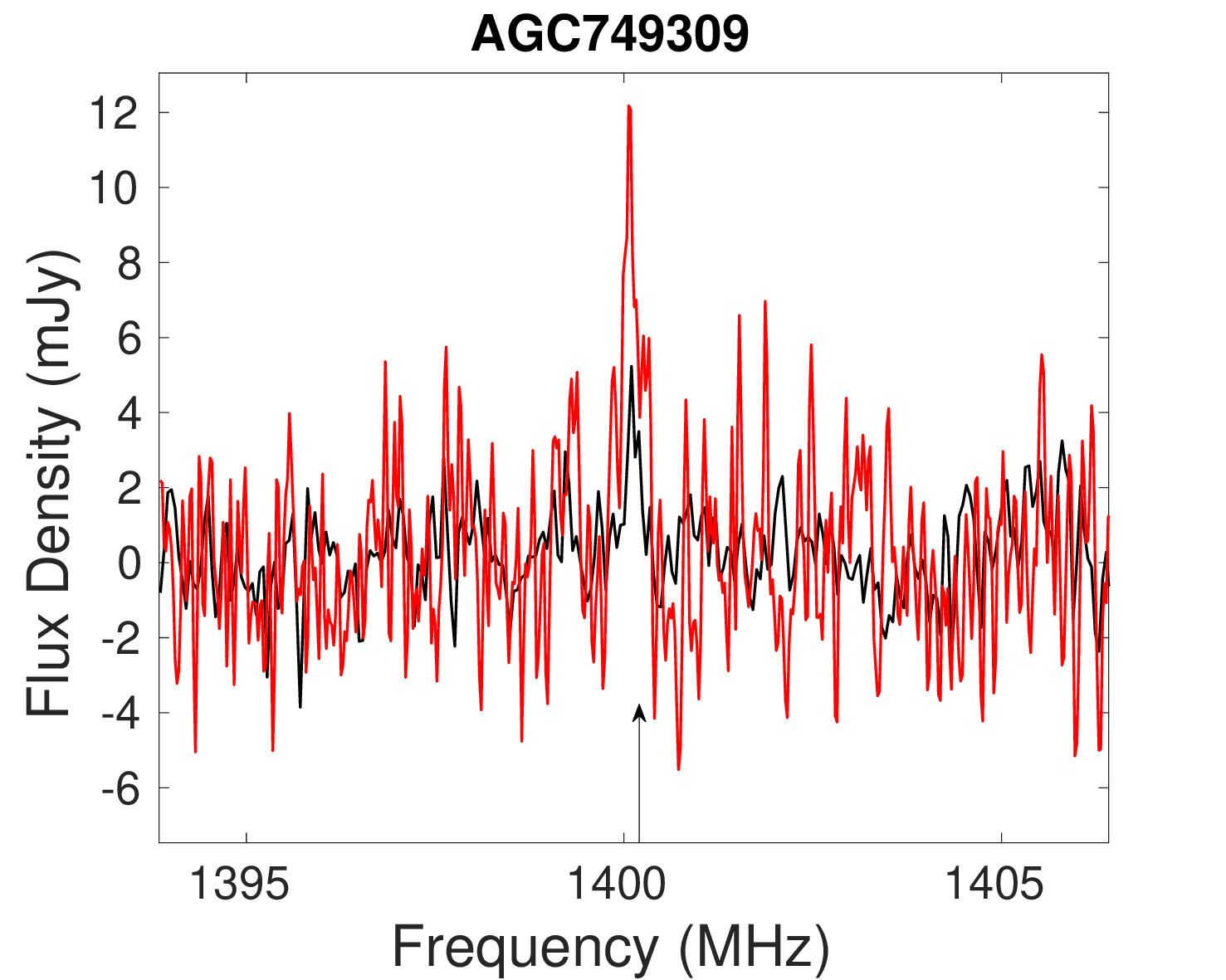}

  \includegraphics[width=0.4\textwidth]{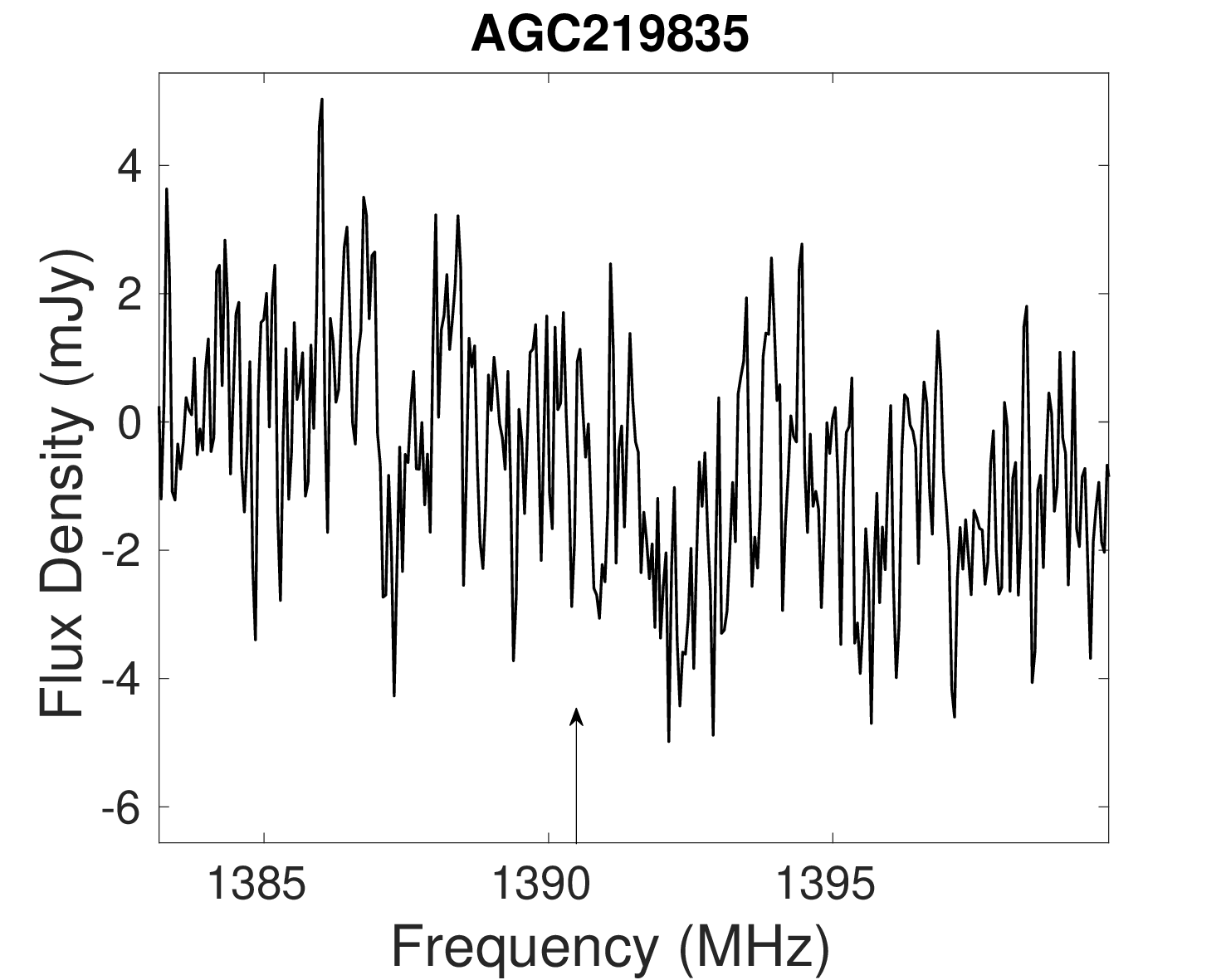}
 \includegraphics[width=0.4\textwidth]{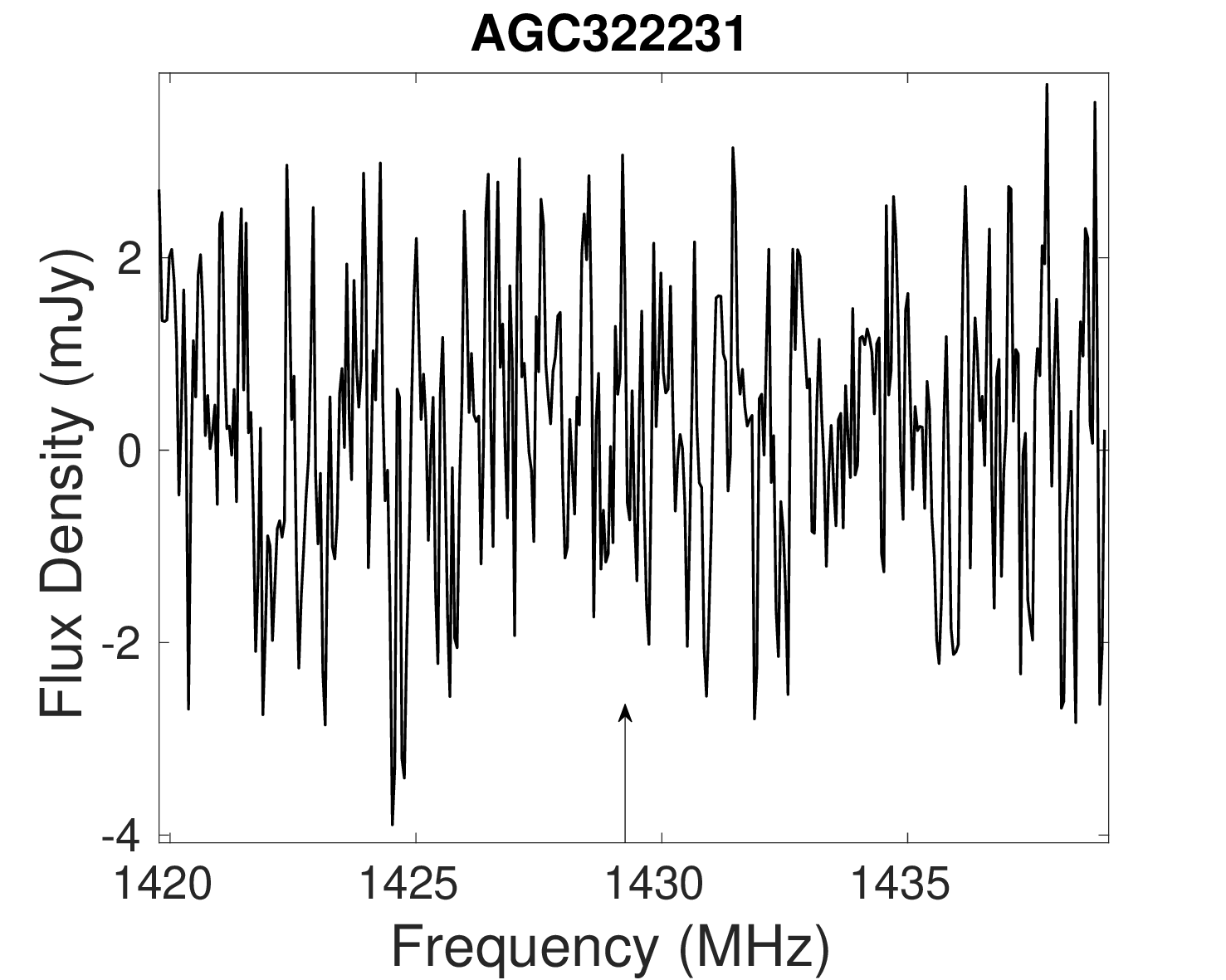}
      \caption{OH line profiles generated using the data from the low-resolution GMRT-central. 
      The OH line profiles were extracted from a circular region with sizes of 60"$\times$60" centered at the position given by \cite{2018ApJ...861...49H} (see Table \ref{tablea1} and Fig. \ref{contour}). The arrow represents the peak frequency of the line profiles given by \cite{2018ApJ...861...49H}. The red line profiles of AGC 749309 represent the ALFALFA \textbf{spectra} of this source from \cite{2011AJ....142..170H}. }

    \label{figa1}%
\end{figure*}


\begin{figure*}
   \centering
 \includegraphics[width=0.49\textwidth]{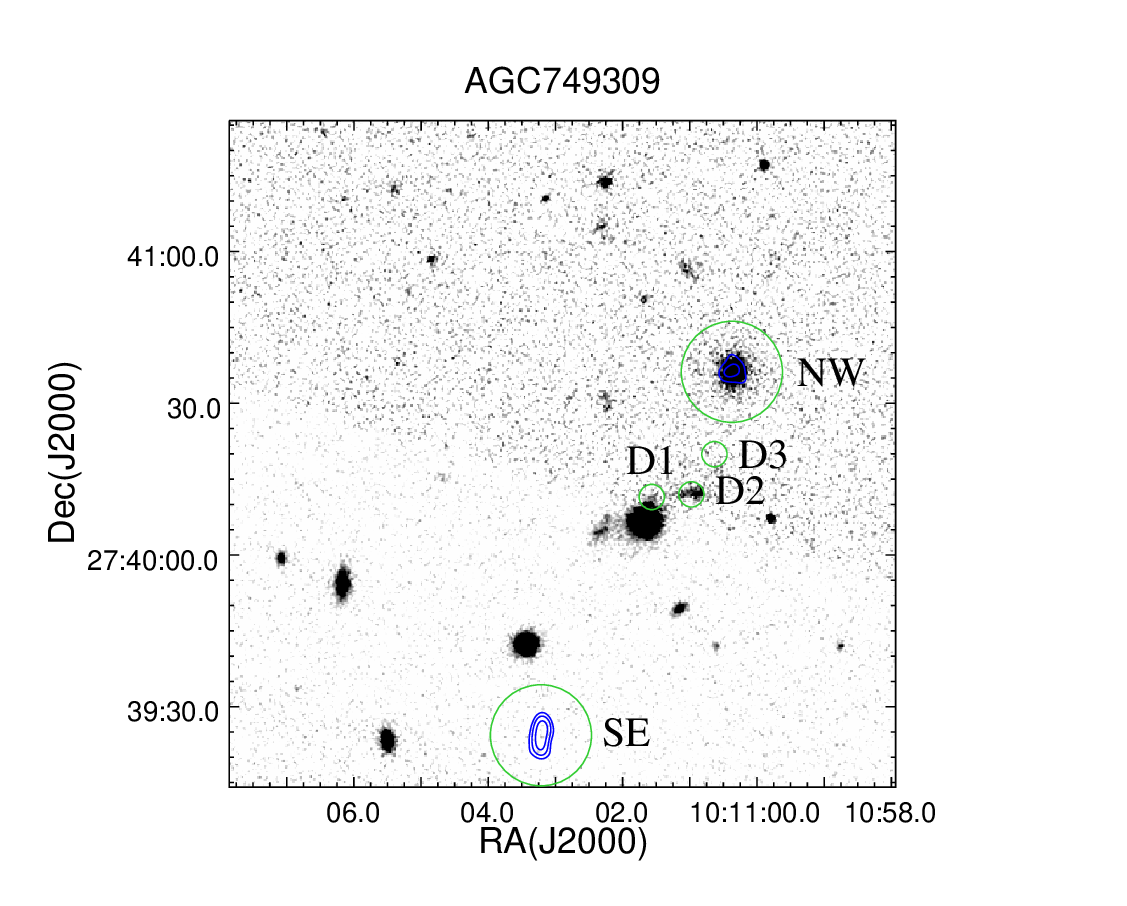}
\includegraphics[width=0.49\textwidth]{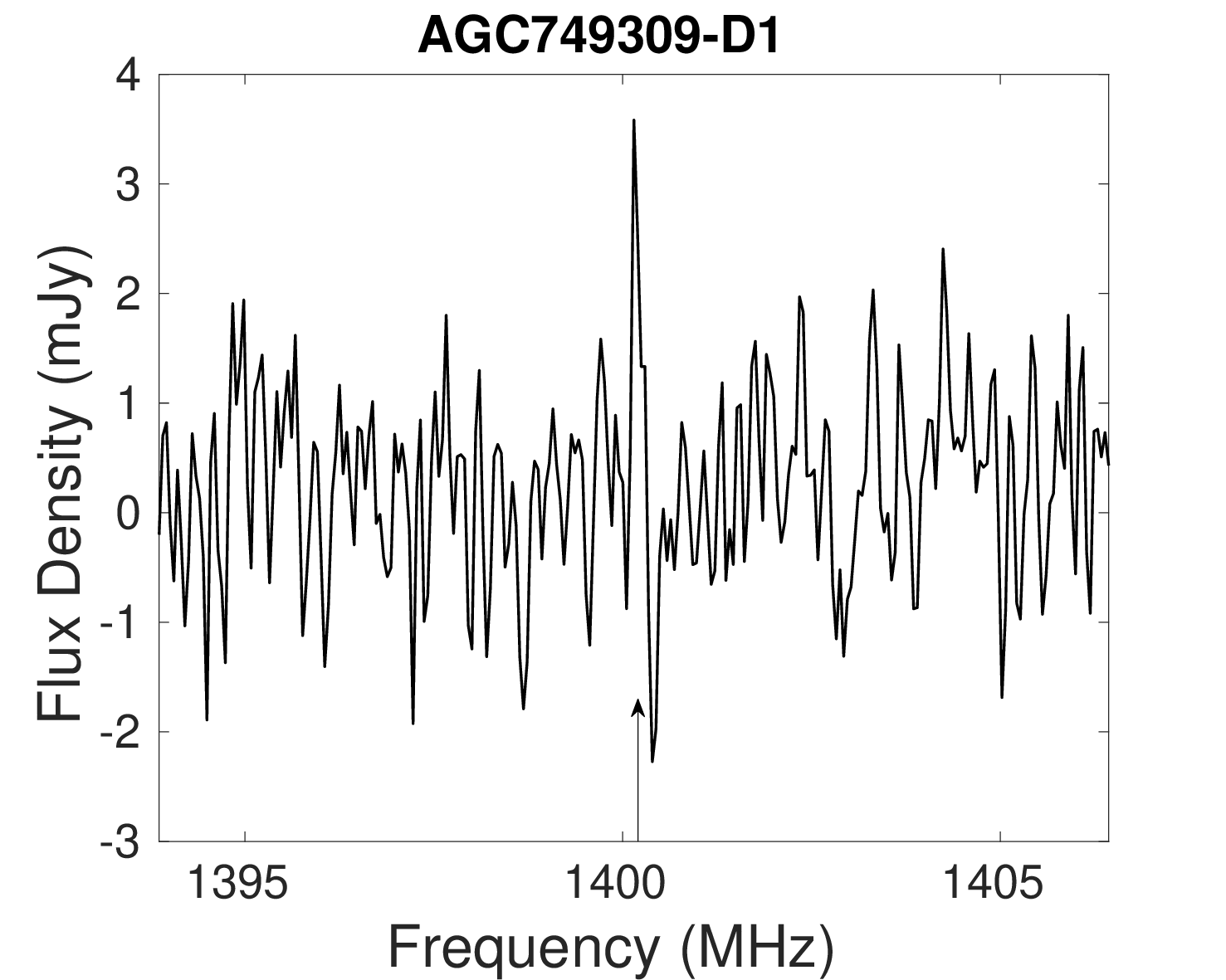}
\includegraphics[width=0.49\textwidth]{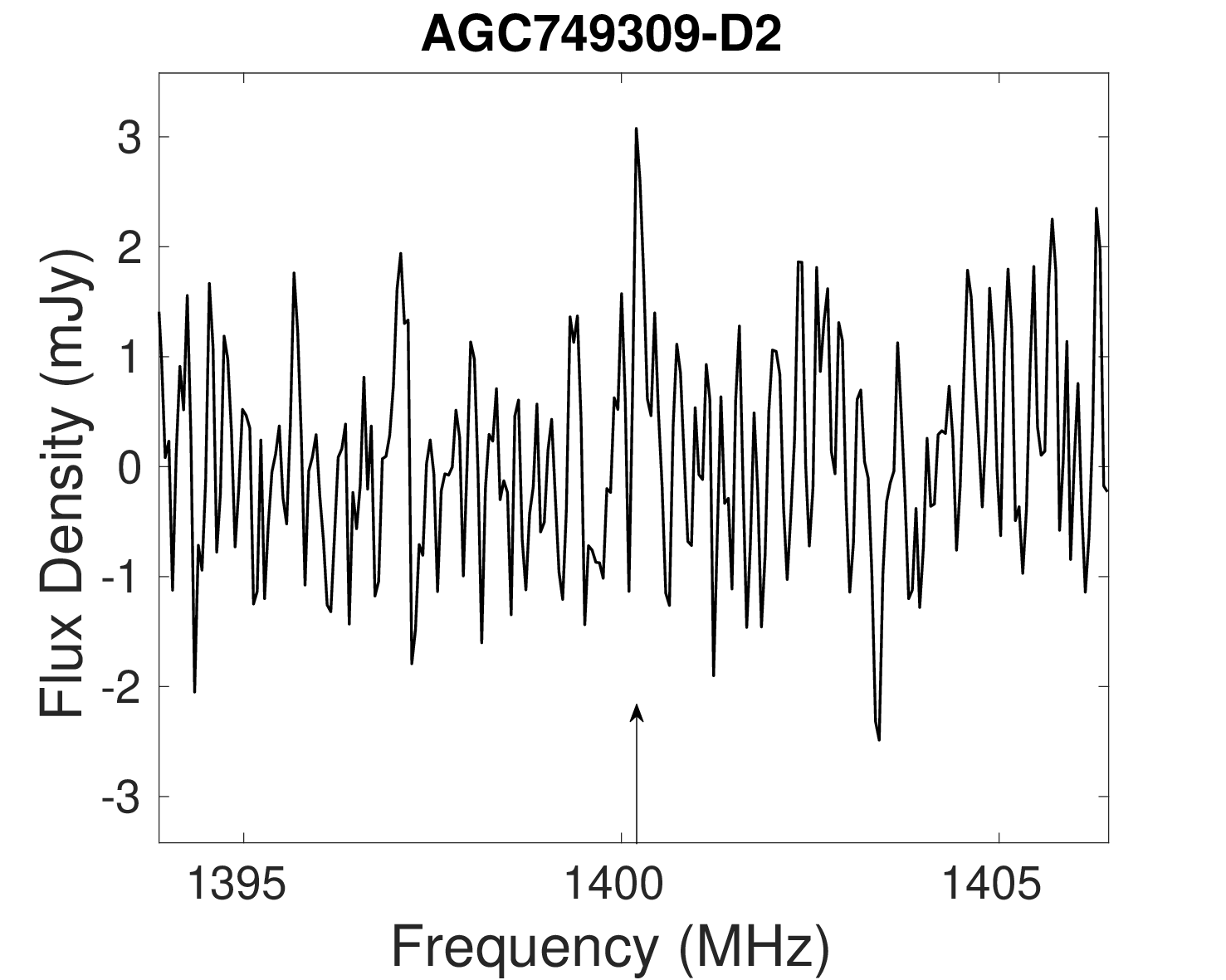}
\includegraphics[width=0.49\textwidth]{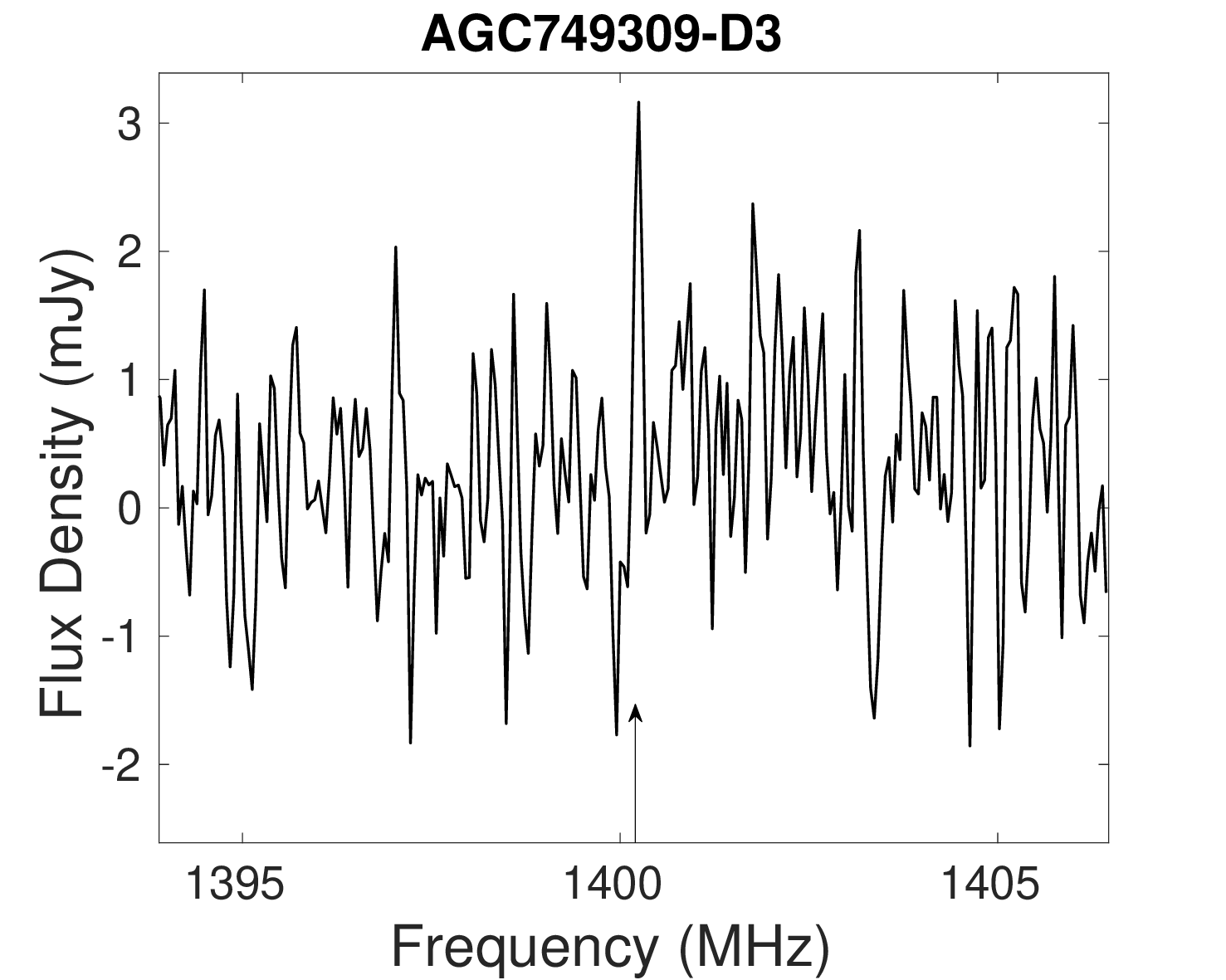}
\includegraphics[width=0.49\textwidth]{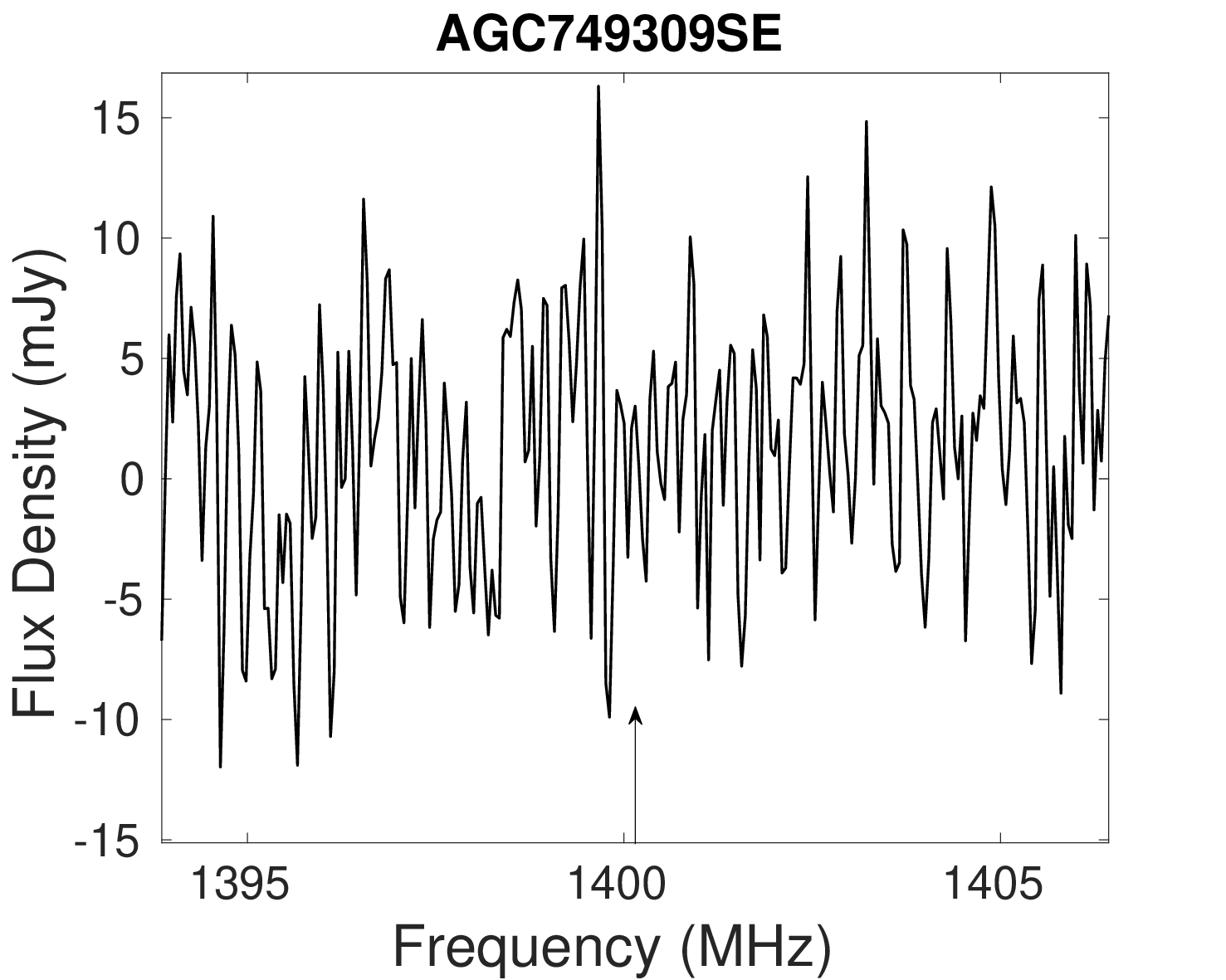}
\includegraphics[width=0.49\textwidth]{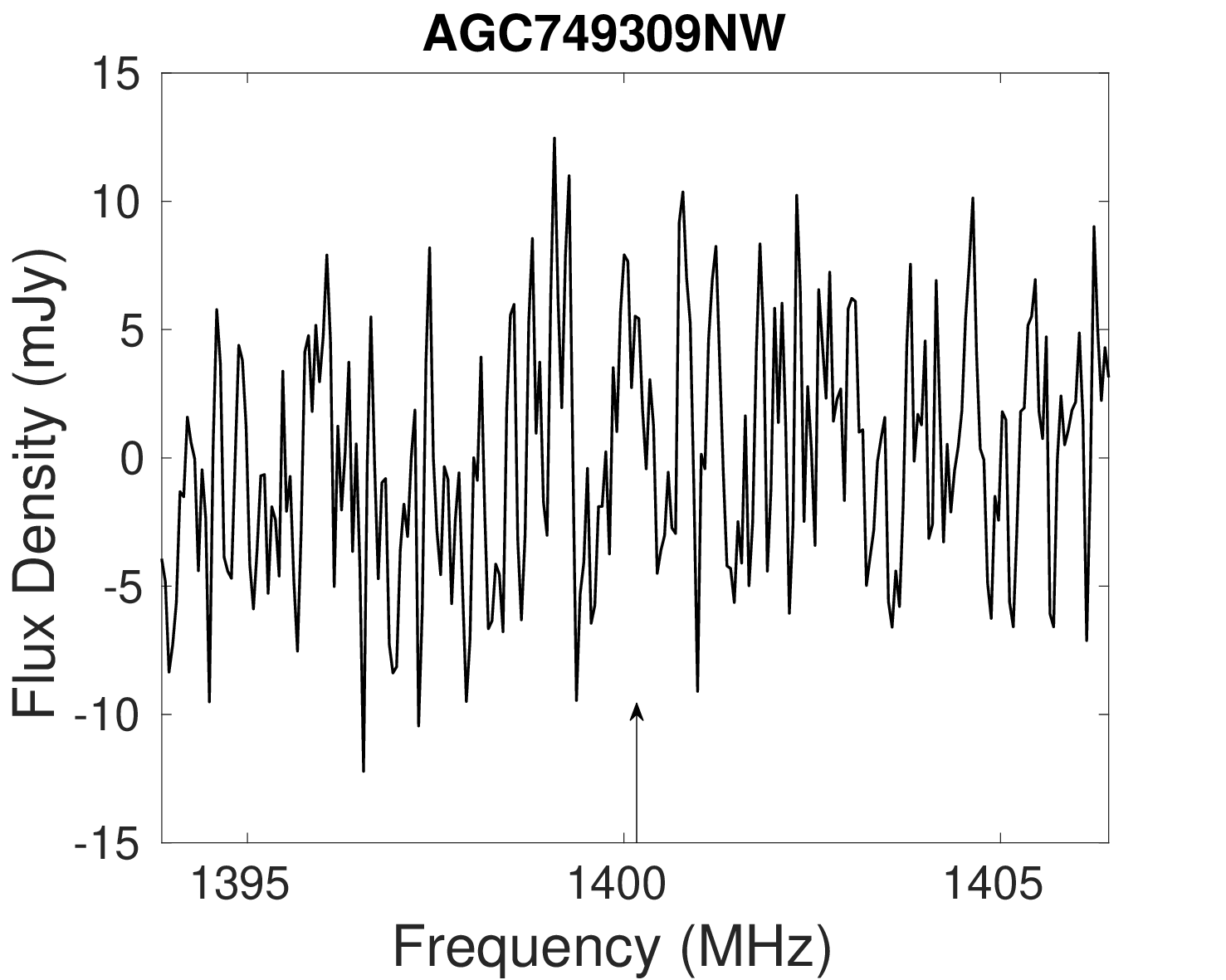}

      \caption{Radio line profiles of AGC 749309 generated using the high-resolution GMRT-full. Top left: the R-band SDSS image of AGC 749309. The green circles stand for the regions where we extract the spectral line emission. The size of the large and smaller cycles are 20"$\times$20" and 5"$\times$5", respectively. The extracted line profiles from these regions are present in other panels of this figure.
      }
      \label{GMRTcontinuum1}
\end{figure*}


\begin{figure*}
   \centering
   \includegraphics[width=0.49\textwidth]{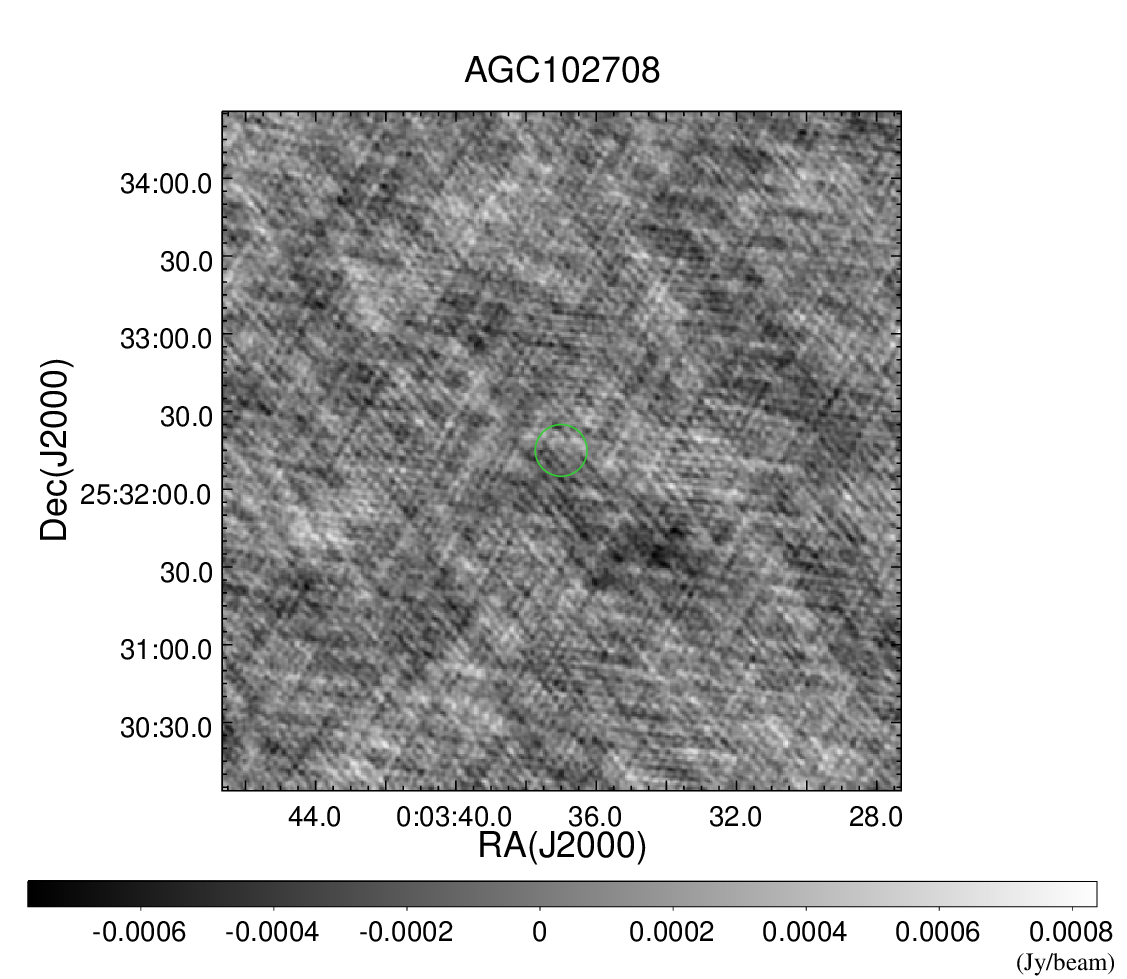}
   \includegraphics[width=0.49\textwidth]{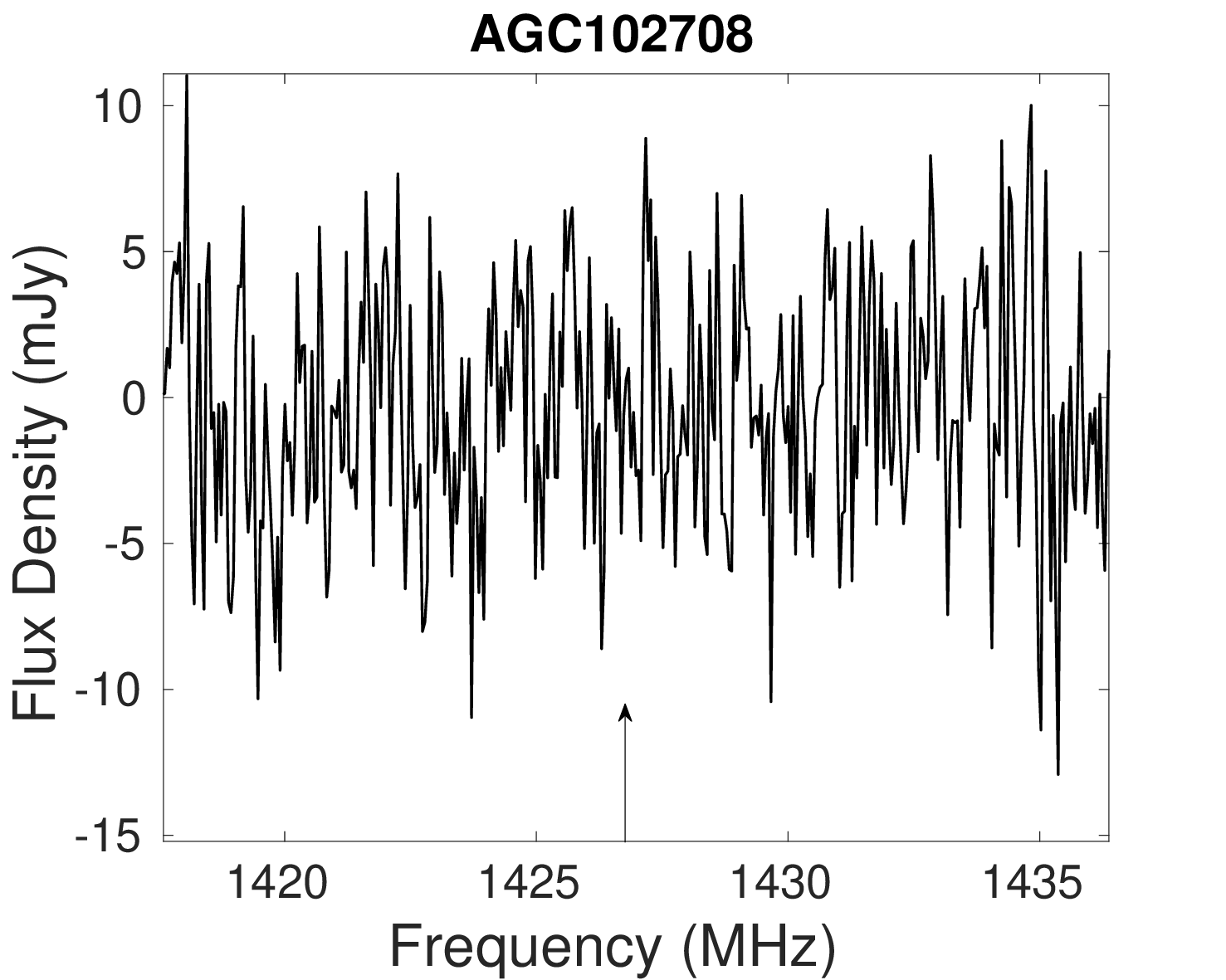}
    \includegraphics[width=0.49\textwidth]{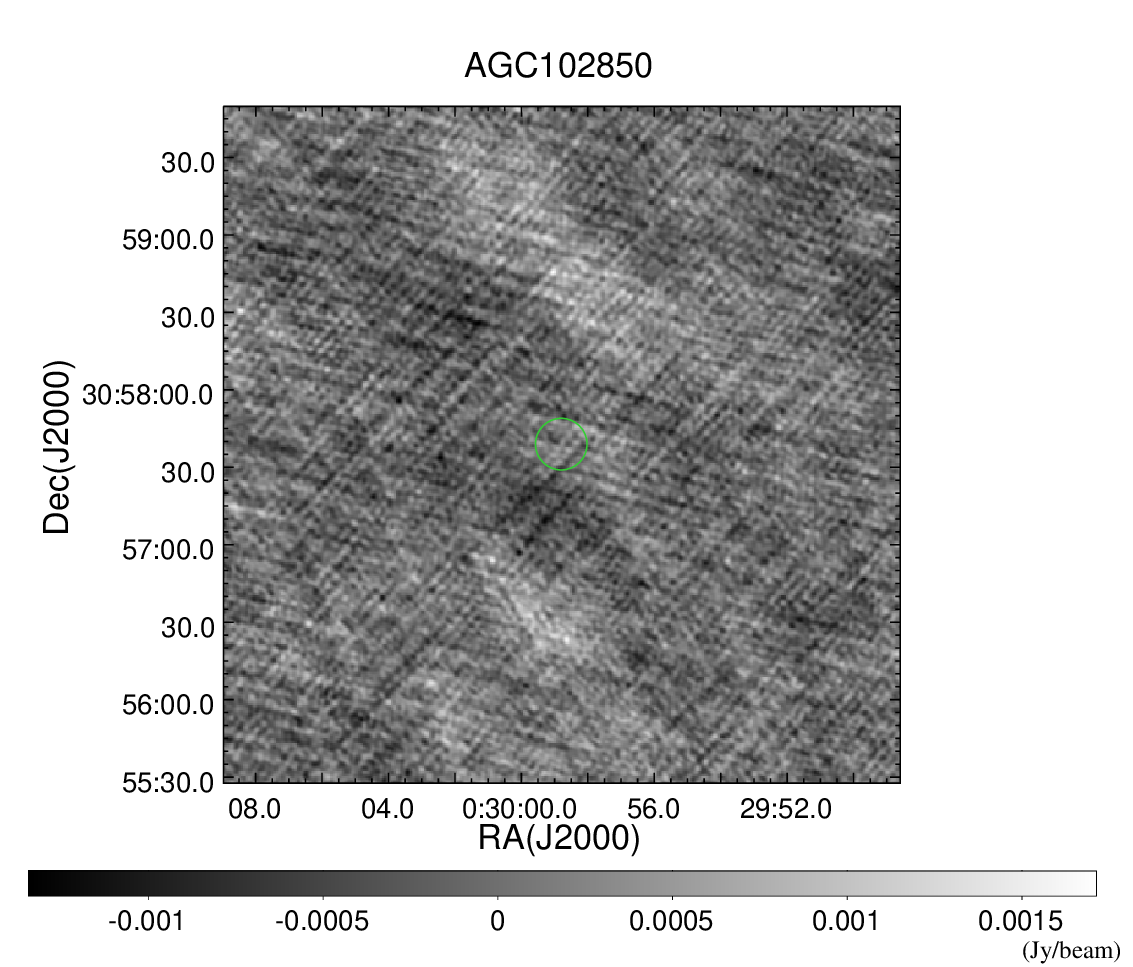}
   \includegraphics[width=0.49\textwidth]{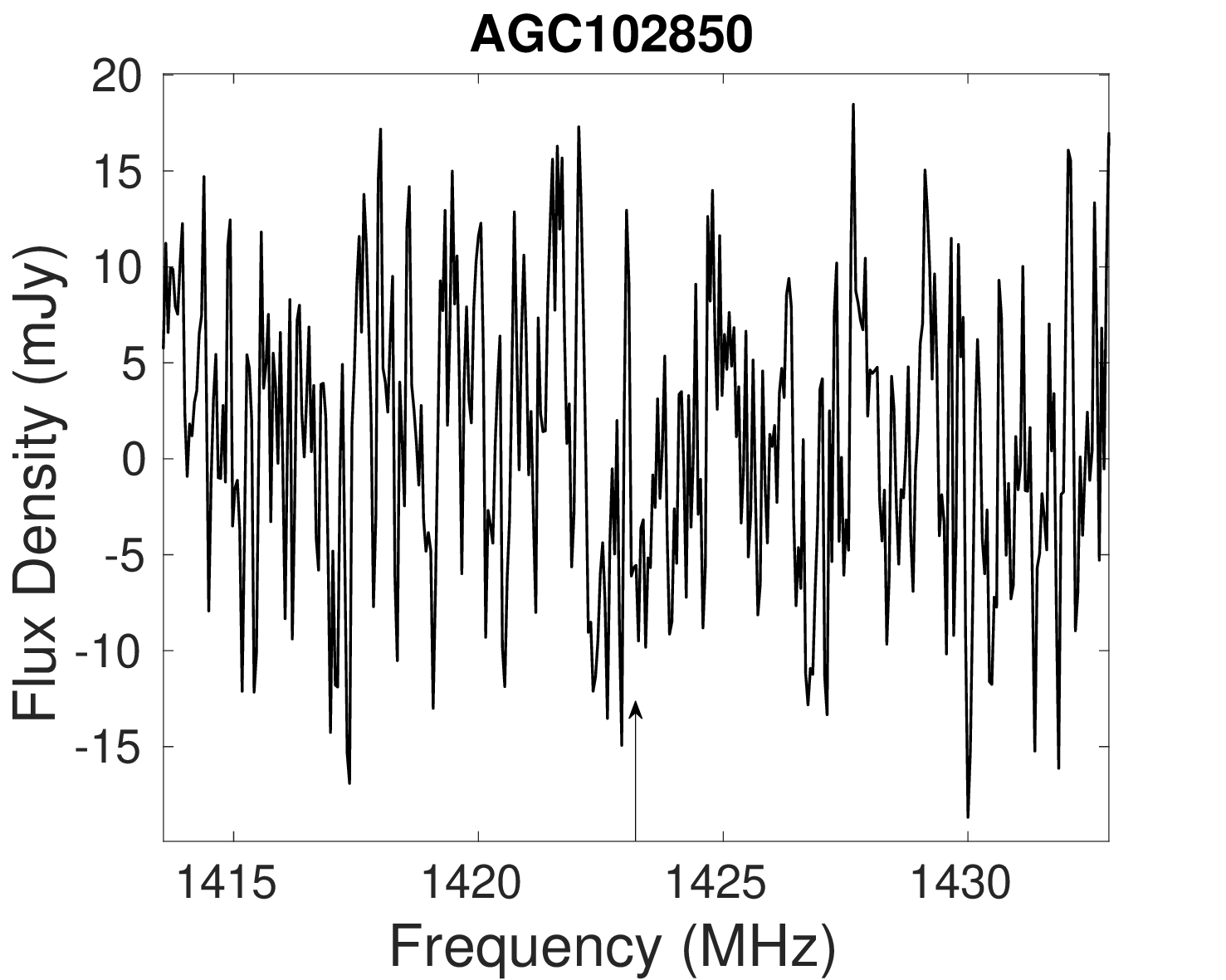}
   \includegraphics[width=0.49\textwidth]{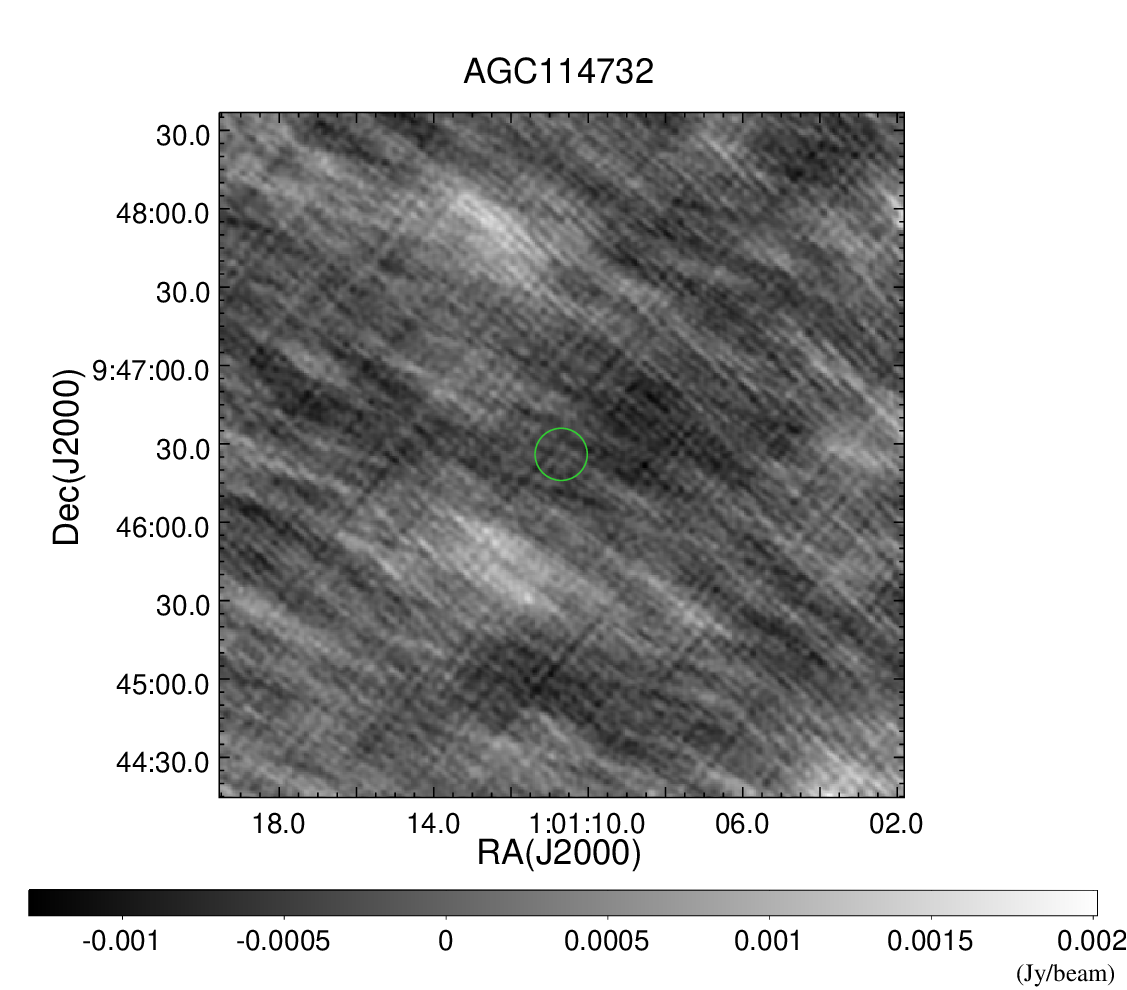}
   \includegraphics[width=0.49\textwidth]{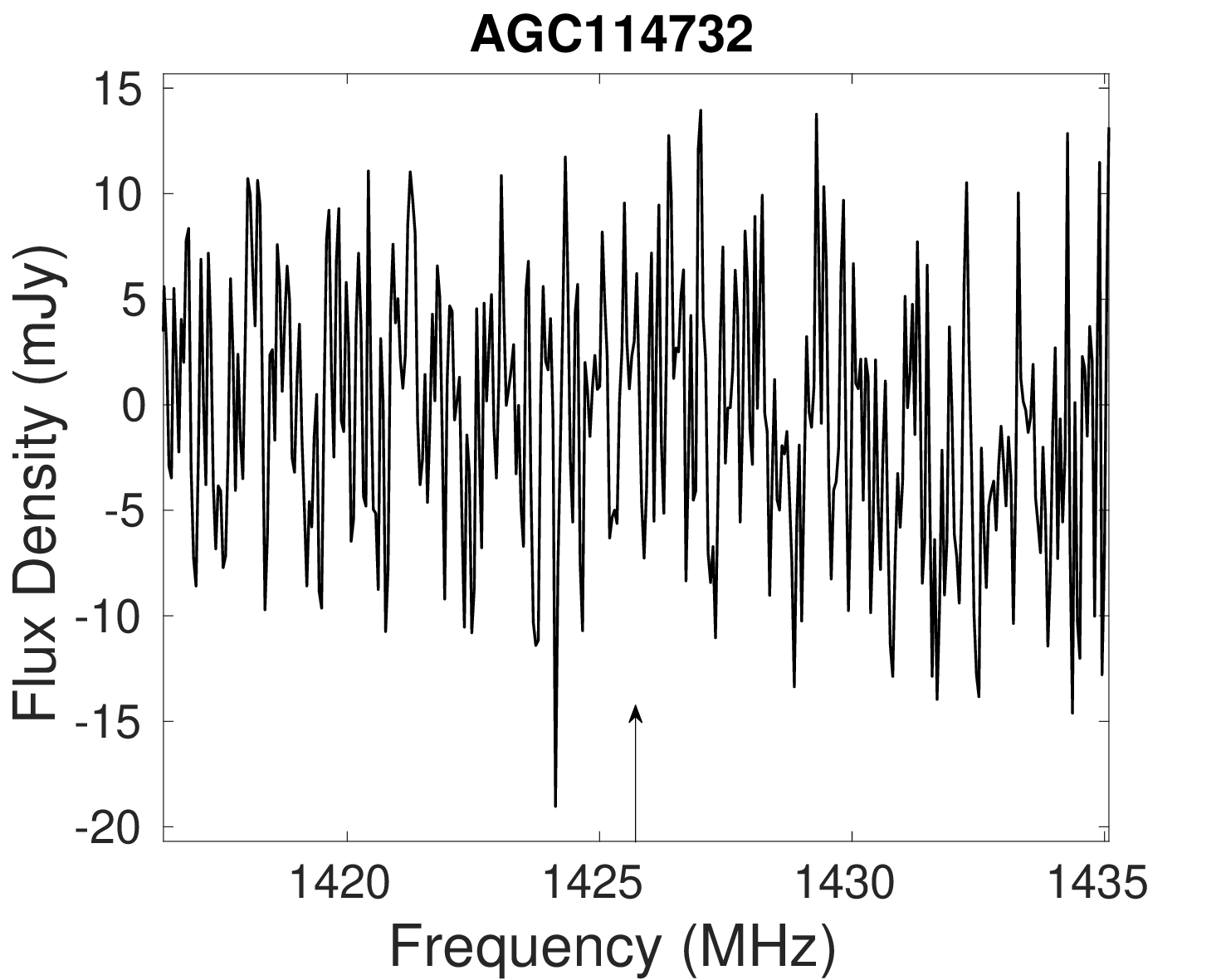}
  
      \caption{OH line profiles generated using the high-resolution GMRT-full. Left: The dirty map of possible OH line emission about 5 MHz bands centered at the peak frequency given by \cite{2018ApJ...861...49H}. The OH line profiles are mostly generated from green circles with size of about 20"$\times$20" except for AGC219835, which is 40"$\times$40". 
      }
      \label{GMRTcontinuum2}
\end{figure*}


\addtocounter{figure}{-1}

\begin{figure*}
   \centering
    \includegraphics[width=0.49\textwidth]{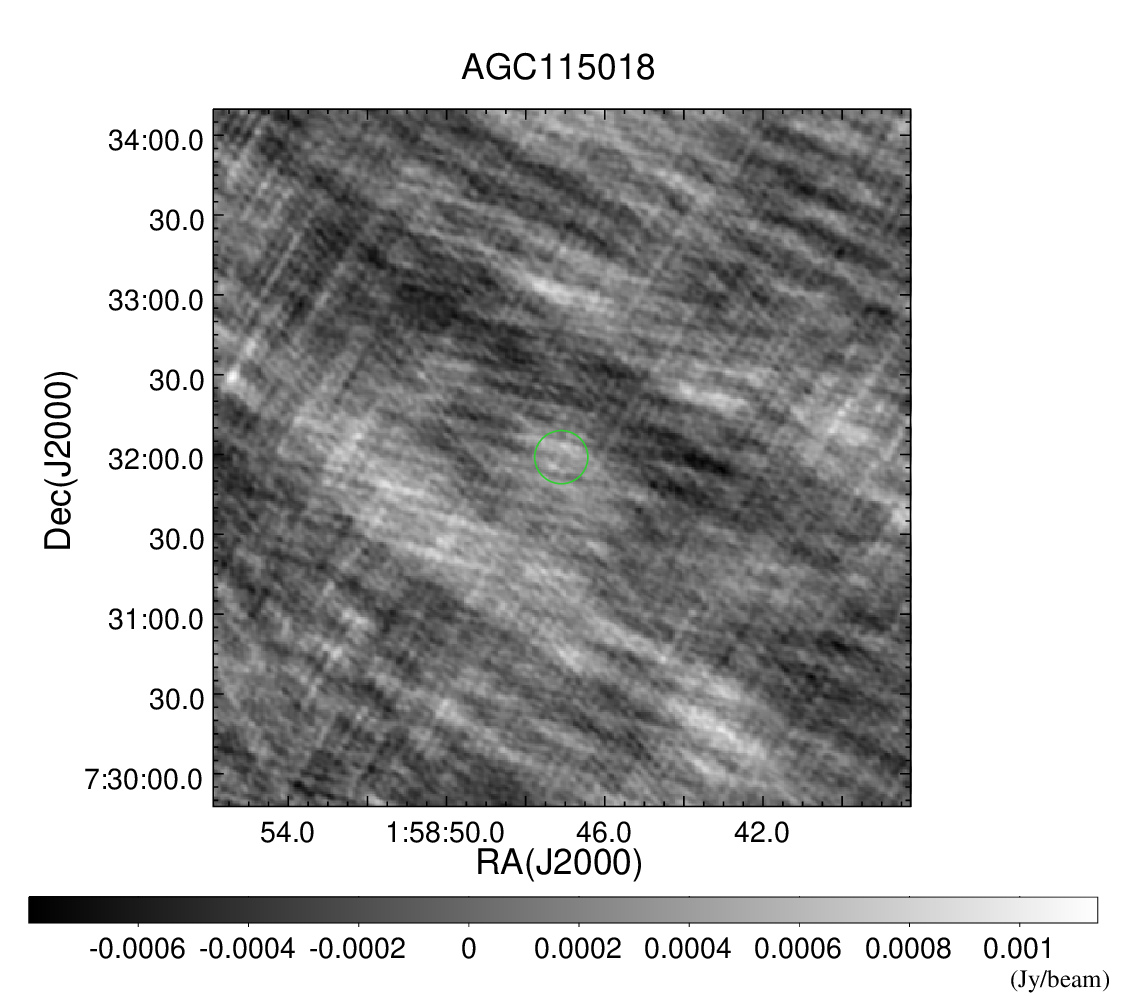}
    \includegraphics[width=0.49\textwidth]{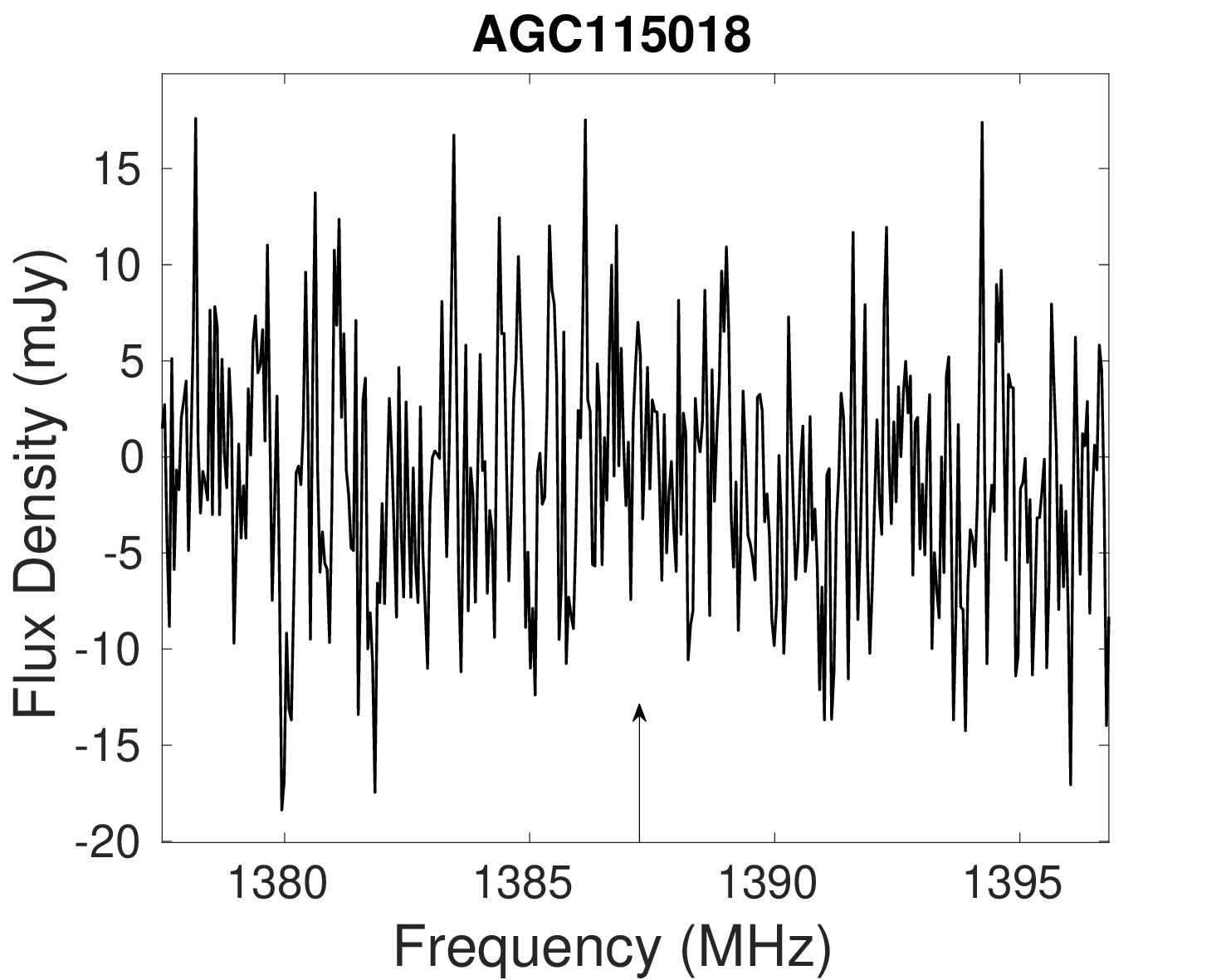}
   \includegraphics[width=0.49\textwidth]{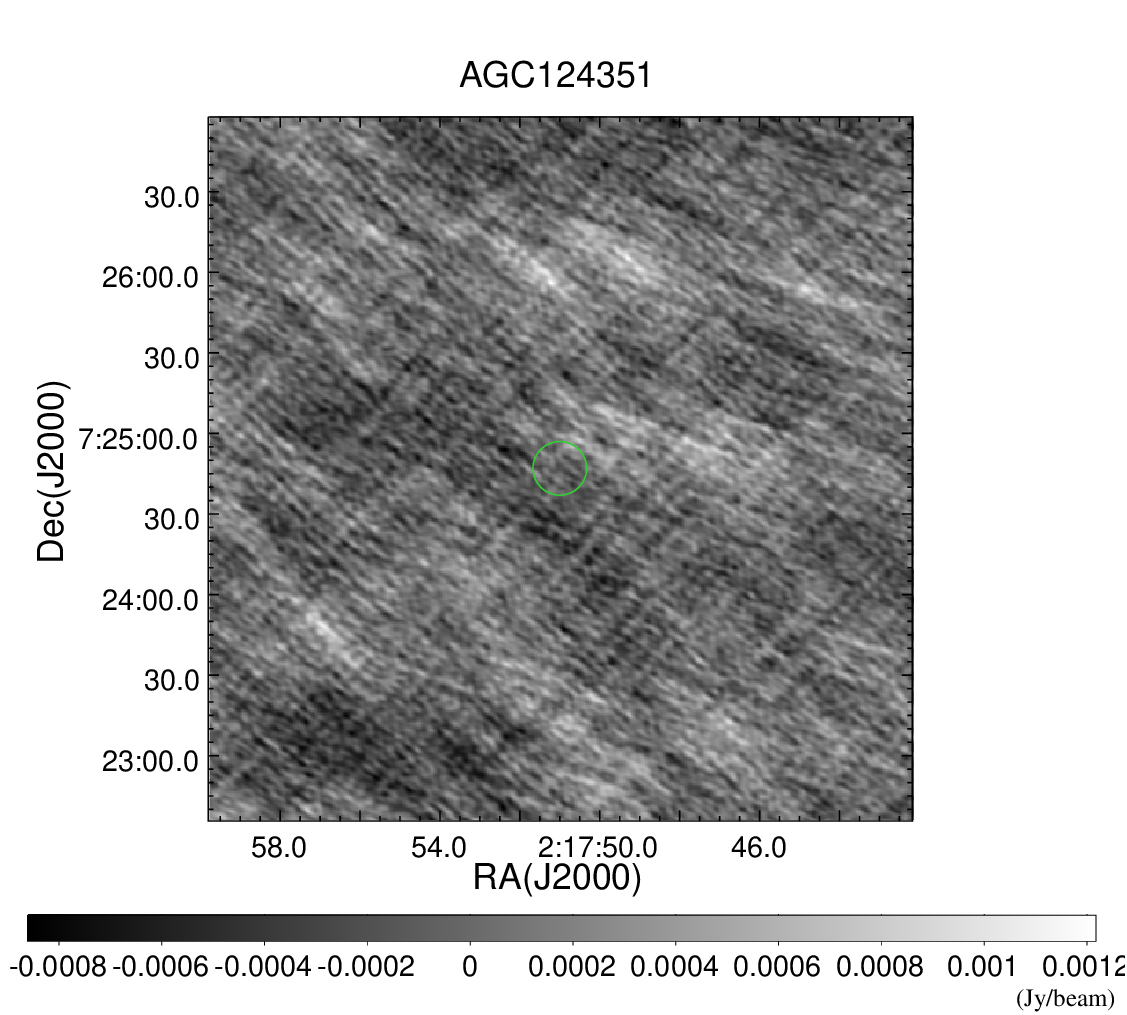}
   \includegraphics[width=0.49\textwidth]{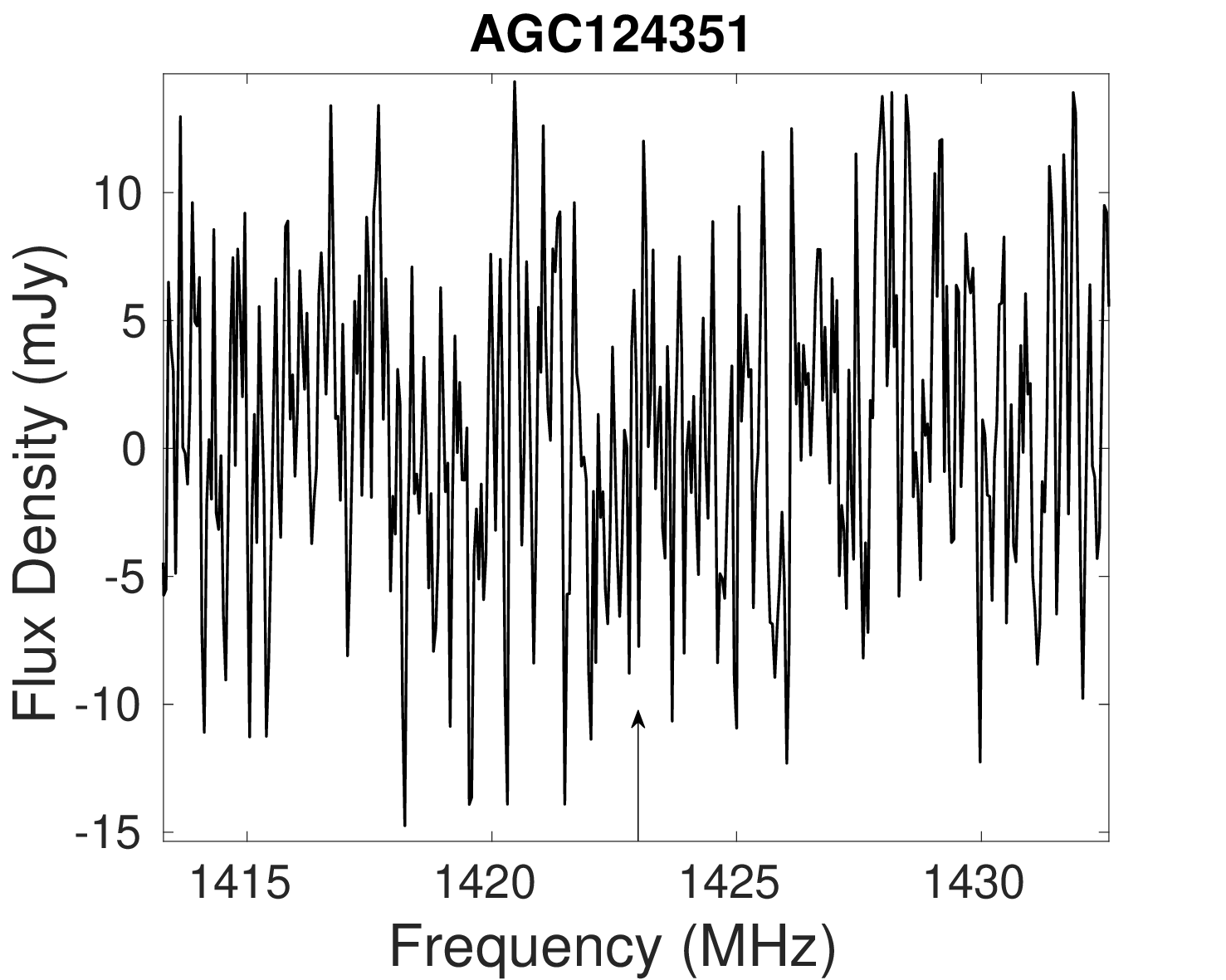}
   \includegraphics[width=0.49\textwidth]{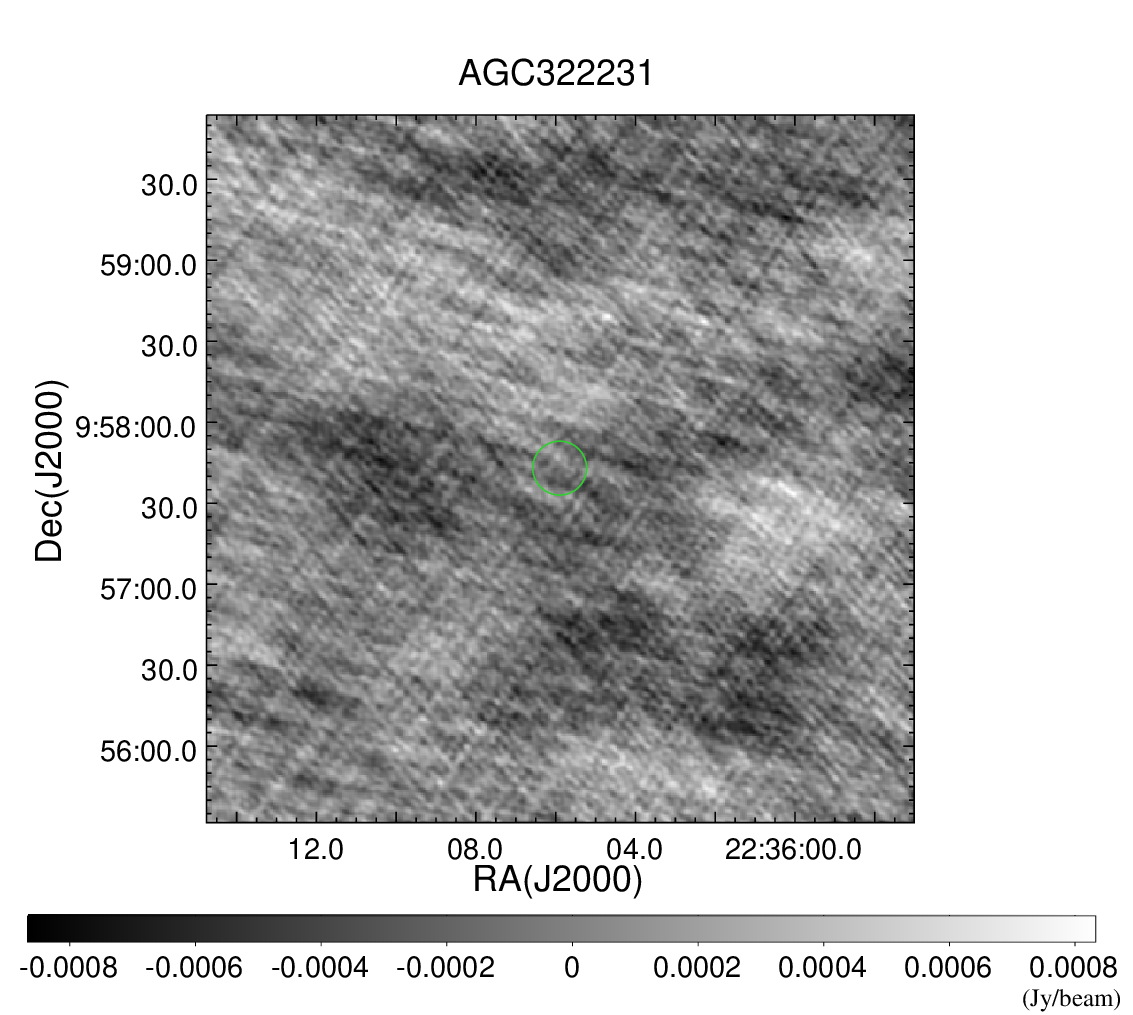}
   \includegraphics[width=0.49\textwidth]{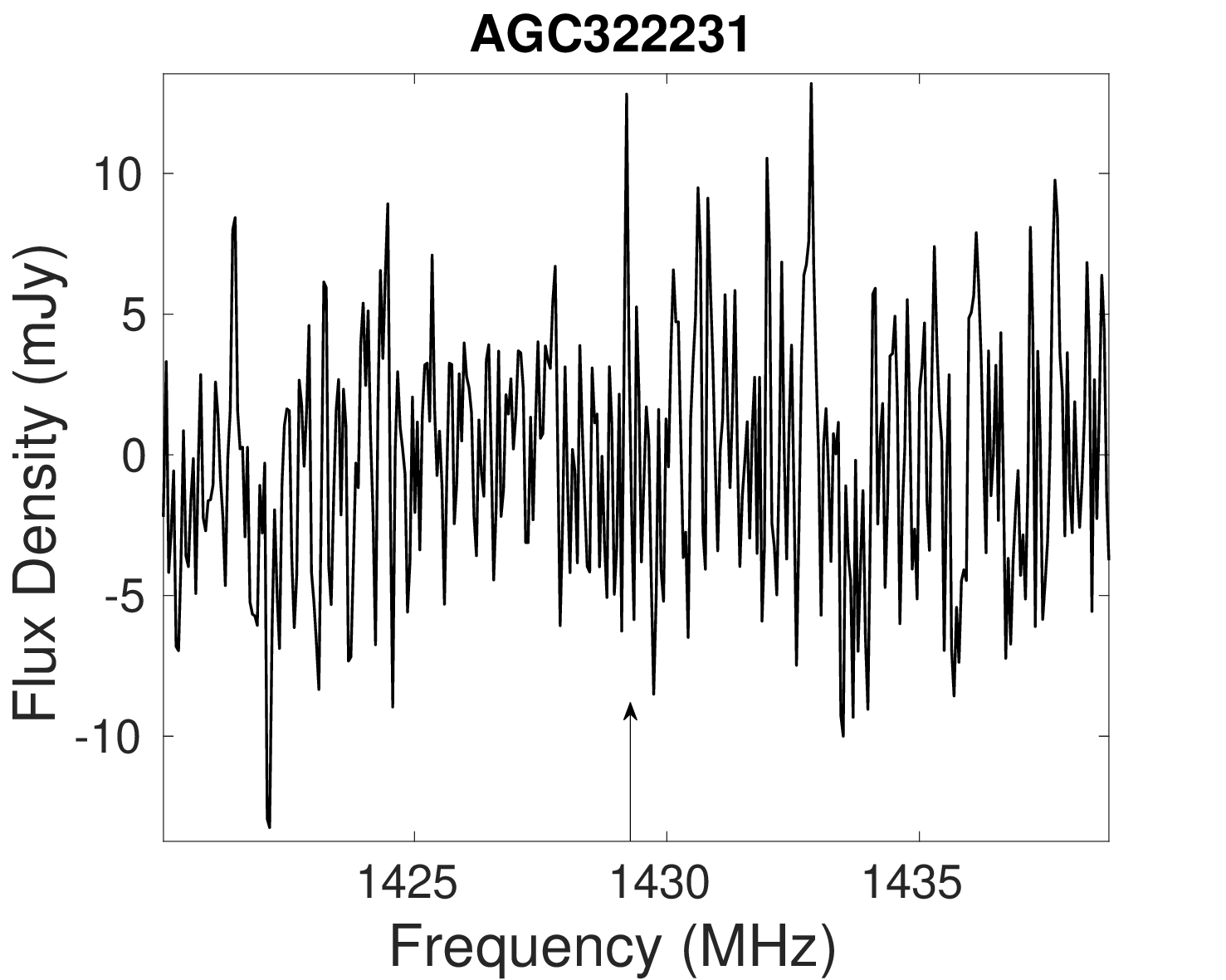}

      \caption{Continued
      }
      \label{GMRTcontinuum2}
\end{figure*}
\addtocounter{figure}{-1}

\begin{figure*}
   \centering

    \includegraphics[width=0.49\textwidth]{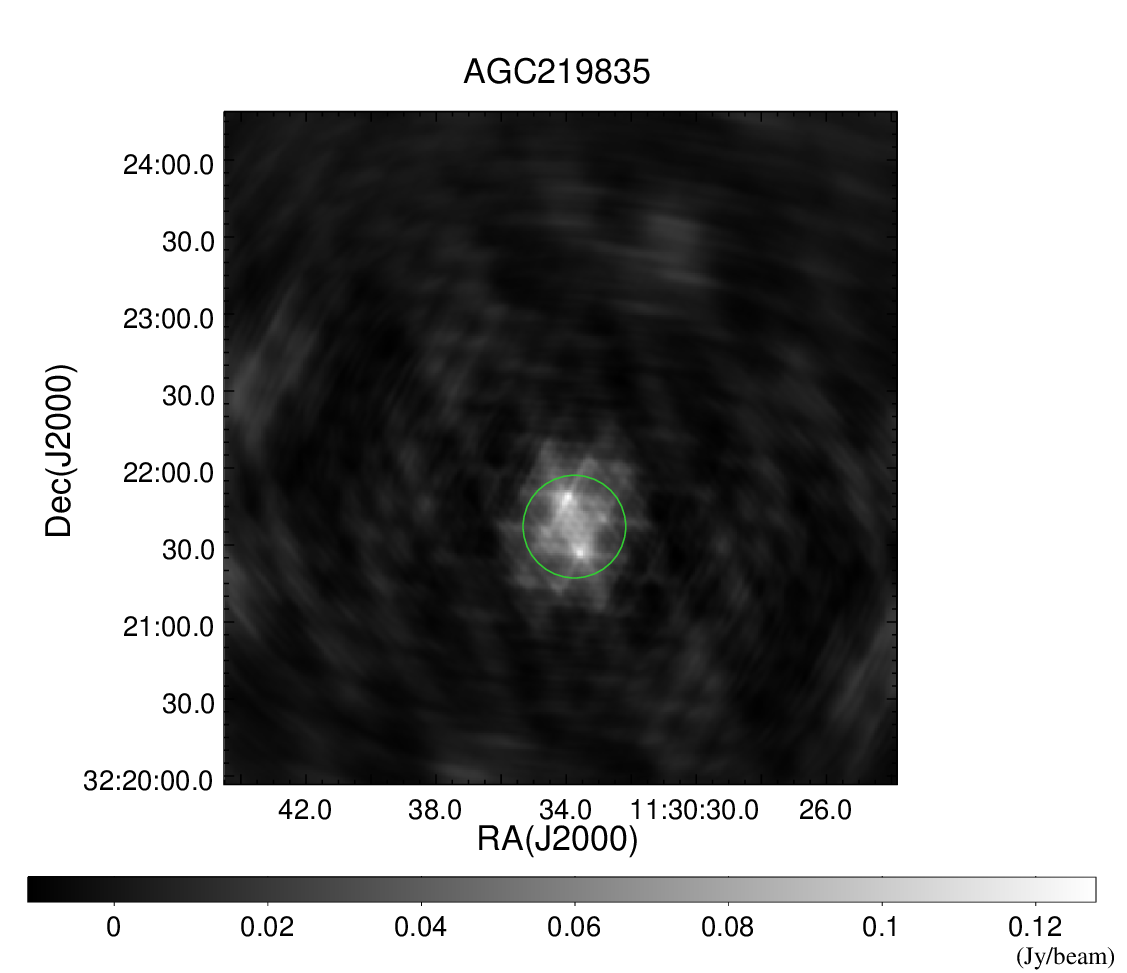}
\includegraphics[width=0.49\textwidth]{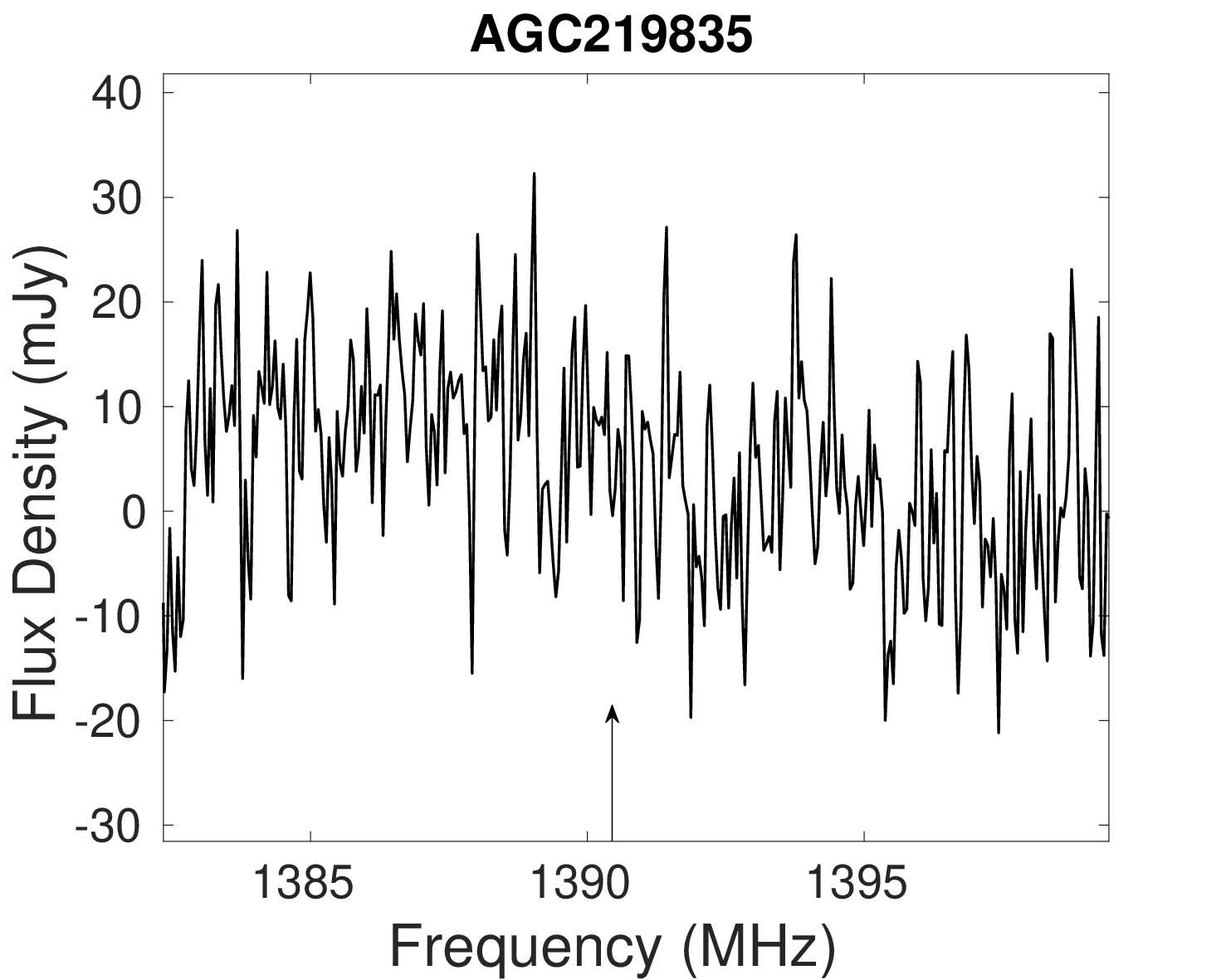}
 
      \caption{Continued. 
      }
      \label{GMRTcontinuum2}
\end{figure*}


\begin{figure*}
   \centering
     \includegraphics[width=0.5\textwidth]{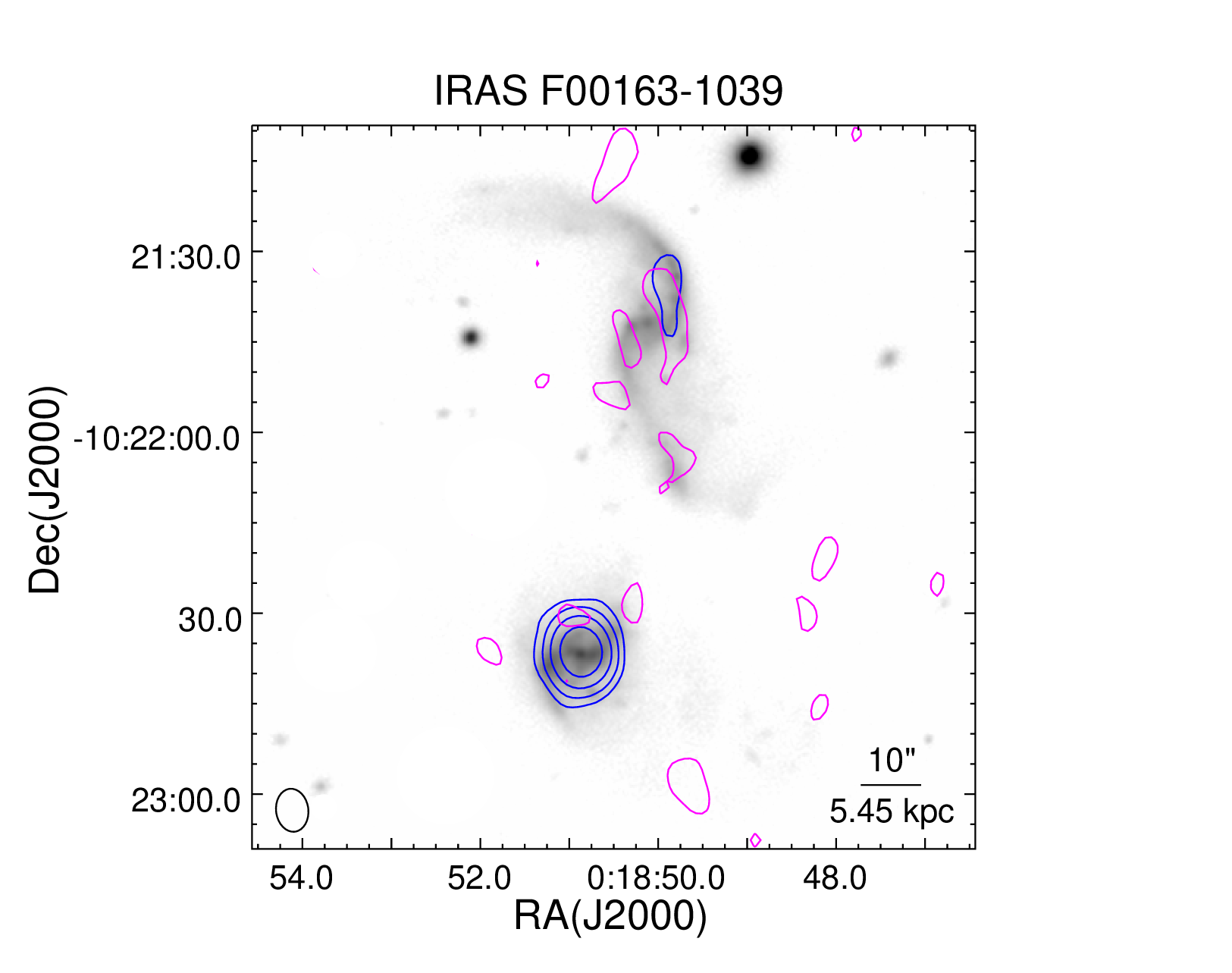}
    \includegraphics[width=0.49\textwidth]{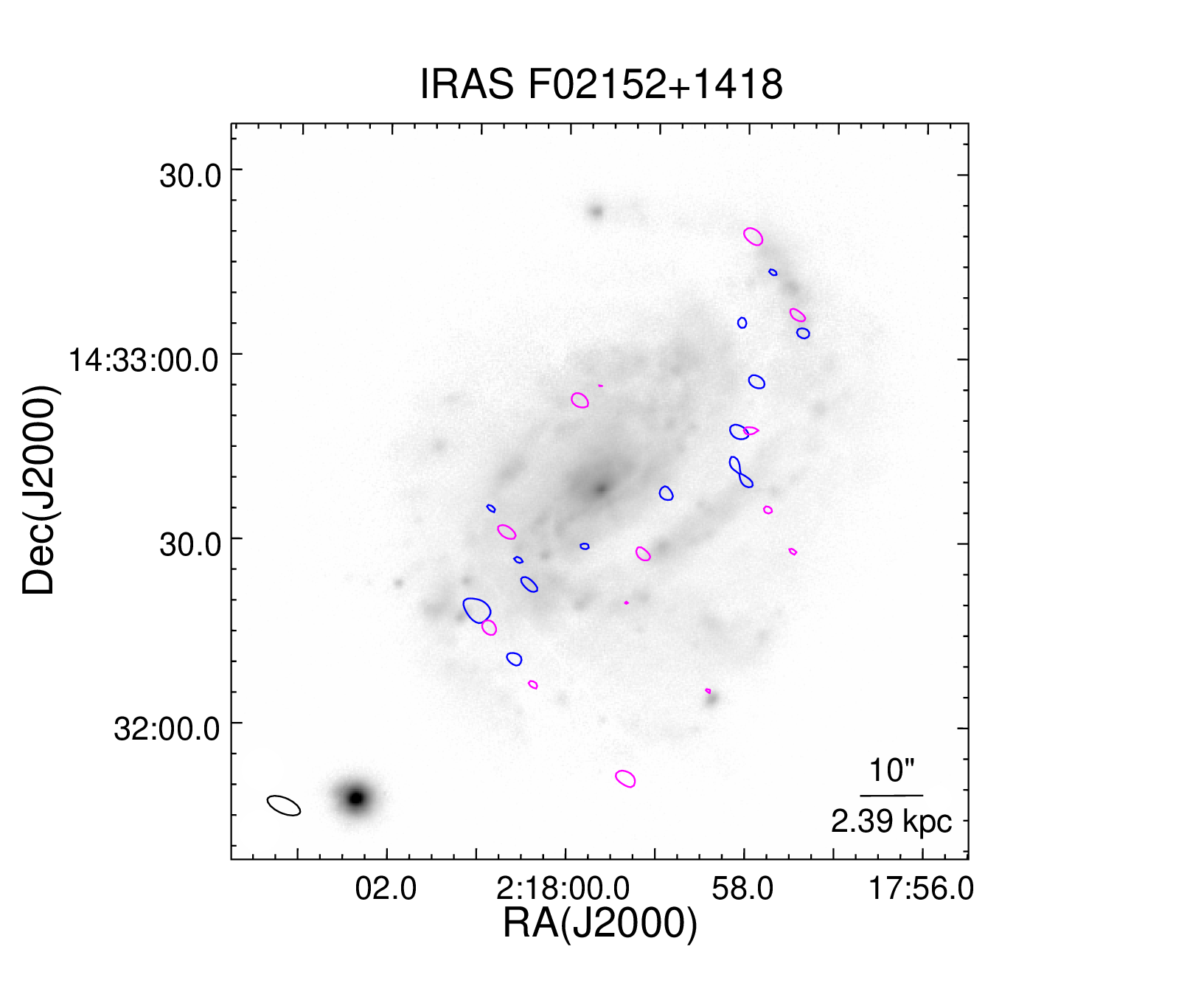}
       \includegraphics[width=0.475\textwidth]{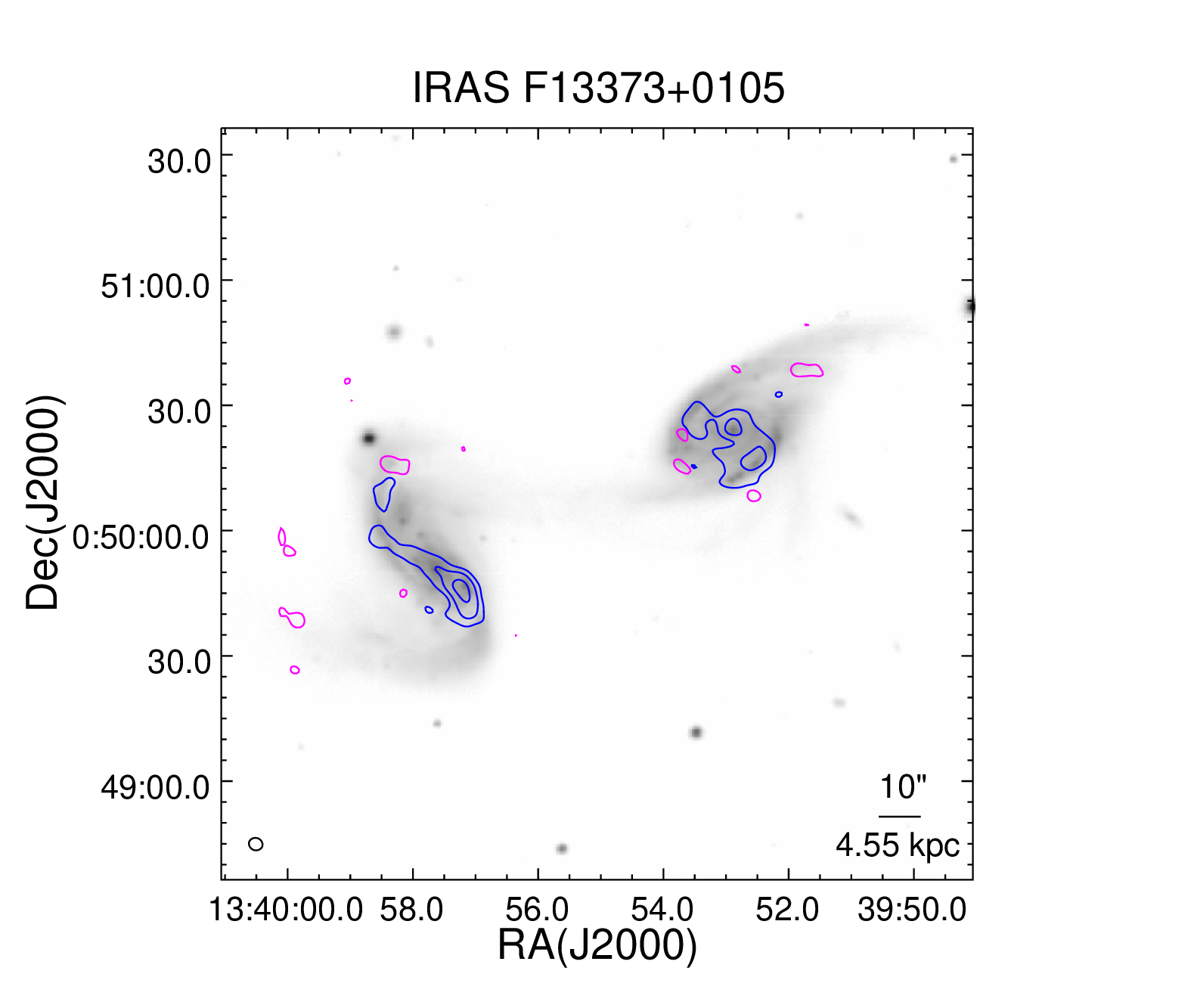}
  \includegraphics[width=0.49\textwidth]{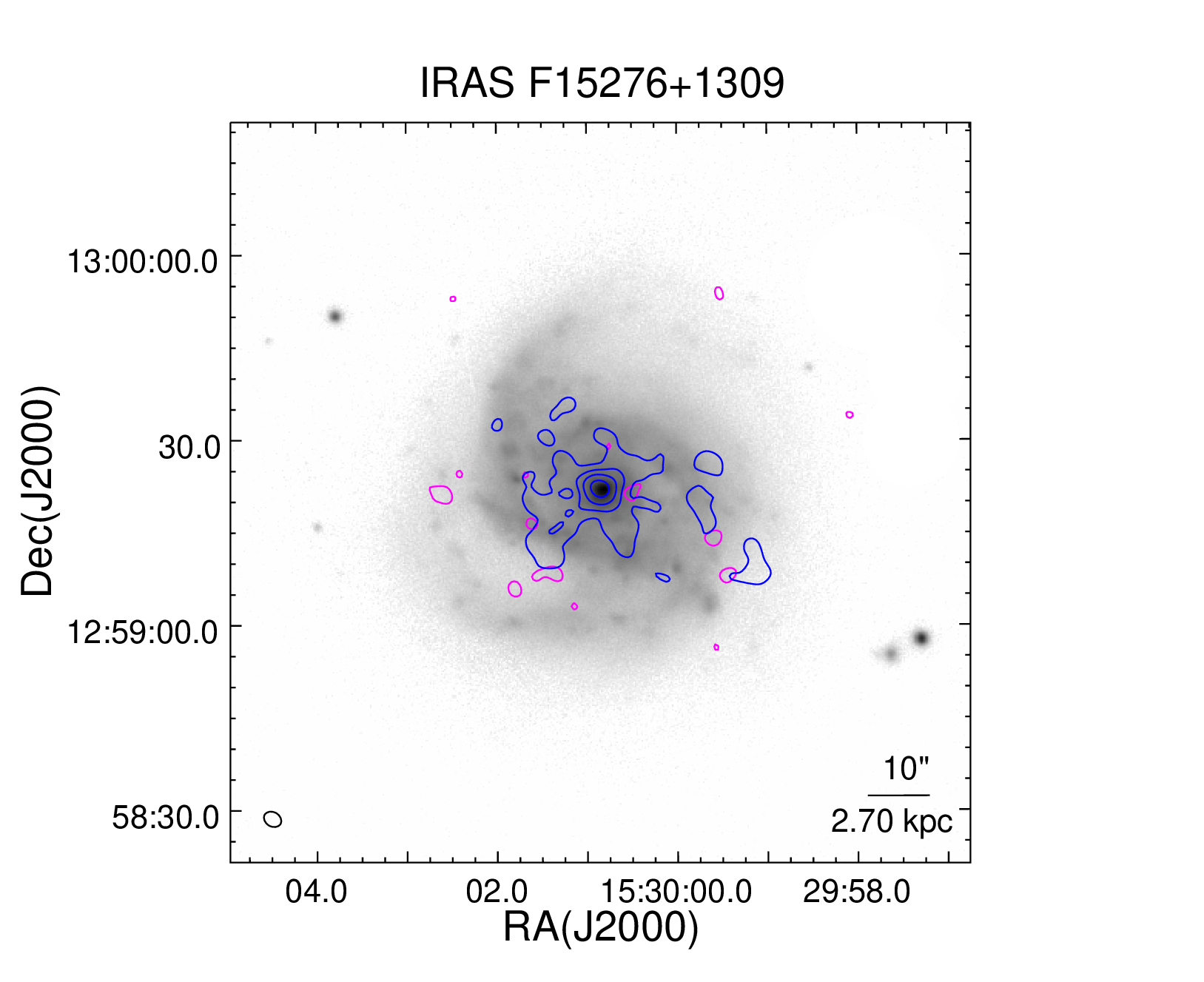}
  \includegraphics[width=0.49\textwidth]{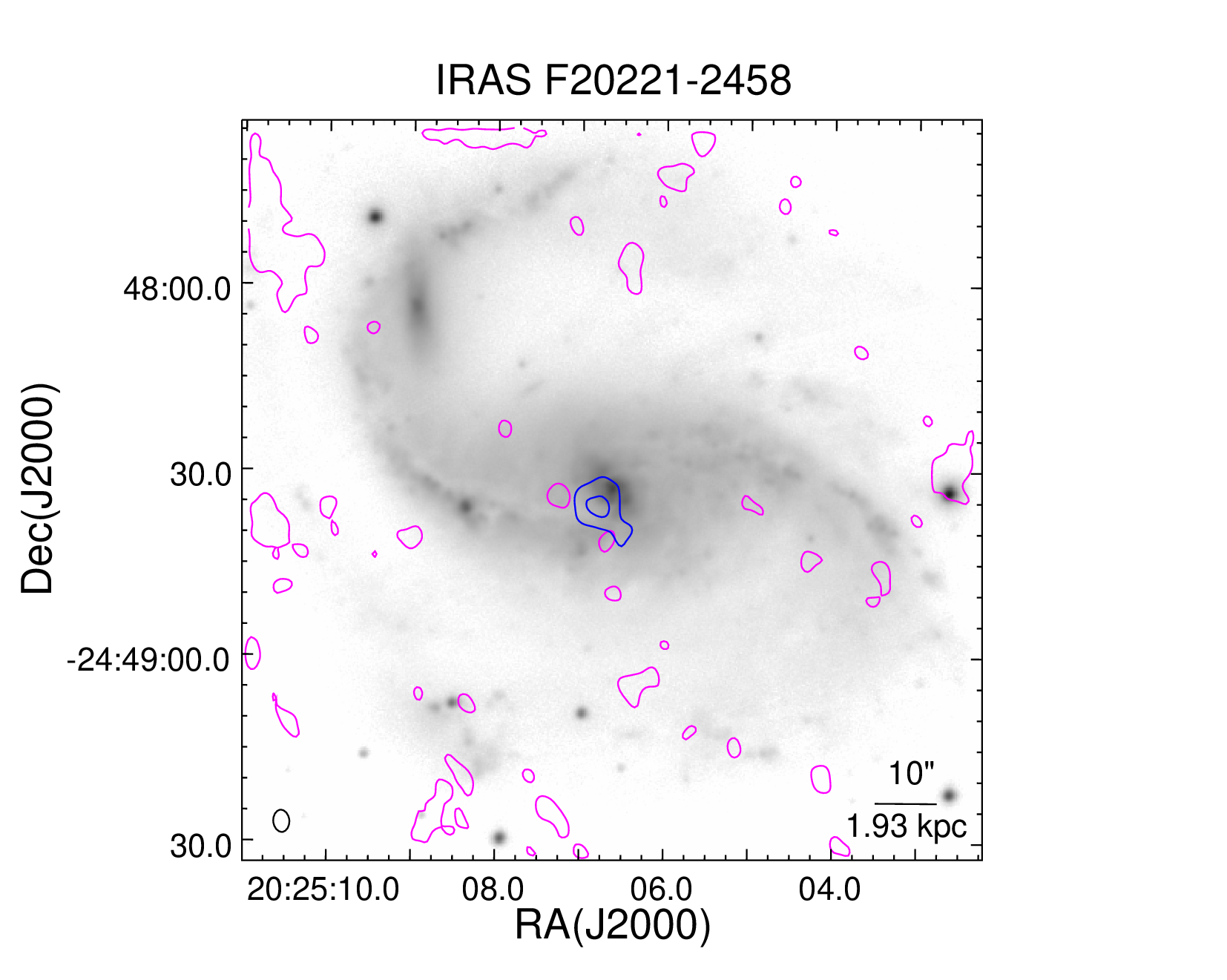}
  \includegraphics[width=0.49\textwidth]{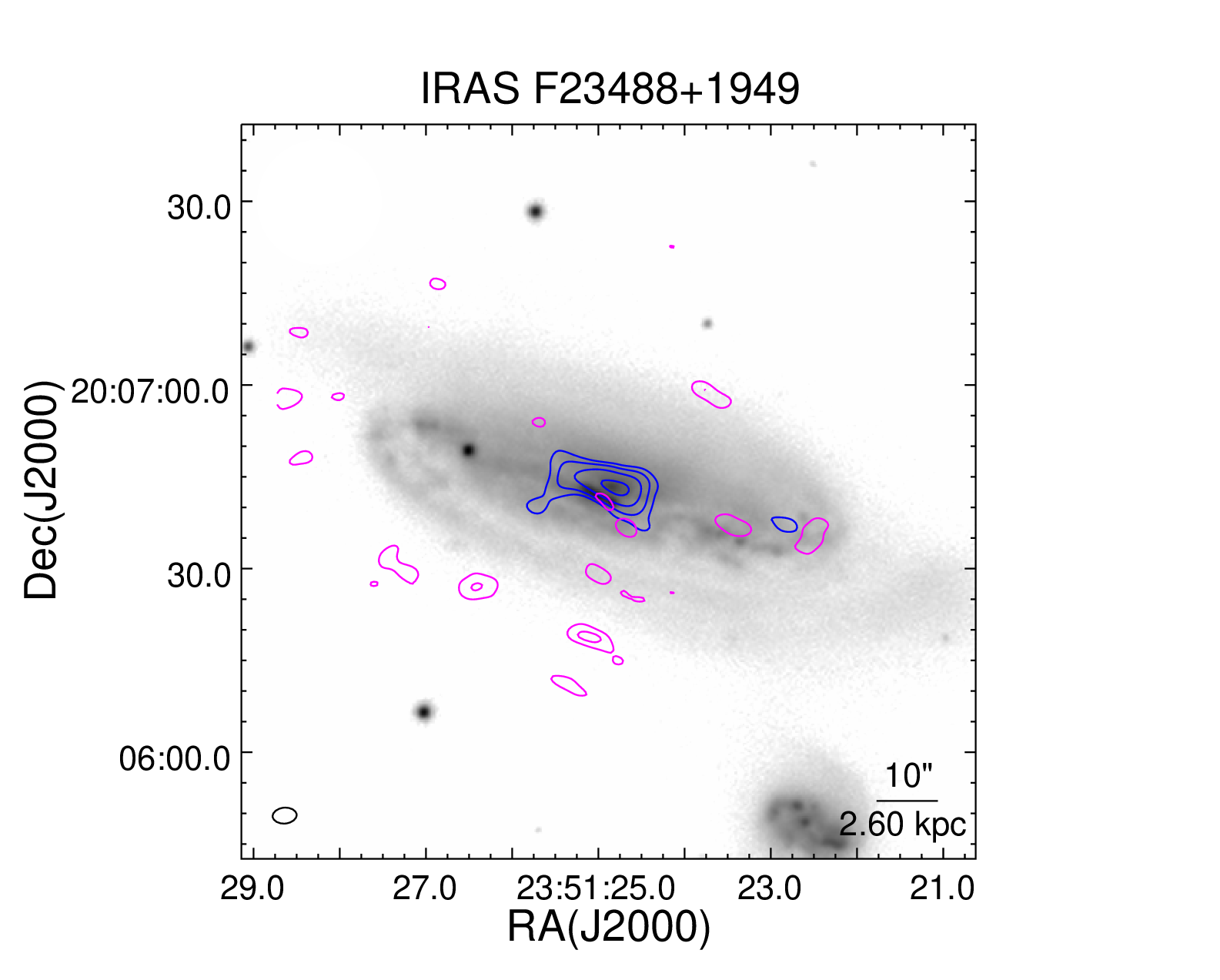}

      \caption{\HI line and radio continuum emission overlaid on an optical R-band grey image (PanSTARRS). The magenta contours represent the integrated \HI line emission in selected velocity ranges, as indicated in Fig. \ref{HIline}. IRAS F00163-1039: 7998.06-8359.56 \kms, IRAS F02152+1418: 3600-4200 \kms, IRAS F13373+0105: 6425.81-6995.61 \kms, IRAS F15276+1309: 3697.75-4277.18 \kms, IRAS F20221-2458: 3011.4-3152.88 \kms, IRAS F23488+1949: 4180.91-4626.35 \kms. The initial contour levels for these sources are at the 2 $\sigma$ level, with values of 0.368, 0.298, 0.154, 0.339, 0.439, and 0.433 mJy/beam, respectively. Only IRAS F23448+1949 has a second contour (2$\sigma$ $\times$2), while other sources have only the first contour. In blue, the emission of the L-band radio continuum is depicted, with the first contour levels set at level 3 $\sigma$ -specifically, 1.062, 0.992, 0.672, 0.765, 1.431, and 2.121 mJy/beam, respectively. The contour levels follow a pattern (1 2 4 8,...).  The ellipses at the bottom left of each panel indicate the beam size of the interferometric observations, as depicted in Table \ref{table4}.}
      
      \label{HIcontour}
\end{figure*}


\begin{figure*}
   \centering
   \includegraphics[width=0.49\textwidth]{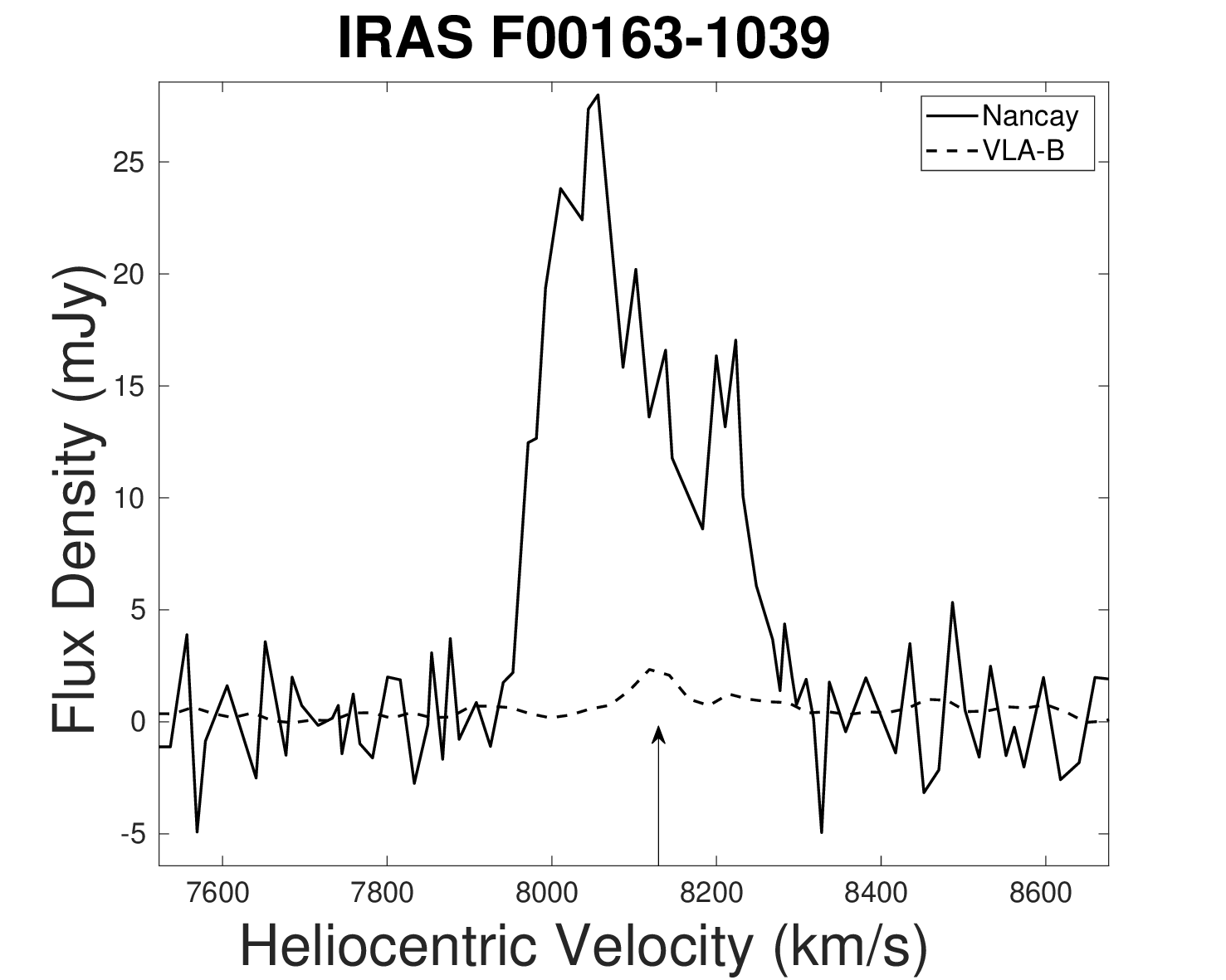}
    \includegraphics[width=0.49\textwidth]{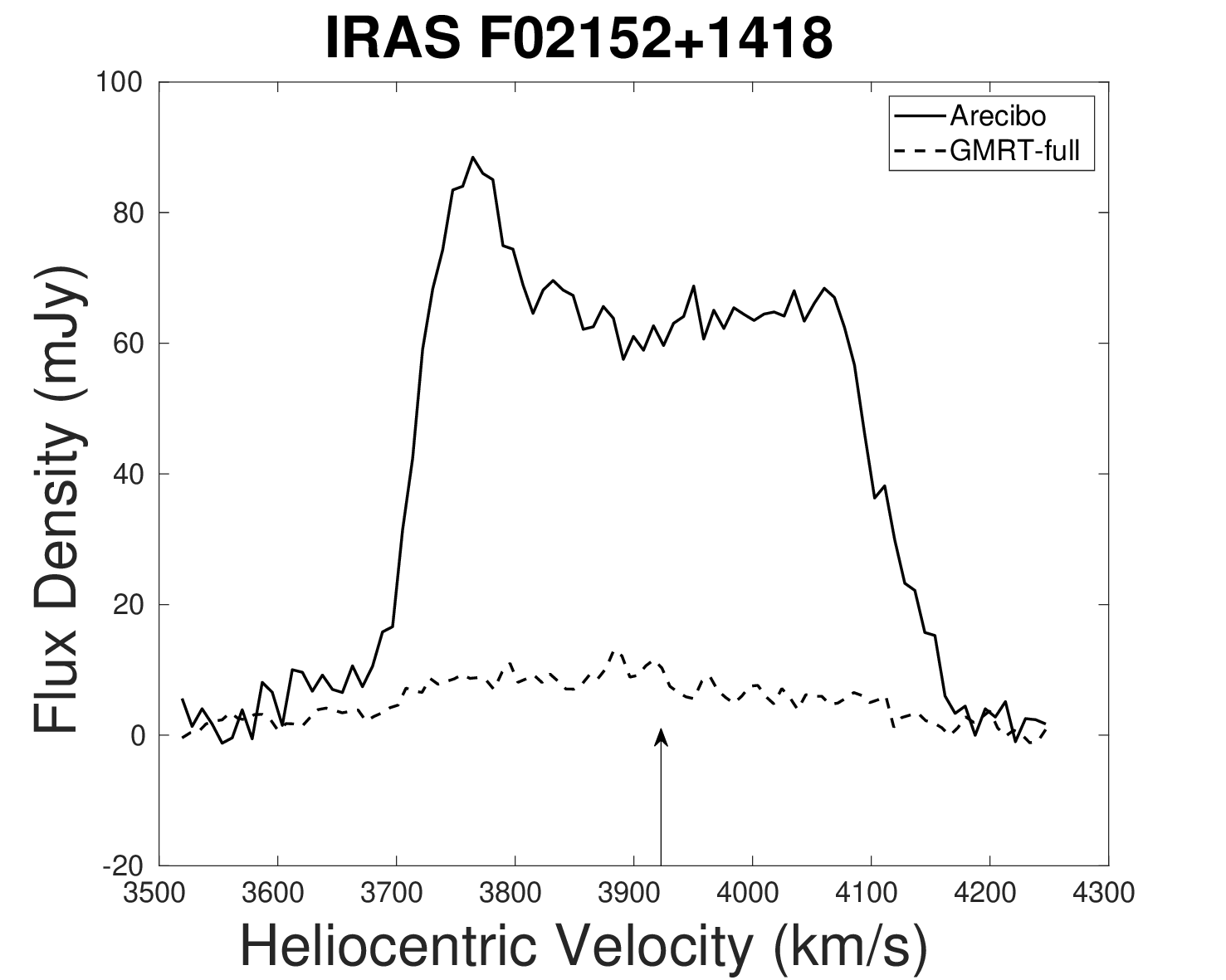}
    \includegraphics[width=0.49\textwidth]{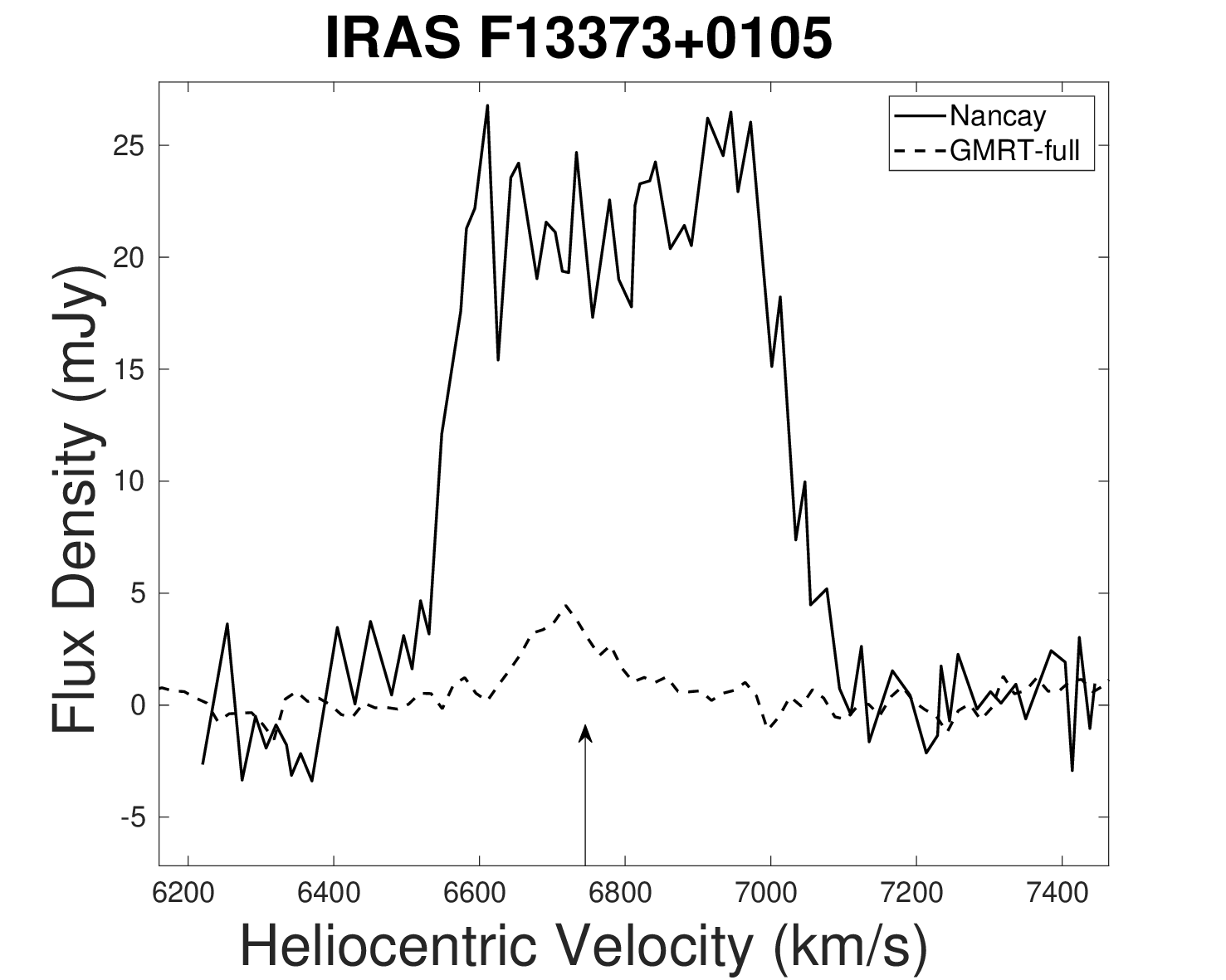}
    \includegraphics[width=0.49\textwidth]{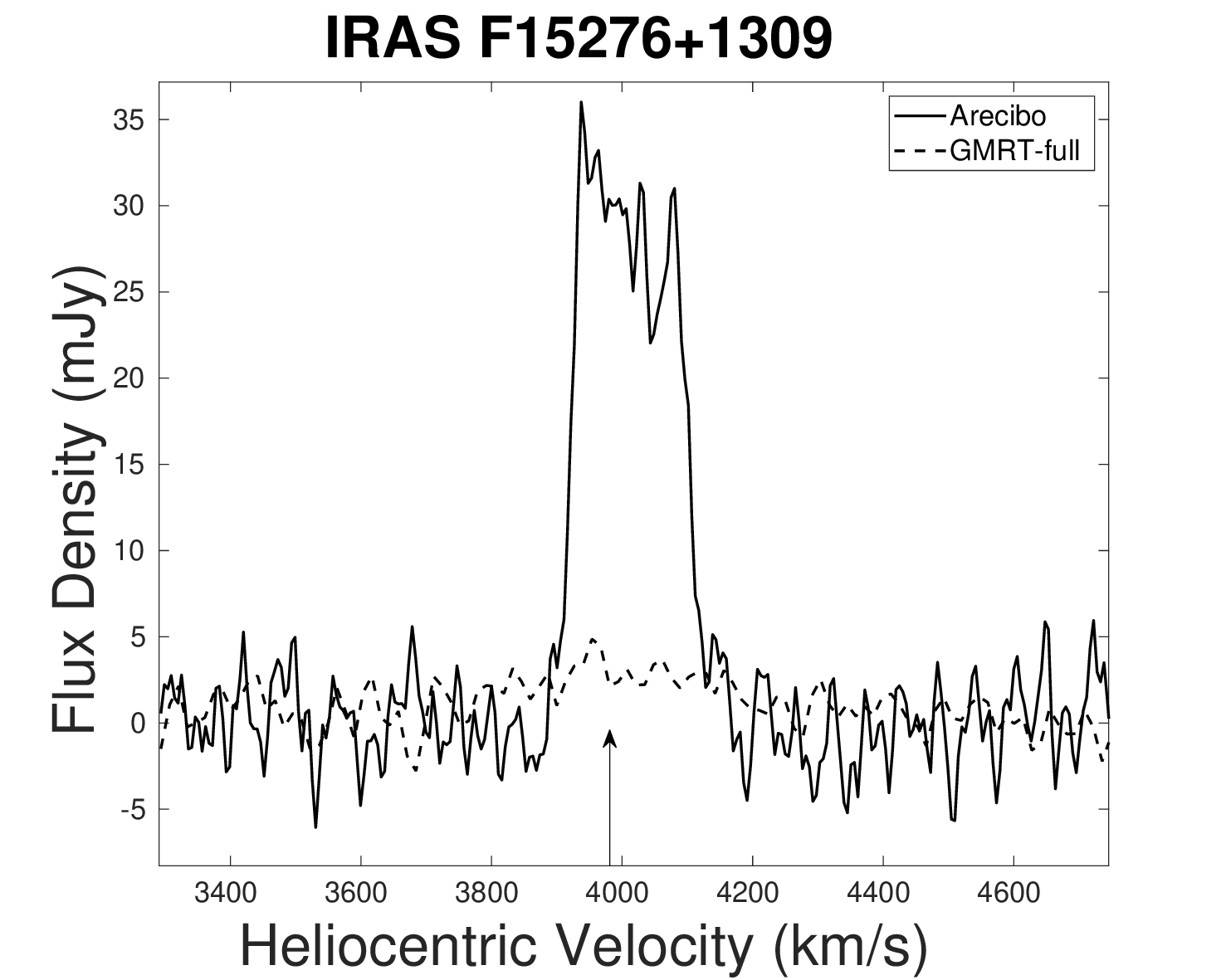}
    \includegraphics[width=0.49\textwidth]{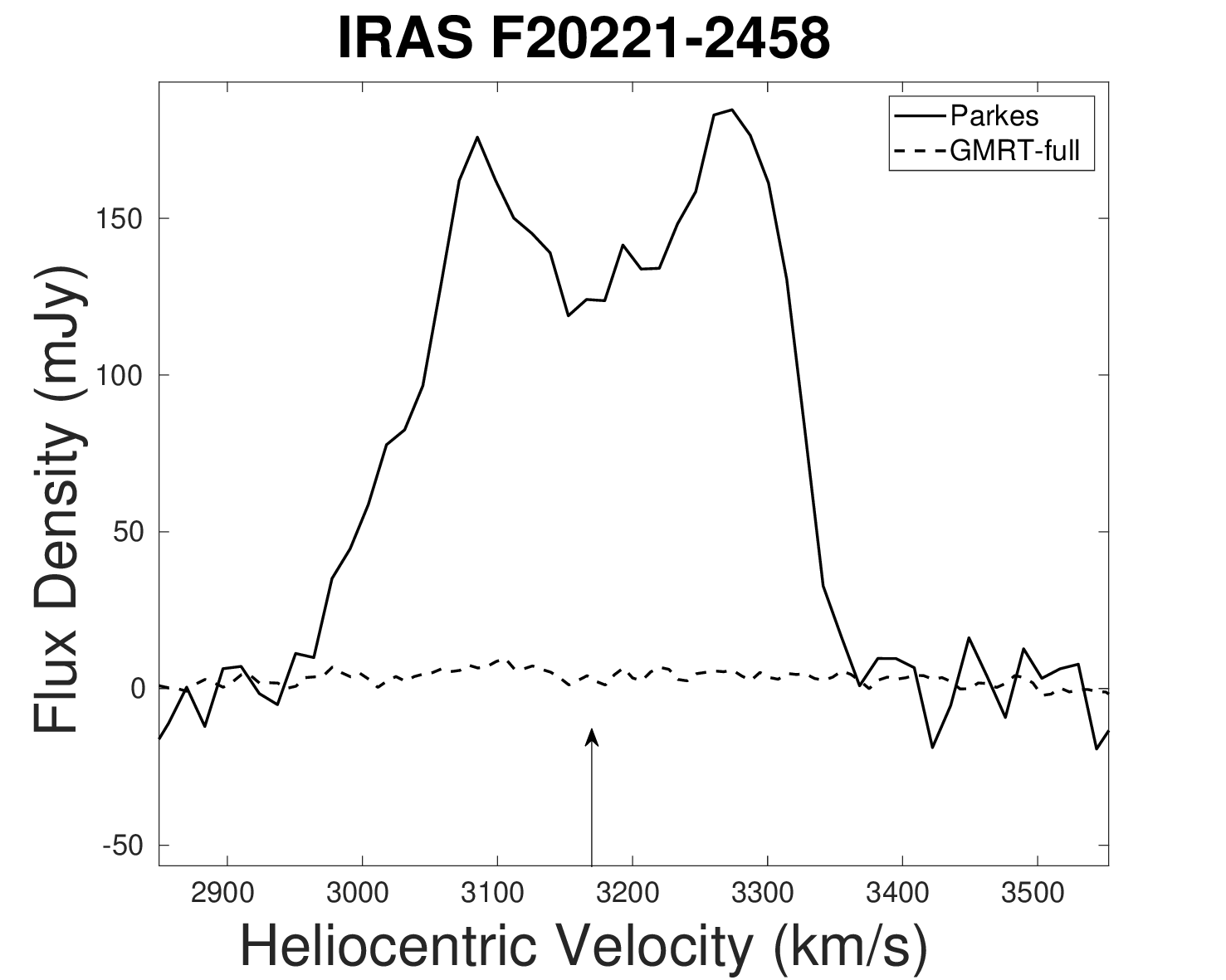}
    \includegraphics[width=0.49\textwidth]{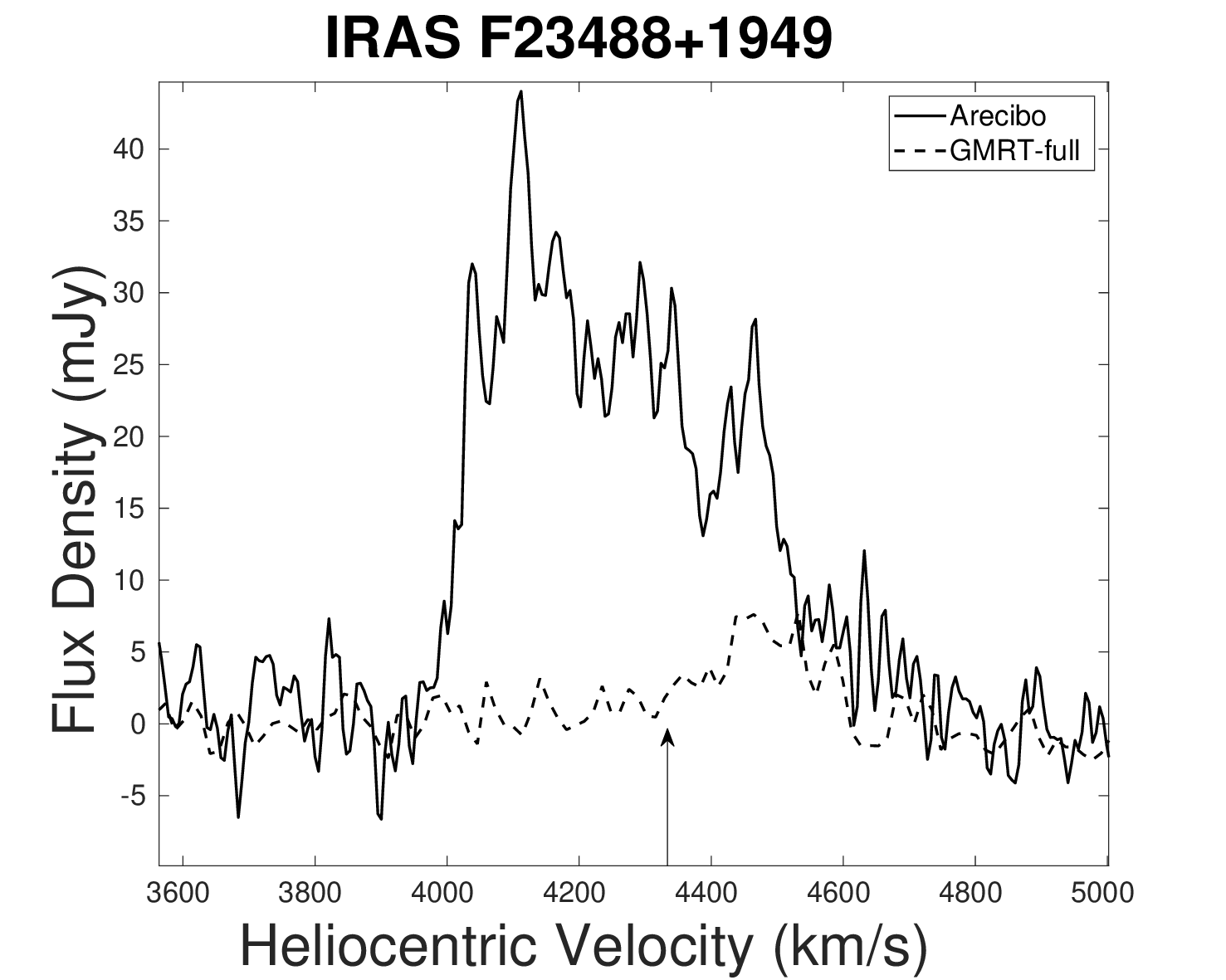}
     
      \caption{\HI line profiles of (U)LIRGs.The arrow in the plot indicates the systemic velocity. For the interferometric array, the \HI line profiles are produced within a circular region of approximately 10"$\times$10", centered around the nuclei center of these galaxies, as illustrated in Fig. \ref{HIcontour}. Exceptionally, in the cases of IRAS F00163-1039 and IRAS F13373+0105, each possessing two nuclei, the line profiles are generated from the south and southeast nuclei of the two sources, respectively. The references for the single-dish line profiles can be found in Table \ref{table4}.} 
    
      \label{HIline}
\end{figure*}


\end{appendix}

\end{document}